\documentclass[12pt,a4paper]{article}

\usepackage{epsfig,graphicx}
\usepackage{amsmath,amsfonts,amssymb}
\usepackage{citesort}

\newcommand{\unit}{\leavevmode\hbox{\small1\kern-3.6pt\normalsize1}}

\def\neumass{m_{\tilde\chi_1^0}}
\newcommand{\crosssec}{\sigma_{\tilde\chi^0_1-p}}

\def\neut{\tilde\chi_1^0}

\def\ee{e^+e^-\to h^0Z}

\def\lsim{\raise0.3ex\hbox{$\;<$\kern-0.75em\raise-1.1ex\hbox{$\sim\;$}}}
\def\gsim{\raise0.3ex\hbox{$\;>$\kern-0.75em\raise-1.1ex\hbox{$\sim\;$}}}
\def\nmh{{\sc nmhdecay}}

\newcommand{\captions}{\sf\caption}

\begin{document}

\thispagestyle{empty}
\begin{flushright}
DESY 04-129\\
IFIC/04-44\\
FTUAM 04/17\\
IFT-UAM/CSIC-04-42\\
hep-ph/yymmddd\\
\vspace*{5mm}{August 7, 2004}
\end{flushright}

\begin{center}
{\Large \textbf{ Theoretical predictions for the direct detection of
    neutralino dark matter in the NMSSM
} }

\vspace{0.5cm}
\hspace*{-1mm}D.~G.~Cerde\~no $^{a}$,  C.~Hugonie $^{b}$, D.E.~L\'opez-Fogliani
$^{c}$, C.~Mu\~noz $^{c}$ and A.~M.~Teixeira $^{c}$\\[0.2cm]
{$^{a}$\textit{II. Institut f\"ur Theoretische Physik, Universit\"at Hamburg\\
Luruper Chaussee 149, D-22761 Hamburg, Germany}} \\[0pt]
{$^{b}$\textit{AHEP Group, Instituto de F\'\i sica Corpuscular -
CSIC/Universitat de Val\`encia\\
Edificio Institutos de Investigaci\'on, Apartado de Correos 22085,\\
E-46071 Val\`encia, Spain}}\\[0pt]
{$^{c}$\textit{Departamento de F\'{\i}sica Te\'{o}rica C-XI and Instituto
de F\'{\i}sica Te\'{o}rica C-XVI, Universidad Aut\'{o}noma de Madrid,
Cantoblanco, E-28049 Madrid, Spain}}\\[0pt]

\begin{abstract}
We analyse the direct detection of neutralino dark matter in the framework of
the Next-to-Minimal Supersymmetric Standard Model.
After performing a detailed
analysis of the parameter space, taking into account all the available
constraints from LEPII, we compute the
neutralino-nucleon cross section, and compare the results with the
sensitivity of detectors. We find that sizable values for the
detection cross section, within the reach of dark matter detectors,
are attainable in this framework.
For example,
neutralino-proton cross sections compatible with the sensitivity of present
experiments can be obtained
due to the exchange of very light Higgses with
 $m_{h_1^0}\lsim 70$ GeV. 
Such Higgses have 
a significant singlet composition, thus escaping detection and being in
agreement with accelerator data. 
The lightest neutralino in these cases exhibits a large
singlino-Higgsino composition, and a mass in the range
 $50\lsim \neumass\lsim 100$ GeV.

\end{abstract}
\end{center}

\vspace*{15mm}\hspace*{3mm}
\newpage

\section{Introduction}\label{intro}

One of the most important enigmas in physics is the problem of the dark matter
in the Universe. Particle physics, and in particular extensions of the standard
model (SM) offer candidates for dark matter. Among the most interesting ones
are Weakly Interacting Massive Particles (WIMPs), since these can be left over
from the Big Bang in sufficient number to account for a significant fraction of
the observed matter density.

Since 1987, impressive experimental efforts have been carried out for the
direct detection of WIMPs through elastic scattering with nuclei in a detector
\cite{lightreview}. In fact, one of the experiments, the DAMA collaboration,
reported data favouring the existence of a WIMP signal \cite{experimento1}.
Once uncertainties on the halo model are taken into account
\cite{halo,experimento1}, this signal is compatible with WIMP masses smaller
than 500-900 GeV and with
WIMP-nucleon cross sections in the range $\sigma \approx
10^{-7}- 6\times 10^{-5}$ pb.

However, this result has not been confirmed by the other
collaborations. In particular, 
CDMS Soudan \cite{soudan}, EDELWEISS \cite{edelweiss} and ZEPLIN I
\cite{ZEPLINI}
have excluded important regions of the DAMA parameter 
space\footnote{For attempts to show that DAMA and these experiments
might not be in conflict, see Ref.~\cite{conflict}.}.
In the light of these experimental results
more than 20 experiments are running or in preparation around
the world. For example, this is the case of GEDEON \cite{IGEX3}, which will be
able to explore positively a WIMP-nucleon cross section $\sigma \gsim 3\times
10^{-8}$ pb. CDMS Soudan will be able to test in the future 
$\sigma \gsim 2\times 10^{-8}$ pb, and the 
very sensitive detector GENIUS
\cite{HDMS2}, will be able to test a WIMP-nucleon cross section
$\sigma\approx 10^{-9}$ pb.
In fact, already planned detectors working with 1 tonne of Ge/Xe 
are expected to reach cross sections
as low as $10^{-10}$ pb \cite{xenon}.

Given this situation, and assuming that the dark matter is a WIMP, it is
necessary to analyse the theoretical predictions for the WIMP-nucleon cross
section. Obviously, the answer to this question depends on the particular WIMP
considered. The leading candidate in this class of particles is the lightest
neutralino, $\tilde{\chi}^0_1$, which appears in supersymmetric (SUSY)
extensions of the SM \cite{old}. The cross section for the elastic scattering
of $\tilde{\chi}^0_1$ on nucleons has been examined exhaustively in the context
of the Minimal Supersymmetric Standard Model (MSSM) \cite{lightreview}. In
particular, there are regions of the parameter space of the MSSM where the
neutralino-nucleon cross section is compatible with the sensitivity of present
(and future) dark matter detectors.

However, it is well known that the MSSM faces a naturalness problem -- the
so-called $\mu$ problem \cite{mupb} --  arising from the presence of a mass
term for the Higgs fields in the superpotential, $\mu H_1 H_2$. The only
natural values for the $\mu$ parameter are either zero or the Planck scale. The
first is experimentally excluded since it leads to an unacceptable axion once
the electroweak (EW) symmetry is broken, while the latter reintroduces the
hierarchy problem. There exist explanations for an $\mathcal{O}(M_W)$ value for
the $\mu$ term, although all in extended frameworks \cite{mupb,musol}.

The Next-to-Minimal Supersymmetric Standard Model (NMSSM)
\cite{NMSSM1,NMSSM2,NLEP,NLHC,NHIGGS,NMHDECAY,Ndirdet,bk,Nrelden,Bast,Abel1,Abel2}
provides an elegant solution to the $\mu$ problem of the MSSM via the
introduction of a singlet superfield $S$. In the simplest form of the
superpotential, which is scale invariant and contains the $S H_1 H_2$ coupling,
an effective $\mu$ term is generated when the scalar component of $S$ acquires
a vacuum expectation value (VEV) of order the SUSY breaking scale. This
effective coupling is naturally of order the EW scale if the SUSY  breaking
scale is not too large compared with $M_W$. In fact, the NMSSM is the simplest
supersymmetric extension of the standard model in which the EW scale
exclusively originates from the SUSY breaking scale. Another appealing feature
of the NMSSM is related to the ``little fine tuning problem'' of the MSSM, or
equivalently, the non-observation of a neutral CP-even Higgs boson at LEP II.
As shown in \cite{Bast}, in the context of the NMSSM the latter problem becomes
less severe. Although the symmetries of the NMSSM may give rise to the
possibility of a cosmological domain wall problem \cite{Abel1}, this can be
avoided by the introduction of suitable non-normalisable operators \cite{Abel2}
that do not generate dangerously large singlet tadpole diagrams \cite{tadp}.
These additional operators can be chosen small enough as not to alter the low
energy phenomenology.

In addition to the MSSM fields, the NMSSM contains an extra CP-even and CP-odd
neutral Higgs bosons, as well as one additional neutralino. These new fields
mix with the corresponding MSSM ones, giving rise to a richer and more complex
phenomenology \cite{NMSSM2,NLEP,NLHC,NHIGGS,NMHDECAY}. A very
light neutralino may be present \cite{NLEP}. The upper bound on the mass of the
lightest Higgs state is larger than in the MSSM \cite{NHIGGS}. Moreover, a very
light Higgs boson is not experimentally excluded \cite{NMHDECAY,NLHC}. All
these properties may modify the results concerning the neutralino-nucleon cross
section with respect to those of the MSSM.

In fact, in comparison with the MSSM, there are
only a few works in the literature studying
the direct detection of the lightest neutralino in the
NMSSM \cite{Ndirdet,bk},
as well as its relic density \cite{Nrelden}.
Thus,
given the recent experimental results concerning the detection of dark matter,
and in view of the appealing theoretical and phenomenological properties of the
NMSSM, it is important to carry out an up-to-date analysis of the
neutralino-nucleon cross section in this framework.
Moreover, the recently published
FORTRAN code \nmh\ \cite{NMHDECAY} allows a precise calculation of the particle
spectrum in the NMSSM, as well as a complete check of all the available
experimental constraints from LEP, thus enabling a thorough study of $\sigma$
in the allowed parameter space of the NMSSM.

The outline of the paper is as follows: In Section~2 we introduce the model,
discussing in particular its Higgs potential, Higgs and
neutralino mass matrices, and the parameter space. In
Section~3 we examine the relevant effective Lagrangian describing the elastic
$\tilde{\chi}^0_1$-nucleon scattering and its associated cross section.
Section~4 is devoted to the presentation of the results for the
$\tilde{\chi}^0_1$-nucleon cross section in the NMSSM, taking into account the
relevant constraints on the parameter space from accelerator data. Our
conclusions are given in Section~5.

\section{Overview of the NMSSM}\label{nmssm}

In this Section, we review some important features of the NMSSM. In
particular, we discuss the Higgs and neutralino sectors of the
model, presenting the tree-level mass matrices and mixings which are
relevant for our analysis.
We also discuss the theoretical and experimental constraints, and how
these are reflected in the parameter space.

\subsection{Higgs scalar potential}
In addition to the MSSM Yukawa couplings for quarks and leptons, the
NMSSM superpotential contains two additional terms involving the Higgs doublet
superfields, $H_1$ and $H_2$, and the new superfield $S$, a singlet under the
SM gauge group $SU(3)_c \times SU(2)_L \times U(1)_Y$,
\begin{equation}\label{2:Wnmssm}
W=
\epsilon_{ij} \left(
Y_u \, H_2^j\, Q^i \, u +
Y_d \, H_1^i\, Q^j \, d +
Y_e \, H_1^i\, L^j \, e \right)
- \epsilon_{ij} \lambda \,S \,H_1^i H_2^j +\frac{1}{3} \kappa S^3\,,
\end{equation}
where we take $H_1^T=(H_1^0, H_1^-)$, $H_2^T=(H_2^+, H_2^0)$, $i,j$ are
$SU(2)$ indices, and $\epsilon_{12}=1$. In this model, the usual MSSM bilinear
$\mu$ term is absent from the superpotential, and only dimensionless trilinear
couplings are present in $W$. However, when the scalar component of $S$ acquires
a VEV, an effective interaction $\mu H_1 H_2$ is generated, with $\mu \equiv
\lambda \langle S \rangle$.

As mentioned in the Introduction, the superpotential in Eq.~(\ref{2:Wnmssm})
is scale invariant, and the EW scale will only appear through the soft SUSY
breaking terms in $\mathcal{L}_{\text{soft}}$, which in our conventions is
given by
\begin{align}\label{2:Vsoft}
-\mathcal{L}_{\text{soft}}=&\,
 {m^2_{\tilde{Q}}} \, \tilde{Q}^* \, \tilde{Q}
+{m^2_{\tilde{U}}} \, \tilde{u}^* \, \tilde{u}
+{m^2_{\tilde{D}}} \, \tilde{d}^* \, \tilde{d}
+{m^2_{\tilde{L}}} \, \tilde{L}^* \, \tilde{L}
+{m^2_{\tilde{E}}} \, \tilde{e}^* \, \tilde{e}
 \nonumber \\
&
+m_{H_1}^2 \,H_1^*\,H_1 + m_{H_2}^2 \,H_2^* H_2 +
m_{S}^2 \,S^* S \nonumber \\
&
+\epsilon_{ij}\, \left(
A_u \, Y_u \, H_2^j \, \tilde{Q}^i \, \tilde{u} +
A_d \, Y_d \, H_1^i \, \tilde{Q}^j \, \tilde{d} +
A_e \, Y_e \, H_1^i \, \tilde{L}^j \, \tilde{e} + \text{H.c.}
\right) \nonumber \\
&
+ \left( -\epsilon_{ij} \lambda\, A_\lambda S H_1^i H_2^j +
\frac{1}{3} \kappa \,A_\kappa\,S^3 + \text{H.c.} \right)\nonumber \\
&
- \frac{1}{2}\, \left(M_3\, \lambda_3\, \lambda_3+M_2\, \lambda_2\, \lambda_2
+M_1\, \lambda_1 \, \lambda_1 + \text{H.c.} \right) \,.
\end{align}
In our subsequent analysis we assume that the soft breaking parameters are free
at the EW scale. In addition to terms from $\mathcal{L}_{\text{soft}}$, the
tree-level scalar Higgs potential receives the usual $D$ and $F$ term
contributions:
\begin{align}\label{2:Vfd}
V_D = & \, \,\frac{g_1^2+g_2^2}{8} \left( |H_1|^2 - |H_2|^2 \right)^2 +
\frac{g_2^2}{2} |H_1^\dagger H_2|^2 \, , \nonumber \\
V_F = & \, \,|\lambda|^2
\left( |H_1|^2 |S|^2 + |H_2|^2 |S|^2 + |\epsilon_{ij} H_1^i H_2^j|^2 \right)
+ |\kappa|^2 |S|^4
\nonumber \\
&
-\left( \epsilon_{ij} \lambda \kappa^* H_1^{i} H_2^{j}S^{*2} + \mathrm{H.c.}
\right) \,.
\end{align}

\subsection{Minimization of the tree level scalar potential}
\label{Sec:minimization}
Once the EW symmetry is spontaneously broken, the neutral Higgs scalars develop
the following VEV's:
\begin{equation}\label{2:vevs}
\langle H_1^0 \rangle = v_1 \, , \quad
\langle H_2^0 \rangle = v_2 \, , \quad
\langle S \rangle = s \,.
\end{equation}
One has to ensure the absence of non-vanishing VEV's for the charged Higgs
fields, which would induce the appearance of charge breaking minima. By means
of an $SU(2)_L \times U(1)_Y$ transformation, one can take, without loss of
generality, $\langle H_2^+ \rangle = 0$ while $\langle H_2^0 \rangle = v_2 \in
\mathbb{R}^+$. The condition to have $v_1^- = \langle H_1^- \rangle =0$ as a
global minimum is quite involved; still, imposing that $v_1^- =0$ is a local
minimum is equivalent to requiring
that the physical charged Higgses have positive mass squared.

Bringing together all the terms in Eqs.~(\ref{2:Vsoft}) and (\ref{2:Vfd}), we
then obtain for the tree-level neutral Higgs potential:
\begin{align}\label{2:pot}
\langle V_{\mathrm{neutral}}^{\mathrm{Higgs}} \rangle =
&\,\,
\frac{g_1^2+g_2^2}{8} \left( |v_1|^2 - |v_2|^2 \right)^2
\nonumber \\
&+
|\lambda|^2 \left(
|s|^2 |v_1|^2 + |s|^2 |v_2|^2 + |v_1|^2 |v_2|^2 \right) +
|\kappa|^2 |s|^4 \nonumber \\
& + m_{H_1}^2 |v_1|^2 + m_{H_2}^2 |v_2|^2 + m_S^2 |s|^2
\nonumber \\
& + \left(-\lambda \kappa^* v_1 v_2 s^{*2}
- \lambda A_\lambda s v_1 v_2 + \frac{1}{3}
\kappa A_\kappa s^3 + \mathrm{H.c.}
\right) \,.
\end{align}
In the following, we assume that $\lambda, \kappa$, as well as the soft SUSY
breaking terms are real. This implies the absence of explicit CP violation in
the scalar sector. Although $v_1$ and $s$ can be complex parameters, the global
$\mathbb{Z}_3$ symmetry exhibited by the superpotential implies that
CP-violating extrema of $V^{\mathrm{Higgs}}_{\mathrm{neutral}}$ are maxima
rather than minima~\cite{Romao:jy}.
In principle, $\lambda$, $\kappa$, and the trilinear soft-breaking
terms, $A_\lambda$ and $A_\kappa$, in Eq.~(\ref{2:pot}) can have both
signs.

Ensuring that the tree-level potential has a minimum with respect to
the phases of the VEV's
directly excludes some combinations of signs for
the parameters.
After conducting this analysis, and
given that the potential is invariant under the symmetries
$\lambda,\,\kappa,\,s\to-\lambda,\,-\kappa,\,-s$ and
$\lambda,\,v_1\to-\lambda,\,-v_1$, we adopt, without loss of
generality, the sign convention where both $\lambda$ and $v_1$
are positive.
We then have
only positive values for $\lambda$ and $\tan \beta$, while
$\kappa$ and $\mu\, (=\lambda s)$, as well as $A_\lambda$ and
$A_\kappa$, can have both signs.

In what follows, we summarise the conditions for
$\kappa$, $A_\lambda$, $A_\kappa$ and $\mu\, (=\lambda s)$ obtained from
the minimization of the potential with respect to the phases of the VEV's.
In particular, for $\kappa > 0$,
one can analytically show that
minima
of $V^{\mathrm{Higgs}}_{\mathrm{neutral}}$
may be obtained for the following three combinations of signs,
provided that in each case the corresponding conditions are fulfilled,
\begin{itemize}
\item[(i)] $\mathrm{sign}(s)=\mathrm{sign}(A_\lambda)=
  -\mathrm{sign}(A_\kappa)$,

  which always leads to a minimum with respect to the phases.

\item[(ii)] $\mathrm{sign}(s)=-\mathrm{sign}(A_\lambda)=
  -\mathrm{sign}(A_\kappa)$,

  with $|A_\kappa| > 3 \lambda v_1 v_2 |A_\lambda|/(-|s
  A_\lambda| +\kappa |s^2|)$, where the denominator has to be
  positive.

\item[(iii)] $\mathrm{sign}(s)=\mathrm{sign}(A_\lambda)=
  \mathrm{sign}(A_\kappa)$,

  with $|A_\kappa| < 3 \lambda v_1 v_2 |A_\lambda|/(|s
  A_\lambda| +\kappa |s^2|)$.
\end{itemize}
Similarly, for $\kappa<0$, minima can only be obtained for the
combination
\begin{itemize}
\item[(iv)] $\mathrm{sign}(s)=\mathrm{sign}(A_\lambda)=
  \mathrm{sign}(A_\kappa)$,

  with
  $|A_\kappa| > 3 \lambda v_1 v_2
  |A_\lambda|/(|s  A_\lambda| -\kappa |s^2|)$, where the
  denominator has to be positive.
\end{itemize}
Numerically, one finds
that these tree-level conditions generally hold even after the inclusion of
higher order corrections.

One can derive three minimization conditions for the
Higgs VEV's and use them
to re-express the soft breaking Higgs masses in terms of $\lambda$, $\kappa$,
$A_\lambda$, $A_\kappa$, $v_1$, $v_2$ and $s$:
\begin{align}\label{2:minima}
m_{H_1}^2 = & -\lambda^2 \left( s^2 + v^2\sin^2\beta \right)
- \frac{1}{2} M_Z^2 \cos 2\beta
+\lambda s \tan \beta \left(\kappa s +A_\lambda \right) \,,
\nonumber \\
m_{H_2}^2 = & -\lambda^2 \left( s^2 + v^2\cos^2\beta \right)
+\frac{1}{2} M_Z^2 \cos 2\beta
+\lambda s \cot \beta \left(\kappa s +A_\lambda \right) \,,
\nonumber \\
m_{S}^2 = & -\lambda^2 v^2 - 2\kappa^2 s^2 + \lambda \kappa v^2
\sin 2\beta + \frac{\lambda A_\lambda v^2}{2s} \sin 2\beta -
\kappa A_\kappa s\,,
\end{align}
where $v^2= v_1^2+v_2^2=2 M_W^2/g_2^2$ and $\tan \beta=v_2/v_1$.

\subsection{Higgs boson mass matrices}
Subsequent to EW symmetry breaking, and after rotating away the CP-odd would-be
Goldstone boson, we are left with five neutral Higgs states and two
charged Higgs
states. Assuming
\begin{align}
H_1^0\equiv v_1+\frac{H_{1R}+i H_{1I}}{\sqrt{2}}\,,
\quad
H_2^0\equiv v_2+\frac{H_{2R}+i H_{2I}}{\sqrt{2}}\,,
\quad
S\equiv s+\frac{S_R+iS_I}{\sqrt{2}}\,,
\end{align}
among the neutral Higgses we find three CP-even states - $H_{1R}, H_{2R} ,
S_R$ and two CP-odd components, $A^0, S_I$, with $A^0$ related to the original
fields as $H_{1(2)I}= \sin \beta (\cos \beta) A^0$. Using the minimization
conditions above, the tree-level mass matrix for the neutral Higgs
bosons can be easily obtained. Since we have made the assumption that there is
no CP-violation on the Higgs sector, CP-even and CP-odd states do not mix, and
the corresponding mass matrices can be written in the respective basis,
$H^0=(H_{1R}, H_{2R}, S_R)$ and $P^0=(A^0, S_I)$. For the CP-even states, we
have
\begin{align}\label{2:Hmass}
\mathcal{M}_{S,11}^2 & = M_Z^2 \cos^2\beta +
\lambda s \tan \beta (A_\lambda+\kappa s) \nonumber \\
\mathcal{M}_{S,22}^2 & = M_Z^2 \sin^2\beta +
\lambda s \cot \beta (A_\lambda+\kappa s) \nonumber \\
\mathcal{M}_{S,33}^2 & = 4 \kappa^2 s^2 + \kappa A_\kappa s+
\frac{\lambda}{s} A_\lambda v_1 v_2 \nonumber \\
\mathcal{M}_{S,12}^2 & = \left( \lambda^2 v^2 - \frac{M_Z^2}{2} \right)
\sin 2\beta - \lambda s \left( A_\lambda+ \kappa s \right)
\nonumber \\
\mathcal{M}_{S,13}^2 & = 2 \lambda^2 v_1 s -\lambda v_2 \left( A_\lambda+
2 \kappa s \right) \nonumber \\
\mathcal{M}_{S,23}^2 & = 2 \lambda^2 v_2 s -\lambda v_1 \left( A_\lambda+
2 \kappa s \right) \,.
\end{align}
The CP-even Higgs interaction and physical eigenstates are related by the
transformation
\begin{equation}\label{2:Smatrix}
h_a^0 = S_{ab} H^0_b\,,
\end{equation}
where $S$ is the unitary matrix that diagonalises the above symmetric
mass matrix, $a,b =
1,2,3$, and the physical eigenstates are ordered as\footnote{Throughout
the paper we always adopt the convention $m_i \lesssim m_j$ for $i<j$.}
$m_{h_1^0} \lesssim m_{h_2^0} \lesssim m_{h_3^0}$. In the pseudoscalar sector,
after rewriting the CP-odd mass terms in the $P^0$ basis, the corresponding
(symmetric) mass matrix reads
\begin{align}\label{2:Amass}
\mathcal{M}_{P,11}^2 & = \frac{2 \lambda s}{\sin 2 \beta}
\left( \kappa s +A_\lambda \right) \nonumber\\
\mathcal{M}_{P,22}^2 & = \lambda \left(2 \kappa +\frac{A_\lambda}{2 s}
\right) v^2 \sin 2 \beta -3 \kappa A_\kappa s \nonumber \\
\mathcal{M}_{P,12}^2 & = \lambda v \left(A_\lambda -2 \kappa s \right)\, ,
\end{align}
and the relation between physical and interaction eigenstates is given by
\begin{equation}\label{2:Pmatrix}
a^0_i = P_{ij} P^0_j\,.
\end{equation}

Regarding the charged Higgs mass, at the tree level it is given by
\begin{equation}
m^2_{H^\pm} = M_W^2 -\lambda^2 v^2 + \lambda (A_\lambda +\kappa s)
\frac{2 s}{\sin 2 \beta}\,.
\end{equation}

\subsection{Neutralino mass matrix}

When compared to the MSSM case, the structure of chargino and squark mass terms
is essentially unaffected, provided that one uses $\mu = \lambda s$.
However, in the neutralino sector, the situation is more involved, since the
fermionic component of $S$ mixes with the neutral Higgsinos, giving rise to a
fifth neutralino state. In the weak interaction basis defined by ${\Psi^0}^T
\equiv \left(\tilde B^0=-i \lambda^\prime, \tilde W_3^0=-i \lambda_3, \tilde
H_1^0, \tilde H_2^0, \tilde S \right)\,, $ the neutralino mass terms in the
Lagrangian are
\begin{equation}
\mathcal{L}_{\mathrm{mass}}^{\tilde \chi^0} =
-\frac{1}{2} (\Psi^0)^T \mathcal{M}_{\tilde \chi^0} \Psi^0 + \mathrm{H.c.}\,,
\end{equation}
with $\mathcal{M}_{\tilde \chi^0}$ a $5 \times 5$ matrix,
{\footnotesize \begin{equation}
  \mathcal{M}_{\tilde \chi^0} = \left(
    \begin{array}{ccccc}
      M_1 & 0 & -M_Z \sin \theta_W \cos \beta &
      M_Z \sin \theta_W \sin \beta & 0 \\
      0 & M_2 & M_Z \cos \theta_W \cos \beta &
      -M_Z \cos \theta_W \sin \beta & 0 \\
      -M_Z \sin \theta_W \cos \beta &
      M_Z \cos \theta_W \cos \beta &
      0 & -\lambda s & -\lambda v_2 \\
      M_Z \sin \theta_W \sin \beta &
      -M_Z \cos \theta_W \sin \beta &
      -\lambda s &0 & -\lambda v_1 \\
      0 & 0 & -\lambda v_2 & -\lambda v_1 & 2 \kappa s
    \end{array} \right).
  \label{neumatrix}
\end{equation}}
The above matrix can be diagonalised by means of a unitary matrix $N$,
\begin{equation}
N^* \mathcal{M}_{\tilde \chi^0} N^{-1} = \operatorname{diag}
(m_{\tilde \chi^0_1}, m_{\tilde \chi^0_2}, m_{\tilde \chi^0_3},
m_{\tilde \chi^0_4}, m_{\tilde \chi^0_5})\,,
\end{equation}
where $m_{\tilde \chi^0_1}$ is the lightest
neutralino mass. Under the above assumptions, the lightest neutralino can be
expressed as the combination
\begin{equation}
\tilde \chi^0_1 = N_{11} \tilde B^0 + N_{12} \tilde W_3^0 +
N_{13} \tilde H_1^0 + N_{14} \tilde H_2^0 + N_{15} \tilde S\,.
\end{equation}
In the following, neutralinos with
$N^2_{13}+N^2_{14}>0.9$, or
$N^2_{15}>0.9$,
will be referred to as Higgsino- or singlino-like,
respectively.

\subsection{NMSSM parameter space}\label{2:parameters}

At the weak scale, the free parameters in the Higgs sector are (at tree level):
$\lambda$, $\kappa$, $m_{H_1}^2$, $m_{H_2}^2$,  $m_{S}^2$, $A_\lambda$ and
$A_\kappa$.  Using the three minimization conditions of the Higgs potential
(including the dominant one- and two-loop corrections), one can eliminate the
soft Higgs masses in favour of $M_Z, \tan \beta$ and $\mu$. We thus consider as
independent parameters the following set of variables
\begin{equation}
\lambda, \, \kappa,\, \tan \beta,\, \mu,\, A_\lambda, \, A_\kappa\,.
\end{equation}
In our study, the soft scalar masses as well as the soft gaugino
soft masses $M_i$ are free parameters at the EW scale.
We scanned over the parameter space using the program \nmh\
\cite{NMHDECAY} and in what follows we overview the most relevant
aspects of the analysis.

For each point in the parameter space, one requires the
absence of Landau singularities for $\lambda$, $\kappa$, $Y_t$ and $Y_b$ below
the GUT scale. For $m_t^{\text{pole}}= 175$ GeV, this translates into
$\lambda \lsim 0.75$, $|\kappa| \lsim 0.65$, and $1.7 \lsim \tan \beta \lsim 54$.
In addition one verifies that the physical minimum is a true one,
in other words, that  it is deeper than the local unphysical minima
with $\langle H_{1,2}^0\rangle=0$ and/or $\langle S\rangle=0$.

One then
computes the scalar, pseudo-scalar and charged Higgs masses and
mixings, taking into account 1- and 2-loop radiative corrections.
The dominant 1-loop corrections to the Higgs masses originate from
top, stop, bottom and sbottom loops, and the corresponding
corrections to $m_{h_1^0}^2$ are of $\mathcal{O}(Y_{t,b}^4)$. Pure
electroweak contributions of $\mathcal{O}(g^2)$ are also taken into
account. Regarding 2-loop corrections to the effective potential, the
dominant ones are associated with top-stop loops, and only the leading
(double) logarithms are included.
The chargino and neutralino masses and mixings are computed and
the couplings of the scalar
and pseudoscalar Higgs to charginos and neutralinos are calculated.

Finally, all available experimental constraints from LEP are checked:

\noindent 1) In the neutralino sector, we check that the lightest neutralino
does not contribute excessively to the invisible width of the $Z$ boson
($\Gamma(Z \to \tilde \chi^0_1
\tilde \chi^0_1) < 1.76$ MeV \cite{LEPneu})
if $m_{\tilde \chi^0_1} < M_Z/2$, and that
$\sigma(e^+e^- \to \tilde \chi^0_1 \tilde \chi^0_i)
< 10^{-2}~{\rm pb}$ if $m_{\tilde \chi^0_1} +
m_{\tilde \chi^0_i} < 209$ GeV ($i>1$)
and $\sigma(e^+e^- \to \tilde \chi^0_i \tilde \chi^0_j) <
10^{-1}~{\rm pb}$ if
$m_{\tilde \chi^0_i} + m_{\tilde \chi^0_j} < 209$ GeV ($i,j>1$)
\cite{LEPneu2}.

\noindent 2)
In the chargino sector, we verify that the lightest chargino is
not too light ($m_{\tilde \chi^+_1} > 103.5$ GeV \cite{LEPchar}).
This leads to a
lower bound on $|\mu| \gsim 100$ GeV.

\noindent 3) In the charged Higgs sector, we impose $m_{H^+} > 78.6$ GeV
\cite{LEPHC}.

\noindent 4) In the neutral Higgs sector, we check the constraints on the
production rates (reduced couplings) $\times$ branching ratios versus the
masses, for all the CP-even states $h^0$
and CP-odd states $a^0$, in all the
channels studied at LEP \cite{LEPH}: $e^+ e^- \to h^0 Z$, independent
of the $h^0$ decay mode (IHDM);  $e^+ e^- \to h^0 Z$, dependent on the
$h^0$ decay mode (DHDM), with the Higgs decaying
via $h^0 \to b \bar b$, $h^0 \to \tau^+ \tau^-$,
$h^0 \to 2\, \text{jets}$
$h^0 \to \gamma \gamma$ and $h^0 \to \text{invisible}$;
associated production modes (APM), $e^+ e^- \to h^0 a^0$, with
$h^0 a^0 \to 4 b$'s, $h^0 a^0 \to 4 \tau$'s and
$h^0 a^0 \to a^0 a^0 a^0 \to 6 b$'s (see \cite{NMHDECAY} for a detailed
discussion).

It is worth noticing here that other available experimental bounds
such as e.g.
the $b\rightarrow s\gamma$ branching ratio,
and the current upper limit on the decay 
$B_s \rightarrow \mu^+ \mu^-$,
might also produce relevant constraints on the parameter space.
The presence of light Higgs states, in particular that of very light pseudoscalars, 
might translate into potentially large contributions to these
processes.
This issue has begun to be addressed in \cite{Gudrun} for 
the large $\tan\beta$ regime.
For instance, the branching ratio $B_s \rightarrow \mu^+ \mu^-$ scales
as $\tan^6\beta/ m_{a^0}^4$, thus producing important constraints \cite{ko}
in non-universal MSSM scenarios
associated with light pseudoscalar Higgs, which favour
large neutralino-nucleon cross sections.
These constraints become specially relevant in the
large  $\tan\beta$ regime ($\tan\beta\gsim 35$).
However, notice, that although in the NMSSM scalar and pseudoscalar
Higgs can indeed be very
light,
large neutralino-nucleon cross sections are in general obtained for
the low $\tan\beta$ regime, as we will discuss in the next sections.
It is important then to carry out a detailed analysis of these
processes.
In particular, $b$, $K$, and $B$ decays, together with the muon anomalous
magnetic moment, may play an important role in further constraining
the NMSSM parameter space. Such a study will be considered
in a forthcoming publication \cite{forthc}.

\section{Neutralino-nucleon cross section}\label{sigma}

The most general supersymmetric
low-energy effective four-fermion Lagrangian that describes the
elastic scattering of the lightest neutralino with the nucleon is given
by~\cite{Jungman:1995df,Falk:1999mq}
\begin{align}
\mathcal{L}_{\mathrm{eff}} =&
\bar{\tilde \chi}^0_1 \, \gamma^\mu \gamma_5 \,{\tilde \chi}^0_1\,
\bar q_i \,\gamma_\mu \left( \alpha_{1i} + \alpha_{2i} \gamma_5
\right) \,q_i +
\alpha_{3i} \,\bar{\tilde \chi}^0_1 \,{\tilde \chi}^0_1 \, \bar q_i\, q_i
\nonumber \\ &
+ \alpha_{4i} \,\bar{\tilde \chi}^0_1 \,\gamma_5 \,{\tilde \chi}^0_1 \,
\bar q_i \, \gamma_5 \, q_i +
\alpha_{5i} \, \bar{\tilde \chi}^0_1 \, {\tilde \chi}^0_1 \,
\bar q_i \, \gamma_5 \, q_i+
\alpha_{6i} \, \bar{\tilde \chi}^0_1 \, \gamma_5 \, {\tilde \chi}^0_1
\, \bar q_i \, q_i \,,
\end{align}
where $i=1,2$ denotes up- and down-type quarks, and the Lagrangian is
summed over the three quark generations.
In the absence of CP-violating phases,
the terms proportional to
$\alpha_5$ and $\alpha_6$ vanish. Moreover, those associated
with $\alpha_1$ and $\alpha_4$ (as well as $\alpha_5$ and $\alpha_6$,
should these be present)
are velocity-dependent, and can be safely neglected for our present
purposes. The cross section associated with the spin-dependent
coefficient ($\alpha_2$) is only non-zero if the target nucleus
has a non-vanishing spin and, contrary to case of the scalar
(spin-independent) term, adds incoherently. For the case of heavy
targets, as those used in the experiments mentioned in the
Introduction, the scalar cross section associated with $\alpha_3$ is
in general substantially
larger, and henceforth we shall focus our discussion on
the latter.
\begin{figure}[htb]
  \begin{center}
    \begin{tabular}{cc}
      \epsfig{file=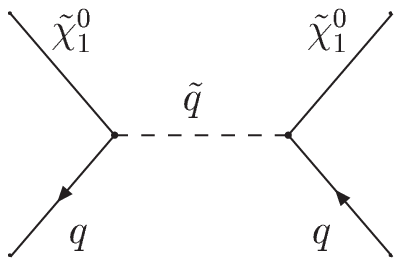,clip=}
      \hspace*{12mm}&\hspace*{12mm}
      \epsfig{file=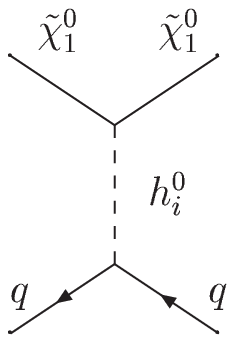,clip=}
      \\ & \\
      (a)\hspace*{8mm}&\hspace*{8mm} (b)
    \end{tabular}
    \captions{Feynman diagrams contributing to the neutralino-nucleon
      scalar cross section:
      (a) squark exchange and (b) scalar Higgs exchange.}
    \label{fig:effL:s}
  \end{center}
\end{figure}

\noindent
We begin by decomposing $\alpha_{3i}$ into two terms, one arising from
squark $s$-channel exchange and the other from the $t$-channel,
neutral Higgs mediated
interaction (Fig.~\ref{fig:effL:s} (a) and (b), respectively).
One obtains\footnote{When compared to the analogous expression of
  Ref.~\cite{bk}, we find some discrepancies in $\alpha_{3i}^{h}$,
  namely a missing singlet-Higgsino-Higgsino term (proportional to
  $\lambda$), and the presence of an additional overall weak coupling
  constant, $g$.}
\begin{align}
\alpha_{3i}^{\tilde q} &=-
\operatornamewithlimits{\sum}_{X=1}^{2}
\frac{1}{4 (m^2_{Xi} - m^2_{\tilde \chi^0_1})}
\operatorname{Re}\left[\left(
C_{R}^{Xi}\right) \left( C_{L}^{Xi}\right)^* \right]\,,
\\
\alpha_{3i}^{h} &=
\operatornamewithlimits{\sum}_{a=1}^{3}
\frac{1}{m^2_{h^0_a}} C_Y^{i} \operatorname{Re}
\left[C_{HL}^{a}\right]\,,
\end{align}
where $X=1,2$ denotes the squark mass eigenstates and
$a=1,2,3$ refers to the scalar Higgs mass eigenstates.
The relevant
NMSSM couplings for the neutralino-squark-quark ($C_{L,R}^{Xi}$),
neutralino-neutralino-Higgs ($C_{HL,R}^{a}$)
and Higgs-quark-quark ($C_Y^{i}$) interactions are given in
Appendix~\ref{app}.

The term $\alpha_{3i}^{\tilde q}$ is formally identical to the MSSM case,
differing only in the new neutralino mixings stemming from the
presence of a fifth component. In particular,
in regions of the NMSSM
parameter space where the singlino component dominates the lightest
neutralino state,
there will be a significant reduction in the Bino- and
Wino-squark-quark couplings, and hence in $\alpha_{3i}^{\tilde q}$.

Regarding the Higgs mediated interaction term
($\alpha_{3i}^{h}$), the situation is
slightly more involved since both vertices and the exchanged Higgs
scalar significantly reflect the new features of the NMSSM.
First, let us recall that in regions of
the parameter space where the lightest Higgs boson has a sizable
singlet component, the
Higgs-quark-quark coupling might be substantially reduced.
Regarding the Higgs-$\tilde \chi^0_1$-$\tilde\chi^0_1$ interaction,
in addition to a new component in the lightest $\tilde \chi^0$ state,
the most important alteration emerges from the
presence of new terms, proportional to $\lambda$ and
$\kappa$ (cf. Appendix~\ref{app}).
Nevertheless, and as already mentioned, light Higgs bosons can be
experimentally allowed in the context of the NMSSM.
Should this occur, and if these states are not pure singlets (thus
displaying a non-vanishing coupling to matter)
the exchange of light Higgs scalars in
the $t$-channel might provide a considerable enhancement to the
neutralino-nucleon cross section.

It is worth mentioning that an enhancement of
$\alpha_{3i}^{h}$ with respect to $\alpha_{3i}^{\tilde q}$
is not an effect unique to the NMSSM.
In fact, it has been already noticed that in the MSSM, and once
the mSUGRA inspired universality for the soft scalar and gaugino masses is
abandoned, the cross section associated with the channels involving
scalar Higgs exchange can be substantially enhanced. Similar to what
will occur in the present model, the MSSM $t$-channel contributions
become larger once
the Higgsino components of
$\tilde\chi^0_1$ are augmented and/or
the Higgs masses are reduced ~\cite{Rosz} (e.g. via
non-universal soft masses at the GUT scale~\cite{cermu}).

The scalar interaction term contributes to the $\tilde
\chi^0_1$-Nucleon cross section as
\begin{equation}
\sigma_{3N}=\frac{4 m^2_r}{\pi} \, f_N^2\,,
\end{equation}
where $m_r$ is the Nucleon-${\tilde \chi_1^0}$ reduced mass,
$m_r=m_N m_{\tilde \chi_1^0}/(m_N +m_{\tilde \chi_1^0})$,
and
\begin{equation}
\frac{f_N}{m_N}\,=\,
\operatornamewithlimits{\sum}_{q=u,d,s} f_{Tq}^{(N)}
\frac{\alpha_{3q}}{m_q} + \frac{2}{27}  f_{TG}^{(N)}
\operatornamewithlimits{\sum}_{q=c,b,t}
\frac{\alpha_{3q}}{m_q} \,.
\end{equation}
In the above,
$m_{q}$ is the quark mass, and the parameters
$f_{Tq}^{(N)}$ are defined as
$\langle N|m_q \bar q q | N \rangle$ = $ m_N f_{Tq}^{(N)}$.
$f_{TG}^{(N)}$ can be derived from $f_{Tq}^{(N)}$ as
$f_{TG}^{(N)}= 1- \operatornamewithlimits{\sum}_{q=u,d,s}
f_{Tq}^{(N)}$. Following~\cite{Ellis:2000ds}, we take the following
values for the hadronic matrix elements:
\begin{align}\label{3:hadronvalues}
f_{Tu}^{(p)}= 0.020 \pm 0.004\,, \ \ \ &
f_{Td}^{(p)}= 0.026 \pm 0.005\,, &
f_{Ts}^{(p)}= 0.118 \pm 0.062\,,
\nonumber \\
f_{Tu}^{(n)}= 0.014 \pm 0.003\,, \ \ \ &
f_{Td}^{(n)}= 0.036 \pm 0.008\,, &
f_{Ts}^{(n)}= 0.118 \pm 0.062\,.
\end{align}
In the numerical analysis of the next section we will use the
central values of the above matrix elements.
Notice that $f_{Ts}^{(n)}=f_{Ts}^{(p)}$ and both are much larger
than $f_{Tq}$ for $u$ and $d$ quarks, and therefore
$f_p$ and $f_n$ are
basically equal.
Thus we will focus on the neutralino-proton cross section,
\begin{equation}
\sigma_{3p}\equiv
\sigma_{\tilde \chi^0_1 -p}= \frac{4 m^2_r}{\pi} \, f_p^2\,,
\end{equation}
with $m_r=m_p m_{\tilde \chi_1^0}/(m_p +m_{\tilde \chi_1^0}) \sim m_p$.

\section{Results and discussion}\label{res}
In this Section
the viability of the detection of the lightest NMSSM neutralino
as a dark matter candidate, will be studied. In particular, we will
compute the theoretical predictions for the
direct detection of neutralinos through their elastic scattering with
nucleons inside a detector.
In our computation we will take into
account relevant constraints on the parameter space from
accelerator data.
On the other hand,
given the complexity of the computation of the relic neutralino density,
we prefer to consider in a forthcoming
publication \cite{omega} the constraints arising from reproducing
the WMAP data \cite{wmap},
$0.094\lsim \Omega_{\mbox{\tiny DM}}h^2\lsim 0.129$,
on our relevant
parameter space for the cross section.
In particular, we will see in this section that light
pseudoscalars are present in interesting regions of the parameter
space, and this
might translate into large contributions to the annihilation cross
section,
implying a reduction in the associated relic density.
Therefore we may expect the WMAP lower bound to play an important role
in those points with a very large neutralino-nucleon cross section. 
For this study we will use the general
analysis of the relic neutralino density, including coannihilations,
that has been carried out for the NMSSM in \cite{omega2}.

As discussed in the Introduction, many experiments for the direct
detection of dark matter are running or in preparation. Thus,
in our analysis, we will be particularly interested in the various
NMSSM scenarios which might potentially lead to values
of $\crosssec$ in the sensitivity range of those detectors.

Although the free parameters in our model have already been presented
in Section~\ref{2:parameters}, it is worth recalling that the Higgs
and neutralino sectors of the theory are specified by
\begin{equation}
\lambda\,, \quad \kappa \,, \quad \mu (=\lambda s) \,,
\quad \tan \beta\, ,\quad
A_\lambda\,, \quad A_\kappa\,, \quad M_1\,, \quad M_2\,.
\end{equation}
As aforementioned, we take these parameters to be free at the
EW scale. Based on an argument of
simplicity\footnote{Since in our analysis of the
neutralino-nucleon cross section the
detection channels mediated by Higgs scalars will be enhanced with respect
to those mediated by squarks, the sensitivity of the results to
variations of the squark parameters will be very small.},
the low-energy squark masses and trilinear couplings, which appear in
the computation of the neutralino-nucleon cross section,
are taken to be degenerate\footnote{
  Regarding the stop mass matrix we will work
  in the maximal-mixing regime, where the off-diagonal term
  takes the form $m_t\,X_t=m_t\,\sqrt{6}\,M_{\text{SUSY}}$.
  Departures from this case would not affect significantly the
  theoretical predictions for the neutralino-nucleon cross section.
  }.
Unless otherwise stated,
the common SUSY scale  will be
$M_{\text{SUSY}} = 1$~TeV.
Having free squark and slepton soft parameters at the EW
scale allows us to ensure that in our analysis the lightest SUSY particle
is indeed the $\tilde\chi^0_1$.
Also led by arguments of naturalness, we shall take a lower bound for
$\lambda$, $\lambda_{\mathrm{min}} \sim \mu/s_{\mathrm{max}}$. Thus,
taking the conservative  range $s \lesssim 10$ TeV,
this translates into $\lambda_{\mathrm{min}} \sim \mu({\rm GeV})
\times 10^{-4}$.

We begin our analysis by taking values for the soft gaugino masses
that mimic at low scale the results from a hypothetical unified value
at the GUT scale. Consequently,
we will choose $M_2=1$ TeV and $M_1=500$ GeV. 
For the gluino mass, the value $M_3=3$~TeV will be taken.
Later on we will
address variations of these values.
In the following we take $|\mu| \geq 110$ GeV,
since in most cases this allows to safely avoid the LEP bound on the
lightest chargino mass.
Throughout this Section, we shall consider
several choices for
the values of $A_\lambda$, $A_\kappa$, $\mu$ and $\tan\beta$,
and for each case, we will study the associated phenomenology.

In order to simultaneously analyse the dark matter predictions and
understand the effect of the experimental constraints on
the parameter space of the NMSSM, it is very illustrative to begin
our study in the plane generated by the Higgs couplings in
the superpotential, $\lambda$ and $\kappa$.
In Section\,\ref{Sec:minimization} we commented on the conditions to
be applied to each of the sign combinations of the
parameters, which
arise from ensuring that the tree-level potential has
a minimum with respect to the phases of the VEV's. In the following it
will be clarifying to
discuss each case separately.
Let us first consider the cases associated with positive values of $\kappa$.

\subsection{$\mu A_\kappa<0$ and $\mu A_\lambda>0$ $(\kappa>0)$}
\label{++-}

As a first
choice, we will consider the two cases where
$\mu A_\kappa<0$ and $\mu A_\lambda>0$, namely
those with $\mu,\,A_\lambda,\,-A_\kappa>0$ and
$\mu,\,A_\lambda,\,-A_\kappa<0$.

In both cases, part of the parameter space can be excluded due to the
occurrence of tachyons in the CP-even Higgs sector.
Namely, it is easy
to see from the expression of the CP-even Higgs matrix (\ref{2:Hmass})
that the off-diagonal terms $|{\cal M}^2_{S,13}|$ or $|{\cal M}^2_{S,23}|$ can
become significantly bigger than ${\cal M}^2_{S,33}$, thus leading to the
appearance of a negative eigenvalue. This will typically happen for
moderate to large values of $\lambda$ and small $\kappa$, for which
$m_{h^0_1}$ is small. Large values of
$|A_\kappa|$ and $\tan\beta$ lead to an increase of the
tachyonic region, as we will later see.
On the other hand, the eigenvalues of the CP-odd Higgs mass matrix
are never negative.
The CP-odd Higgs masses also decrease for
large $\lambda$ and small $\kappa$, but their
minimum value is bounded by the appearance of tachyons in the CP-even
sector.

\begin{figure}
  \epsfig{file=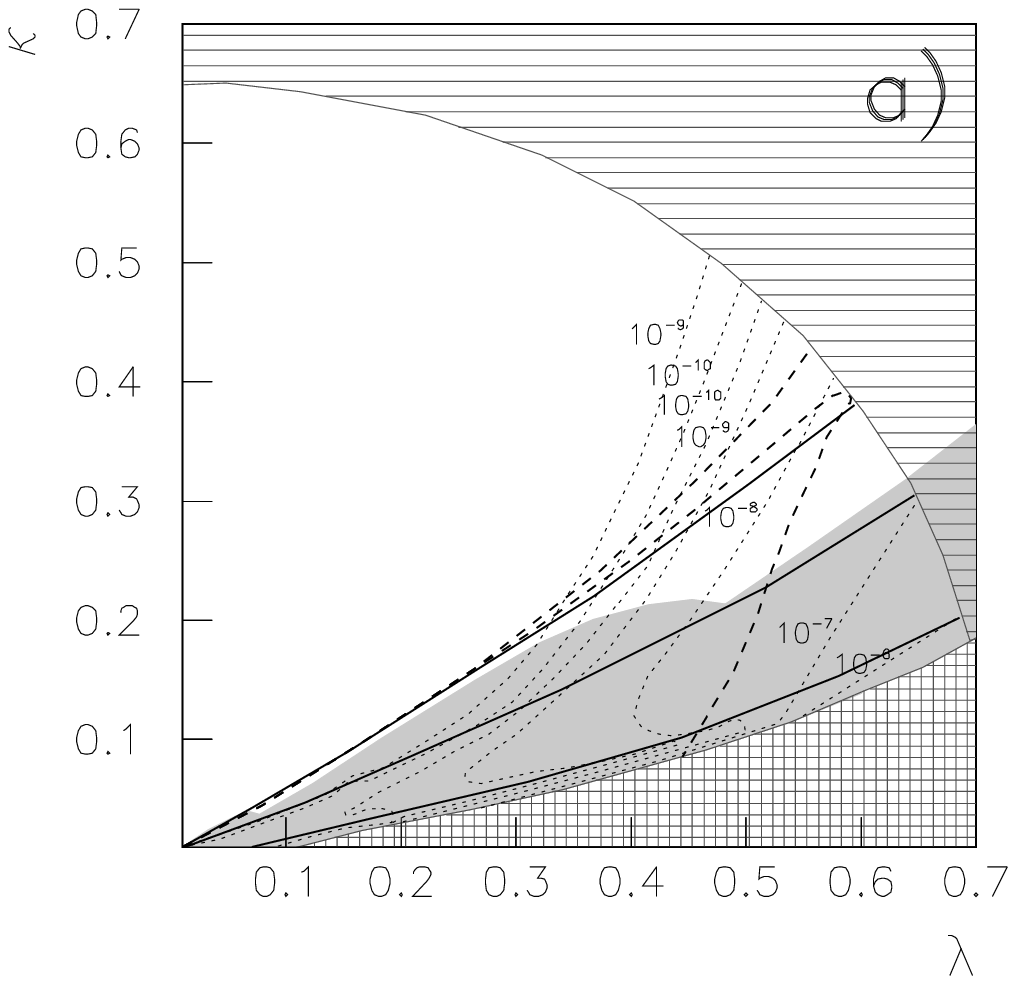,height=8cm}
  \hspace*{-1cm}\epsfig{file=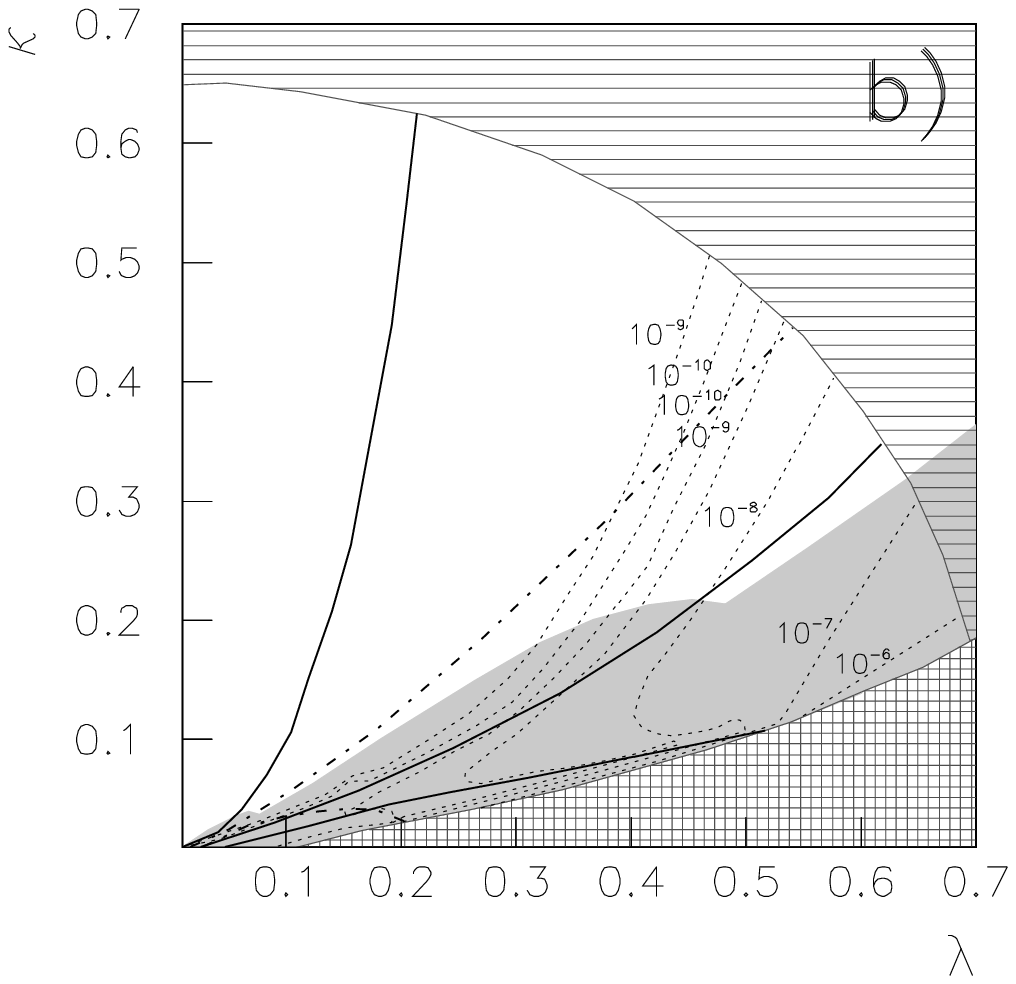,height=8cm}
  \captions{($\lambda , \kappa$) parameter space for
    $\tan\beta=3$, $A_\lambda=200$ GeV, $A_\kappa=-50$ GeV and
    $\mu=110$ GeV.
    In both cases, the ruled area represents points which are excluded
    due to the occurrence of a Landau pole. The grided area is
    excluded because of the appearance of tachyons.
    The grey area is associated to those
    points that do not satisfy the LEP constraints or where
    (at least) the LEP bound on direct neutralino production is
    violated.
    Dotted lines in the experimentally accepted region
    represent contours of scalar
    neutralino-proton cross section $\crosssec$.
    In (a), from top to bottom, solid lines indicate different
    values of lightest Higgs scalar mass,
    $m_{h_1^0}=114,\,75,\,25$ GeV,
    and dashed lines separate the regions where the
    lightest scalar Higgs has a singlet composition given by
    ${S_{13}^{\,2}}=0.1,\,0.9$.
    In (b), from top to bottom,
    solid lines are associated with different values of the
    lightest neutralino mass,
    $m_{\tilde\chi^0_1}=100,\,75,\,50$ GeV,
    while dot-dashed lines reflect the singlino composition
    of the
    lightest neutralino, $N_{15}^2=0.1,\,0.9$. }
  \label{++-a}
\end{figure}

The $(\lambda,\kappa)$ parameter space is shown in
Fig.\,\ref{++-a} for an example with  $\tan\beta=3$, $A_\lambda=200$ GeV,
$A_\kappa=-50$ GeV and  $\mu=110$ GeV. The points which
are excluded due to the occurrence of a Landau pole are indicated, as
well as those not fulfilling the experimental constraints. According
to the discussion above, the tachyons appearing in the lower right
corner are due to the CP-even Higgs sector.
It is worth remarking that in these cases, due to the smallness of the
lightest CP-even Higgs mass, the experimental constraints (see
Section\,\ref{2:parameters}) from $\ee$,
 both IHDM and DHDM
($h^0\to b\bar b$, $h^0\to \tau^+\tau^-$, and $h^0\to 2\,{\rm
  jets}$),
become very important and typically exclude
the regions in the vicinity of those excluded by tachyons.

Dashed lines in Fig.\,\ref{++-a}a indicate the singlet composition
of the
lightest scalar Higgs. Singlet-like Higgses can be found for small
values of $\kappa$, whereas doublet-like Higgses
appear for large $\kappa$. 
This can be qualitatively understood from the expression of the
corresponding mass matrix (\ref{neumatrix}). In particular, the
diagonal term ${\cal M}_{S,33}^2$ becomes very small when $\kappa$
decreases.  
Interestingly, when the singlet
composition is significant,
the reduced coupling can be smaller and thus
Higgses with $m_{h^0_1}\lsim 114$ GeV can escape detection and be in
agreement with experimental data. This opens a new window in the
allowed parameter space, characteristic of the NMSSM,  
which can have 
relevant consequences for dark matter
detection as we will discuss below.

In Fig.\,\ref{++-a}b
the same case is represented, but emphasizing the information on the
neutralino properties. The singlino composition of the lightest
neutralino is shown with dot-dashed lines, while solid lines
correspond to different values of its mass. 
As one would expect from the structure of the neutralino mass matrix
(\ref{neumatrix}), 
for small $\kappa$, the lightest neutralino is essentially a singlino,
with a small mass which can be approximated as
$\neumass \sim 2 \mu
\kappa/\lambda$.
In the present case, singlino-like neutralinos appear for
$\kappa\lsim0.04 $ and $\lambda\lsim0.2$, whereas heavier,
Higgsino-like, 
neutralinos (due to our choice of input values with $\mu<M_1$)
populate the rest of the parameter space.
Regions with small masses of the neutralino may be excluded due to the
bound on direct neutralino production, which becomes quite severe for
light Higgsino-like neutralinos.

In both figures, the different values of the neutralino-nucleon
cross section are represented with dotted lines. As already commented
in Section\,\ref{sigma}, the cross section increases in those regions
with a light CP-even Higgs, as long as it is not a pure singlet. This
behaviour is clearly illustrated in these figures, which feature very
large values for $\crosssec$ in the vicinity of the areas where the
lightest Higgs becomes tachyonic.
On the other hand far from these regions the cross section
stabilizes at $10^{-8}$ pb $>\crosssec> 10^{-9}$ pb.

\begin{figure}[t]
  \epsfig{file=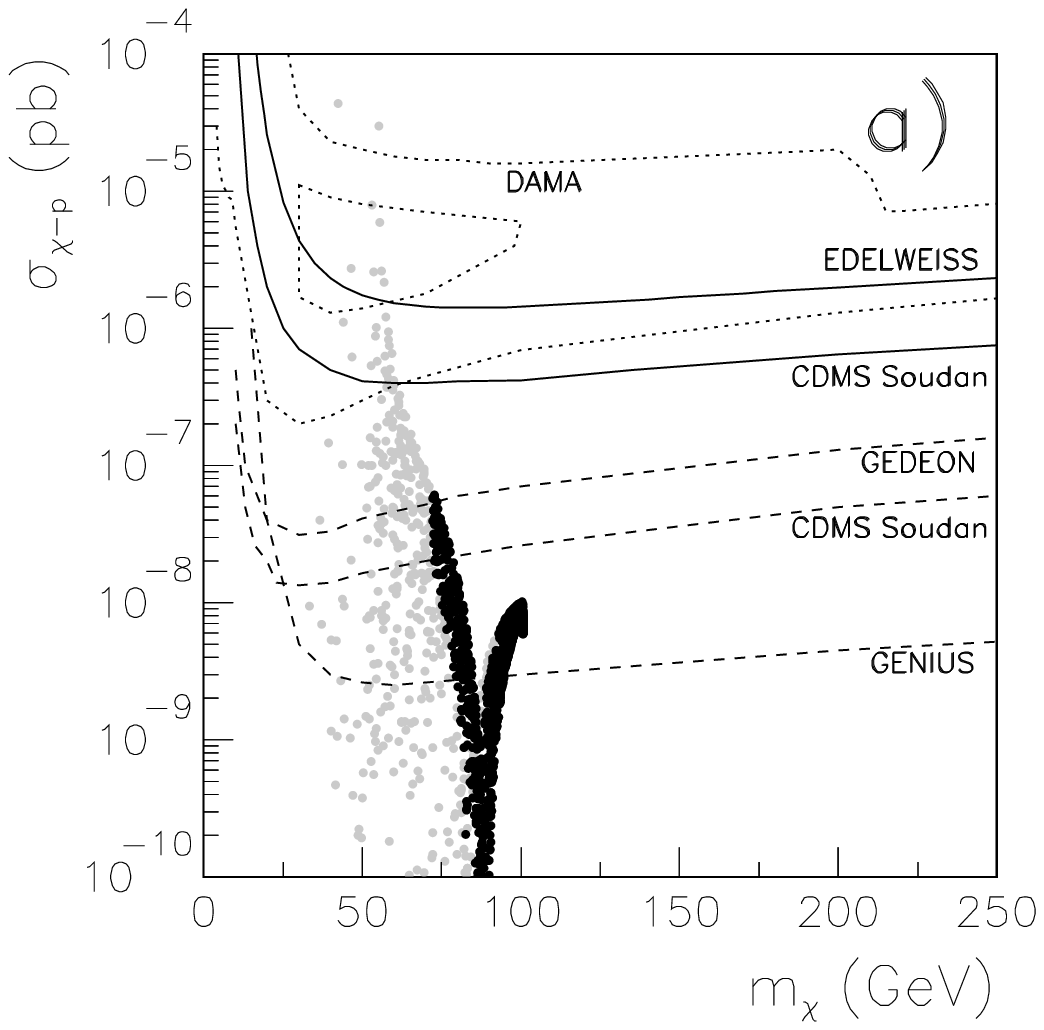,height=8cm}
  \hspace*{-1cm}\epsfig{file=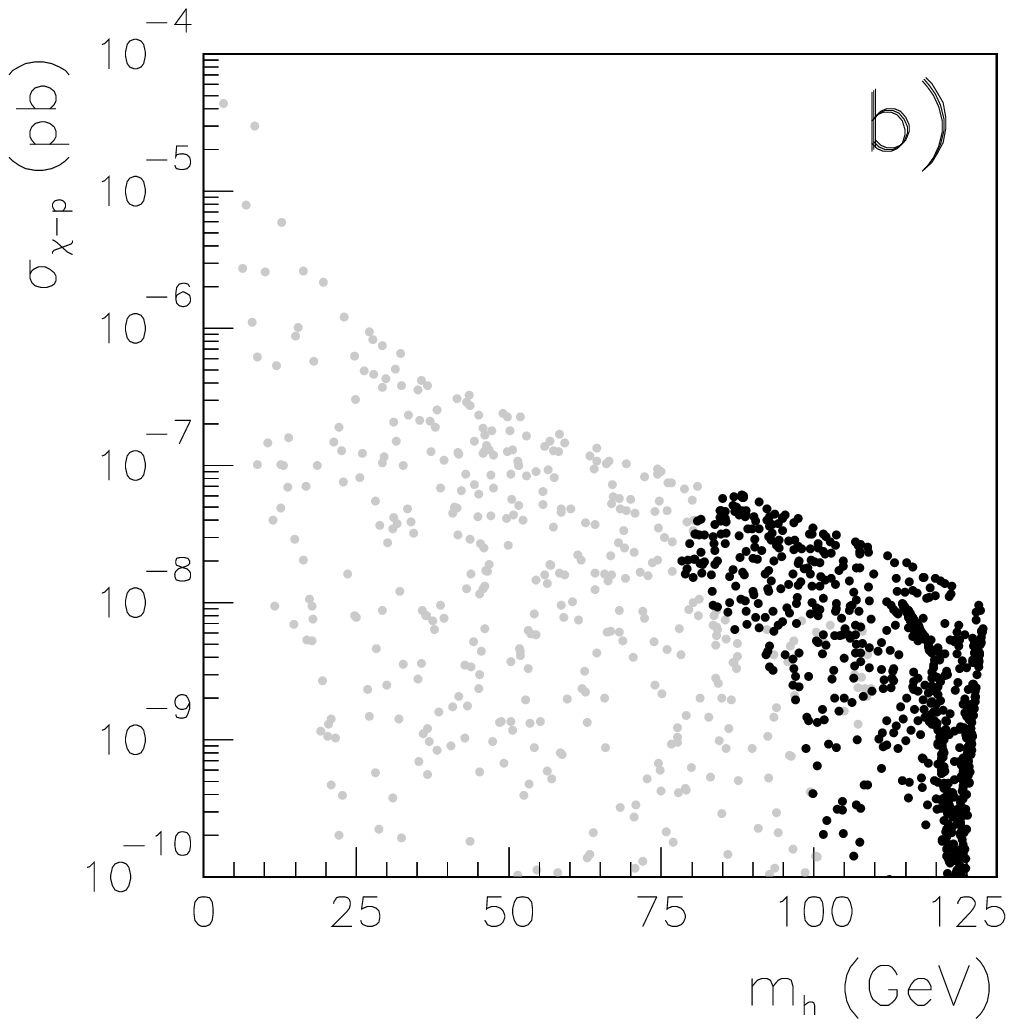,height=8cm}
  \captions{Scatter plot of the scalar neutralino-nucleon cross
    section, $\crosssec$, as a function of (a) the neutralino mass,
    $\neumass$, and (b) the lightest scalar Higgs mass, $m_{h_1^0}$,
    for $A_\lambda=200$ GeV, $\mu=110$ GeV, $A_\kappa=-50$
    GeV, and $\tan\beta=3$.
    Black dots correspond to points fulfilling all the experimental
    constraints, whereas grey dots represent those excluded.
    In (a)
    the sensitivities of present and projected experiments are also
    depicted with solid and dashed lines, respectively. The large
    (small) area bounded by dotted lines is allowed by the DAMA
    experiment when astrophysical uncertainties are (are not) taken
    into account.
  }
  \label{++-crossa}
\end{figure}

In order to illustrate this point in more detail, we have represented
in Fig.\,\ref{++-crossa} the resulting $\crosssec$ versus the lightest
Higgs mass and the neutralino mass. Black dots fulfil all the
experimental constraints, whereas grey dots are those experimentally
excluded
(we do
not plot those regions ruled out due to theoretical arguments, such as
the occurrence of a Landau pole).
The sensitivities of present and projected dark matter experiments are
also depicted as a function of $\neumass$ for comparison.
The small area bounded by dotted lines is allowed by the DAMA
experiment in the simple case of an isothermal spherical
halo model. The larger area also bounded by dotted lines represents
the DAMA region
when uncertainties to this simple model are taken into account.
For the other experiments in the figure only the spherical halo has
been considered in their analyses.
In particular, the (upper) areas bounded by solid lines are
excluded by EDELWEISS\footnote{Since the exclusion area due to
ZEPLIN I is similar to EDELWEISS we have not depicted it here, nor in
any
subsequent Figure.} and CDMS Soudan. 
Finally, the dashed lines represent the
sensitivities of the projected GEDEON, CDMS Soudan, and GENIUS
experiments.

Very large values for the cross section could in principle be
obtained. However, these are associated to very light Higgses and are
therefore 
subject to the strong constraints on $\ee$ discussed above.
Once every constraint is taken into account,
points with $\crosssec\lsim10^{-7}$ pb appear, which
correspond to light scalar Higgses with $m_{h^0_1}\gsim75$~GeV, 
surviving the experimental constraints due to their
important singlet character, $S_{13}^{\,2}\gsim 0.85$. This is a clear
consequence of the NMSSM that we will exploit in subsequent examples,
since it allows for a significant increase in the cross section.
The neutralinos in these regions have  $\neumass\gsim 70$ GeV and
have a mixed
singlino-Higgsino composition ($N_{15}^2\lsim0.3$ and
$N_{13}^2+N_{14}^2\gsim0.7$ in the region with
larger cross section).

Notice that $\crosssec$ displays an important suppression around
$\neumass\approx90$ GeV. This is due to the cancellation of the
contribution of the cross section coming from
neutralino-neutralino-Higgs interaction due to the occurrence of terms
with different signs.
This type of accidental cancellations is analogous to those appearing
in MSSM analyses for $\mu<0$ \cite{Ellis:2000ds}.

\begin{figure}[t]
  \epsfig{file=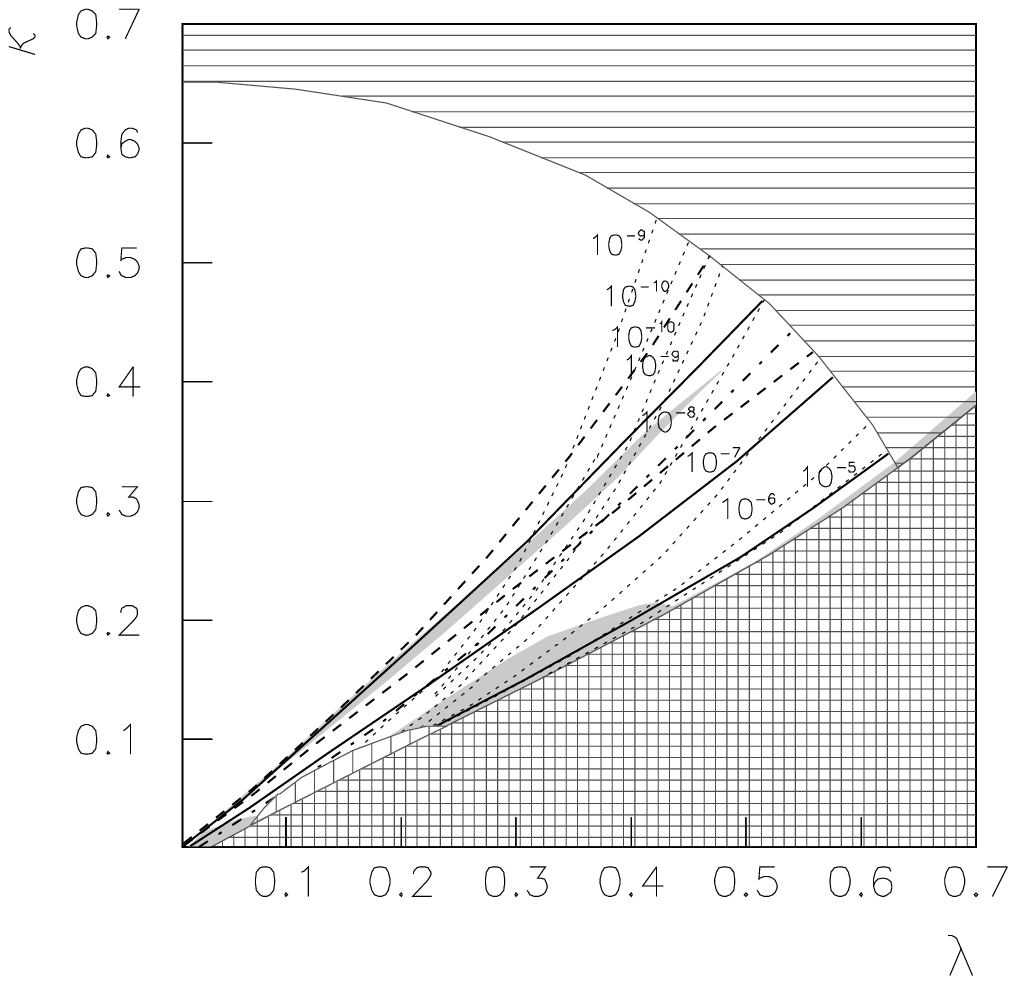,height=8cm}
  \hspace*{-1cm}\epsfig{file=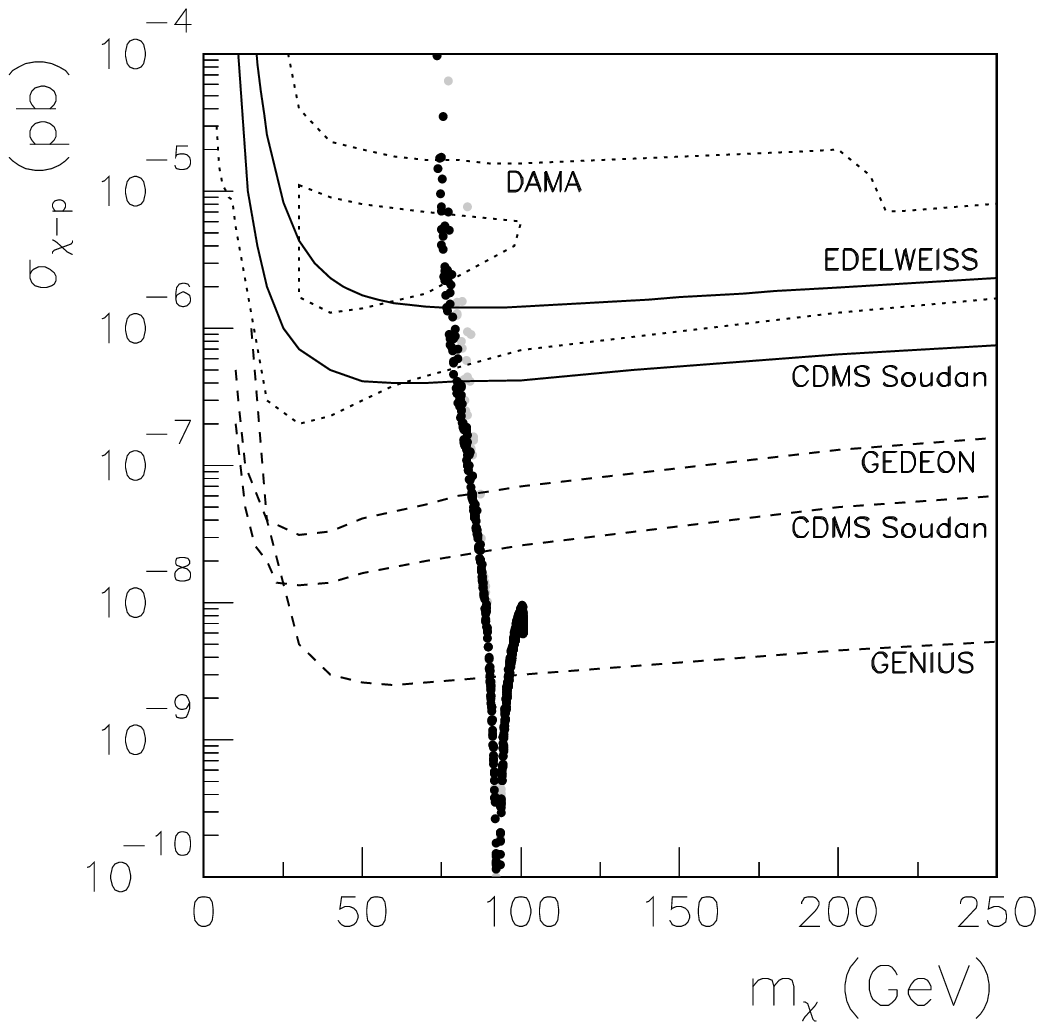,height=8cm}
  \captions{$(\lambda,\kappa)$ parameter space with the corresponding
    constraints and neutralino-nucleon
    cross section as a function of the 
    lightest neutralino mass for the case
    $A_\lambda=200$ GeV, $A_\kappa=-200$ GeV,  $\mu=110$ GeV  and
    $\tan\beta=3$.
    In the $(\lambda,\kappa)$ plane the mass and composition of the
    lightest scalar Higgs,
    the composition of the lightest neutralino (only the line with
    $N_{15}^2=0.1$), and the
    predictions for $\crosssec$ are represented with the same line
    conventions as in Fig.\,\ref{++-a}, and the new ruled area (vertical
    lines)
    is excluded due to the occurrence of unphysical minima.
    The colour convention for the plot $\crosssec$ versus $\neumass$
    is as in Fig.\,\ref{++-crossa}.
    }
  \label{++-ka}
\end{figure}

These results are also sensitive to variations in the rest of
the input parameters ($A_\kappa$,
$\tan\beta$, $\mu$, and $A_{\lambda}$).
For instance, increasing $|A_\kappa|$ (i.e., making it more negative)
leads
to a further decrease in ${\cal M}^2_{S,33}$ in the CP-even Higgs
mass matrix, and therefore lighter Higgses can be obtained with a
larger singlet composition.
Although this translates into an enlargement of the regions where one
has a tachyonic scalar Higgs,
one may nevertheless find a larger
$\crosssec$ in
the allowed areas. Choosing
$\tan\beta=3$, $A_\lambda=200$ GeV and  $\mu=110$ GeV, but with
$A_\kappa=-200$ GeV, one can obtain $\crosssec\gsim10^{-4}$ pb (points which
in fact are already excluded by direct dark matter searches). The
corresponding
$(\lambda,\kappa)$ parameter space, as well as $\crosssec$ versus the
neutralino mass, are represented in Fig\,\ref{++-ka}.
Remarkably, very light Higgses are allowed in this
case ($m_{h^0_1}\gsim 20$~GeV) due to their significant singlet
character
($0.9\lsim S_{13}^{\,2}\lsim0.95$).
Once again, the lightest neutralino exhibits a large
singlino-Higgsino composition ($N_{15}^2\lsim0.3$ and
$N_{13}^2+N_{14}^2\gsim0.7$).
For these reasons, one hardly finds experimentally excluded regions:
only narrow stripes, mostly due to direct production of $\tilde\chi^0$
and $h^0\to b\bar b$. Also, for small values of $\lambda$ and
$\kappa$, a very thin region excluded by the existence of false minima
(see Section\,\ref{2:parameters}) appears.
Conversely, decreasing $|A_\kappa|$ helps reducing tachyonic
regions. In the particular case where $A_\kappa=0$, no tachyons emerge
from the CP-even sector.
The implications of this variation in the value of $\crosssec$ are minimal.

\begin{figure}
  \epsfig{file=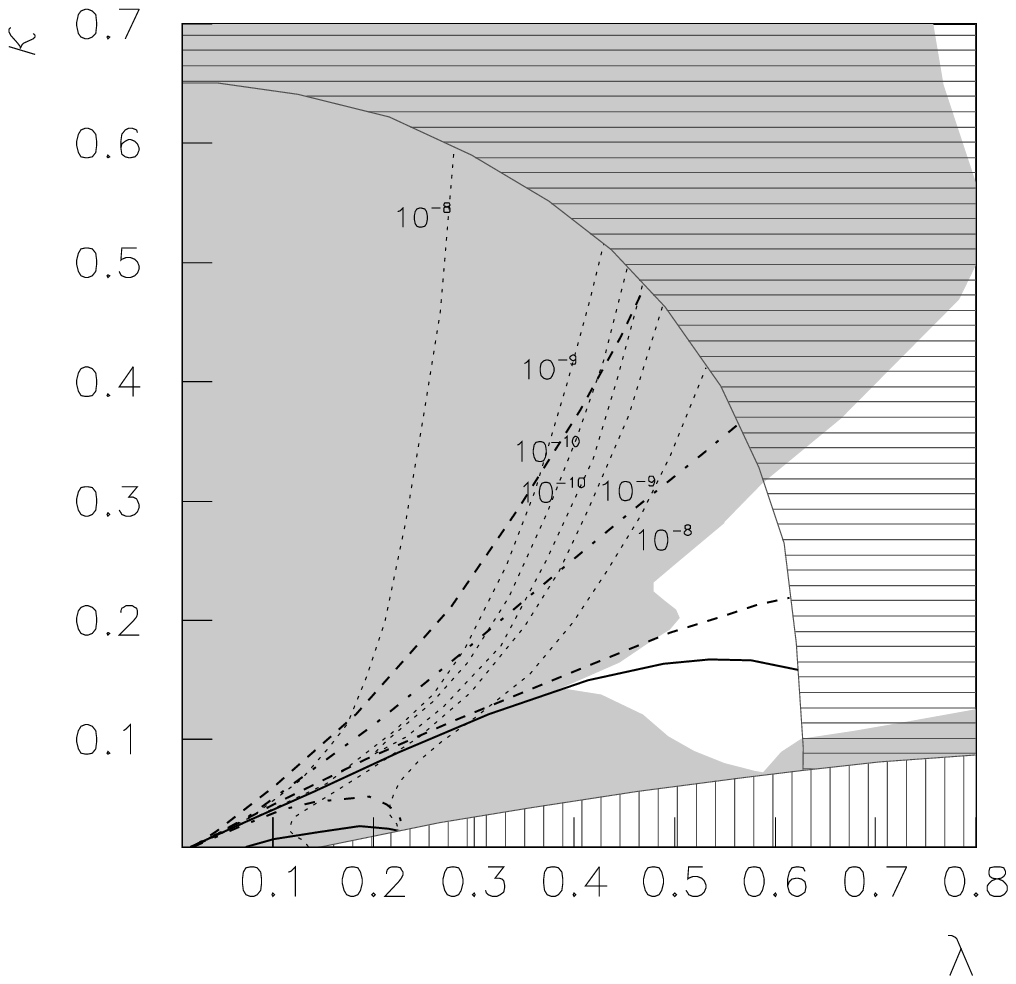,height=8cm}
  \hspace*{-1cm}\epsfig{file=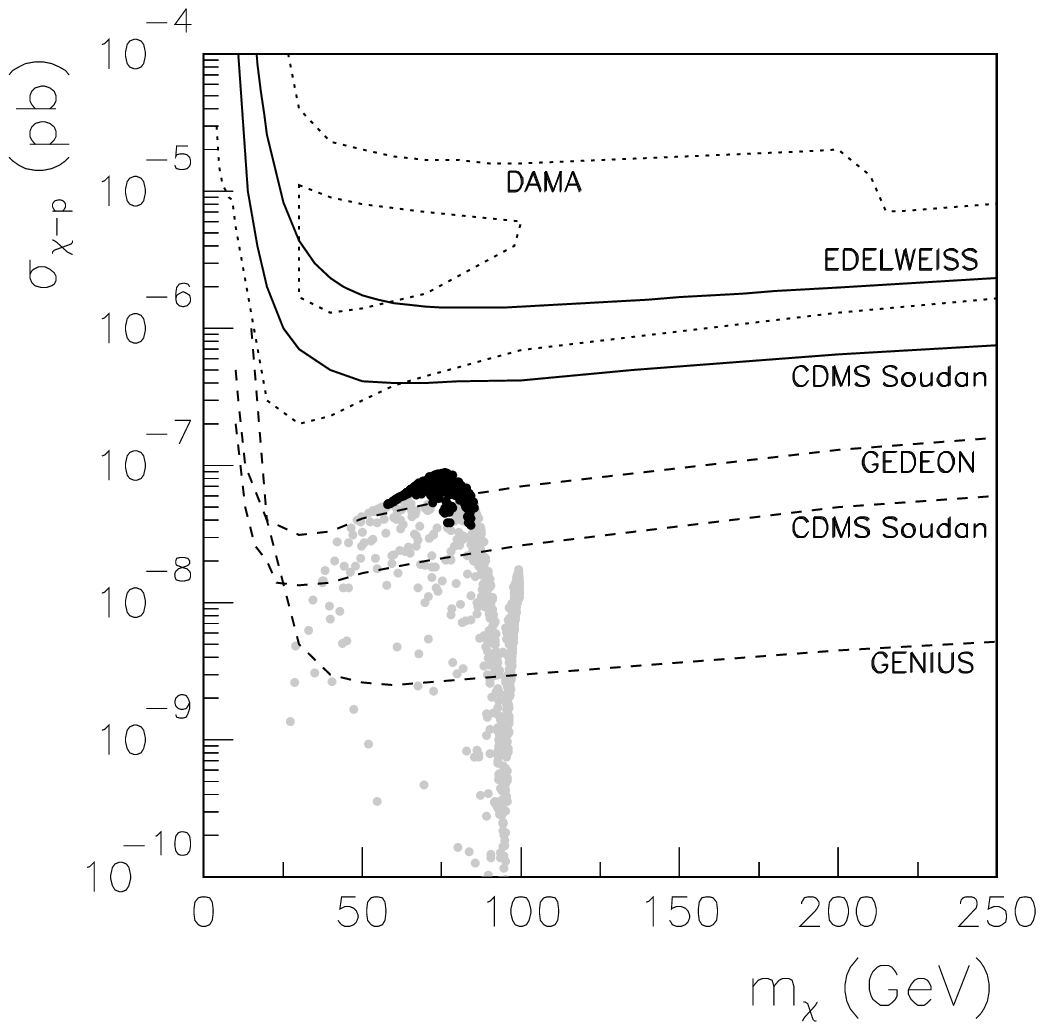,height=8cm}
  \\[-3ex]
  \epsfig{file=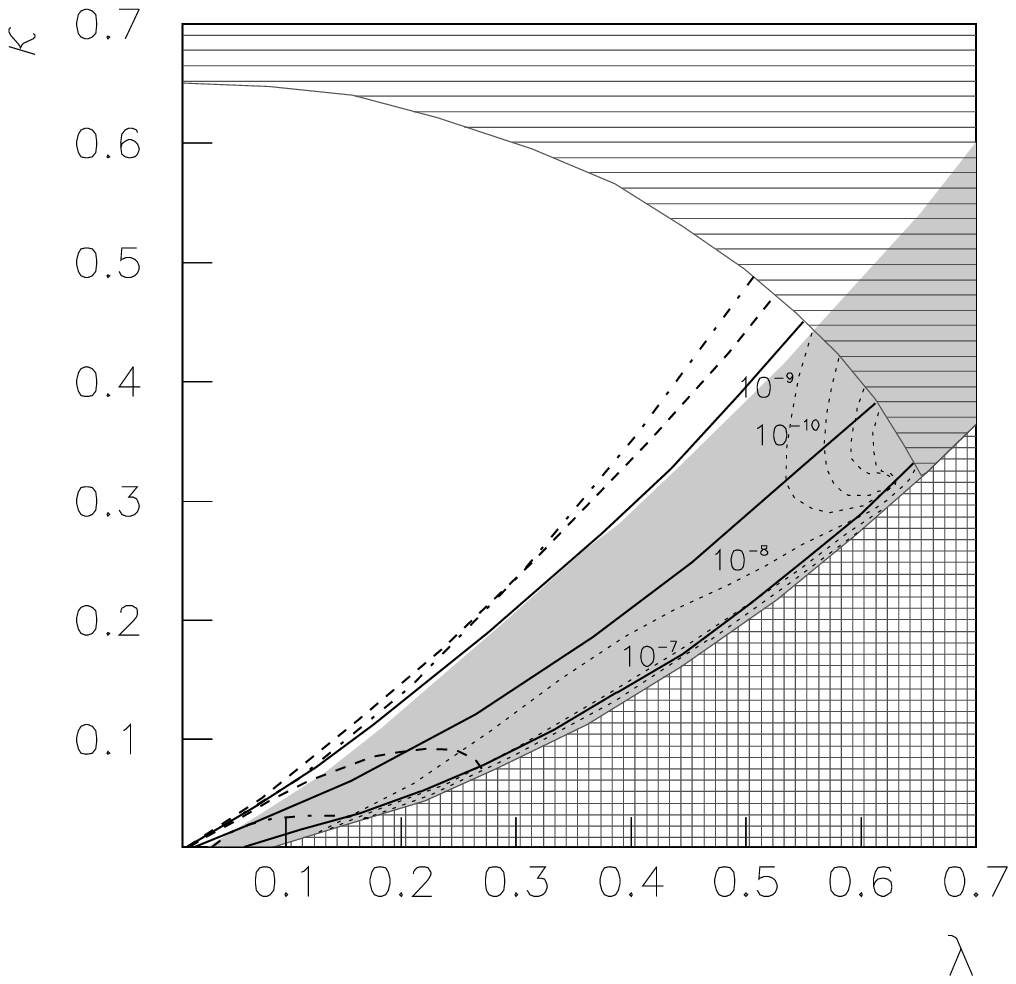,height=8cm}
  \hspace*{-1cm}\epsfig{file=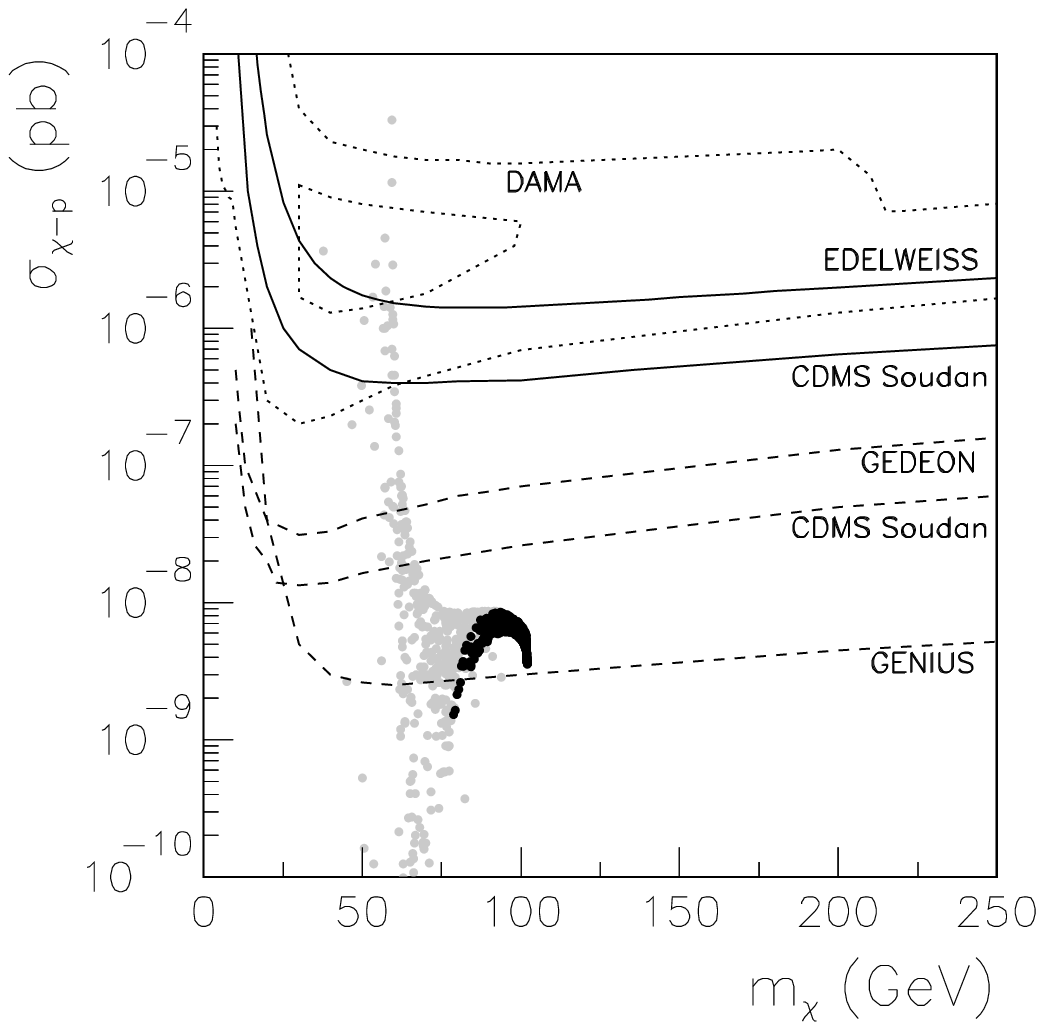,height=8cm}
  \\[-3ex]
  \epsfig{file=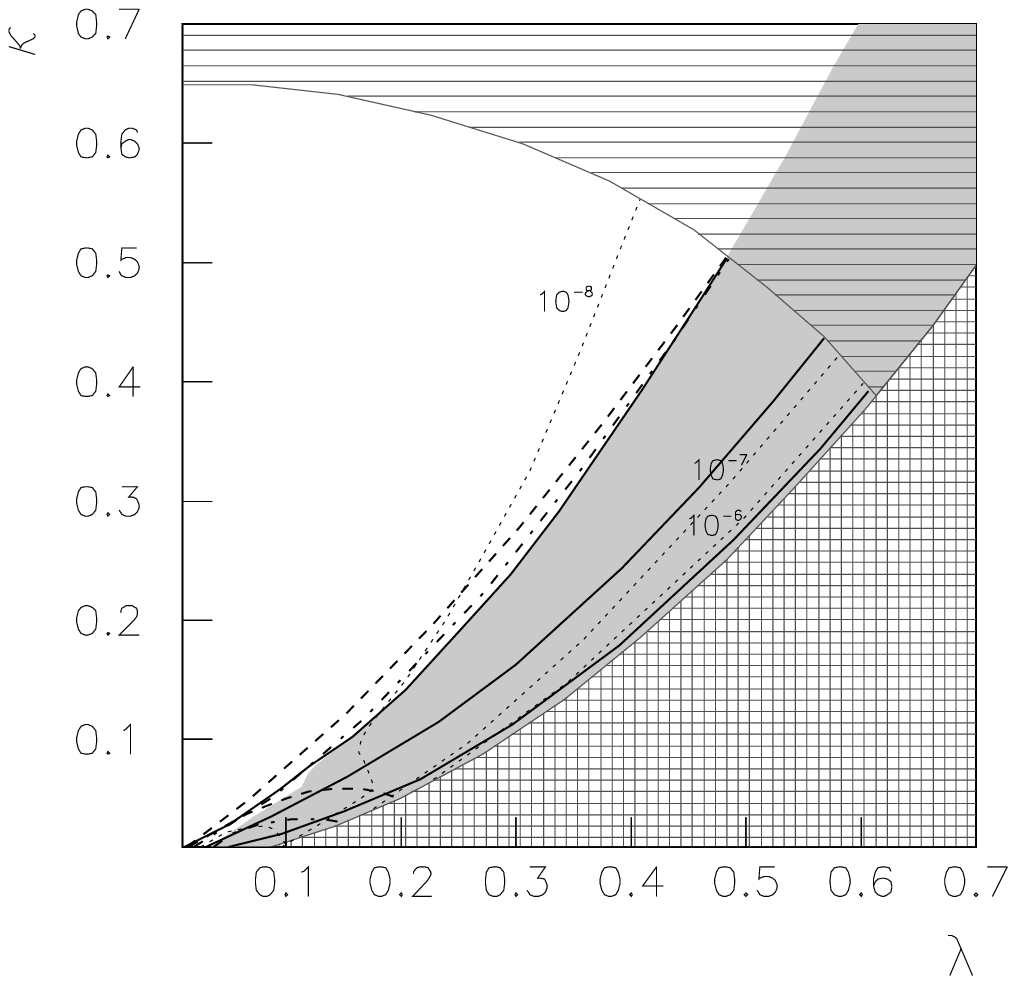,height=8cm}
  \hspace*{-1cm}\epsfig{file=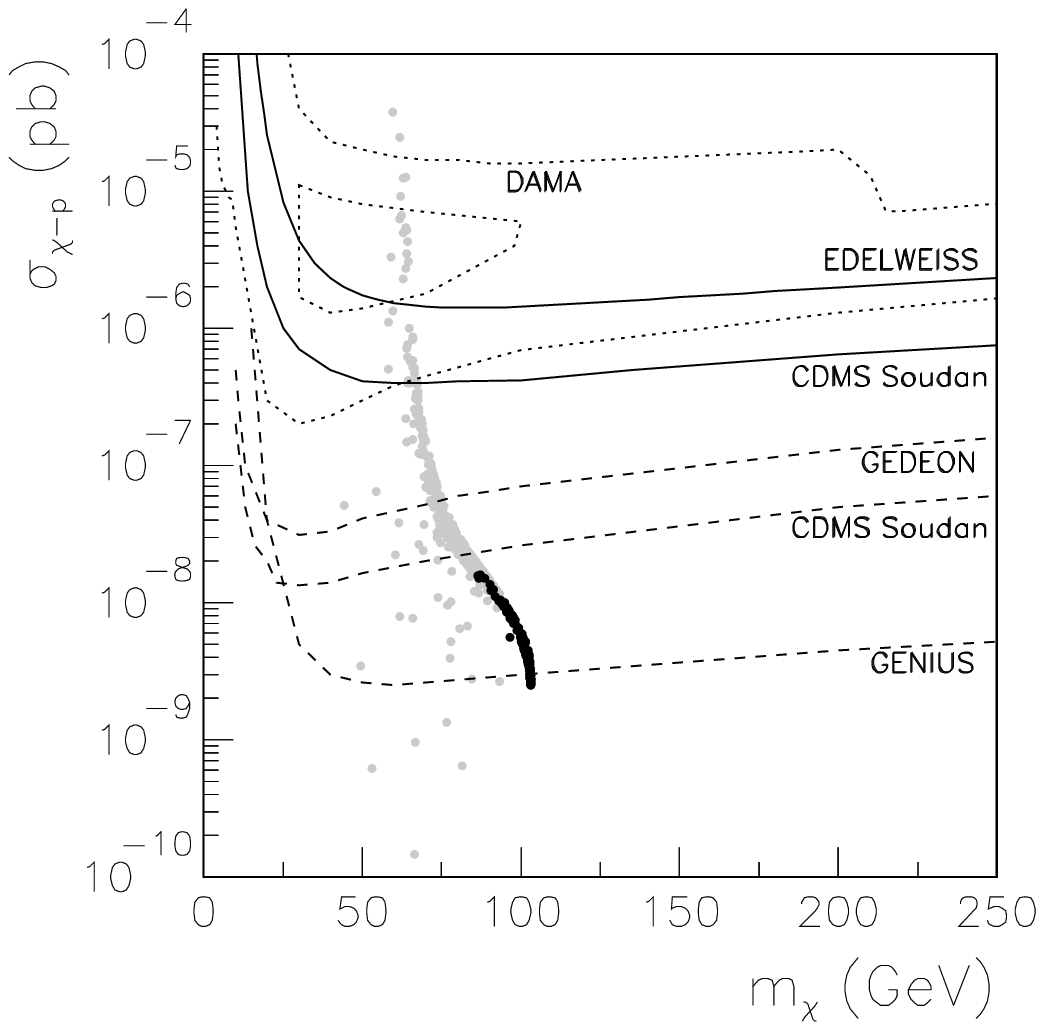,height=8cm}
  \vspace*{-1cm}
  \captions{The same as in Fig.\,\ref{++-ka} but for the cases
    $A_\lambda=200$ GeV, $A_\kappa=-50$ GeV, 
    $\mu=110$ GeV, and
    $\tan\beta=2,\,5,\,10$, from top to bottom.
    In the case with $\tan\beta=2$, only the lines with
    $m_{h_1^0}=75,\,25$~GeV are represented, since $m_{h_1^0}\lsim110$
    GeV.
    }
  \label{++-tgb}
\end{figure}

Changing $\tan\beta$ has an important impact in the
analysis, mainly due to the effect on the Higgs sector.
The tachyonic regions become larger as $\tan\beta$
increases (extending towards higher values of $\lambda$ and
$\kappa$). As a consequence, the neutralino is never a pure singlino
and its mass increases due to the larger mixing with Higgsinos. For this
reason the exclusion due to direct neutralino production becomes
larger. In the end,
not only the allowed region is reduced, but also the predictions for
$\crosssec$ are smaller.
Also, for very small values of $\tan\beta$ very light Higgsino-like
neutralinos can be found in large regions of the parameter space.
The experimental constraints
are, nevertheless, more important and only
small areas survive. Fig.\,\ref{++-tgb} illustrates these
properties.

\clearpage
\thispagestyle{empty}

\clearpage

\begin{figure}
  \epsfig{file=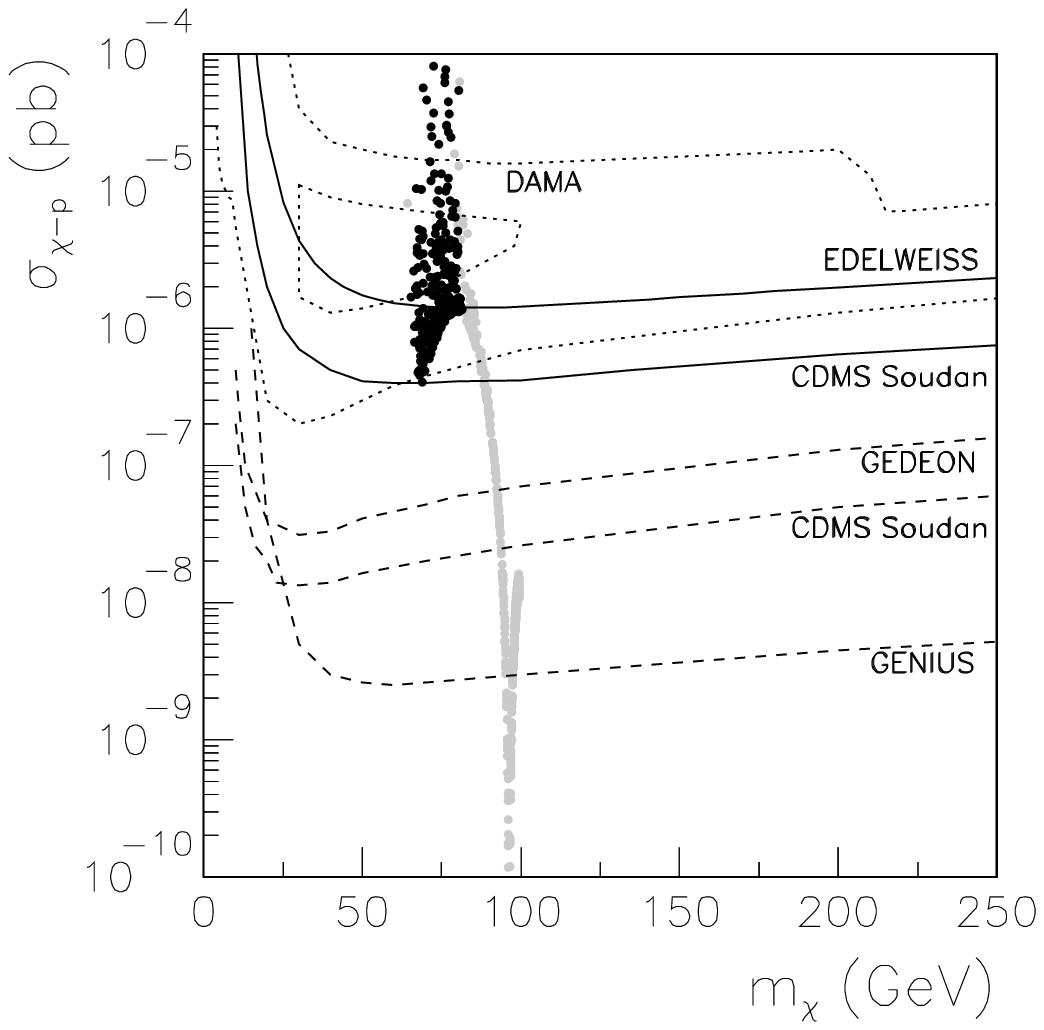,height=8cm}
  \hspace*{-1cm}\epsfig{file=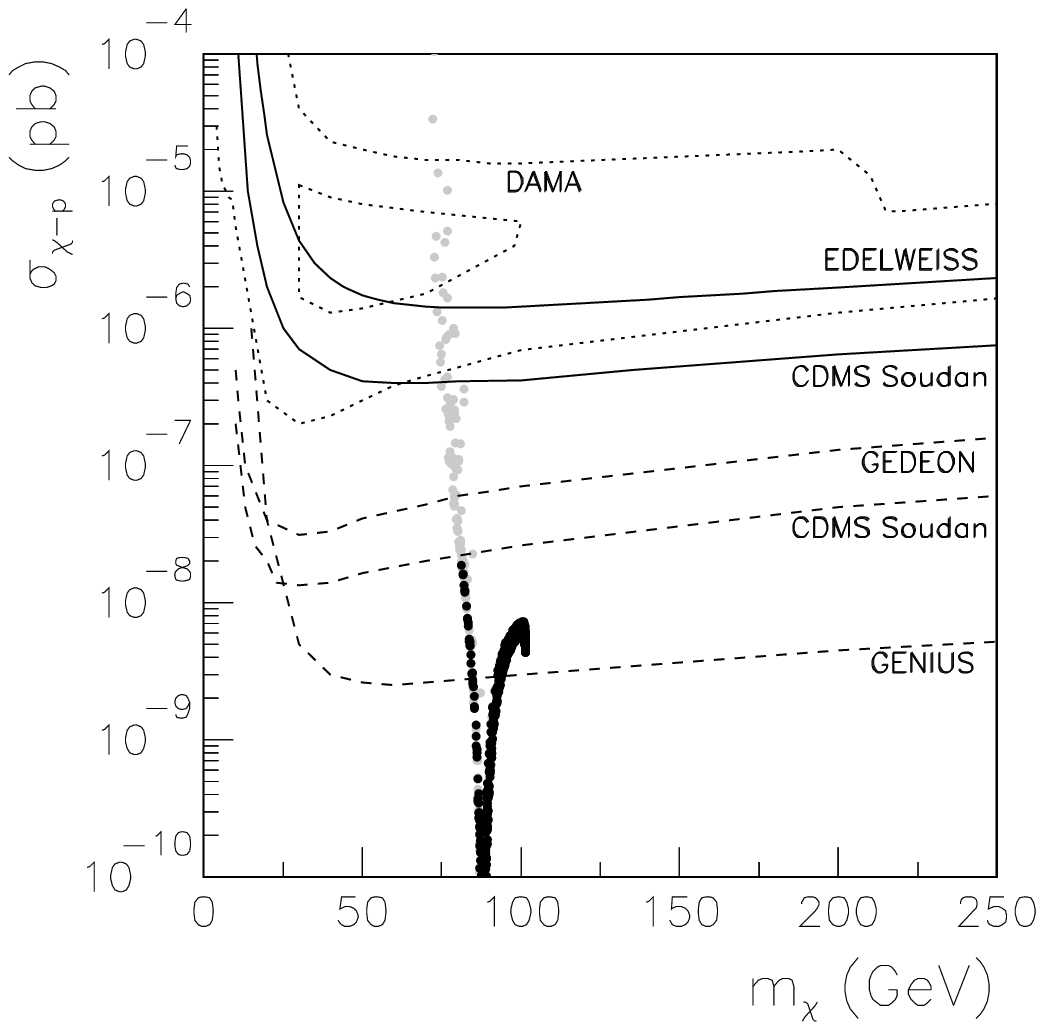,height=8cm}
  \\[-3ex]
  \hspace*{3.5cm}\epsfig{file=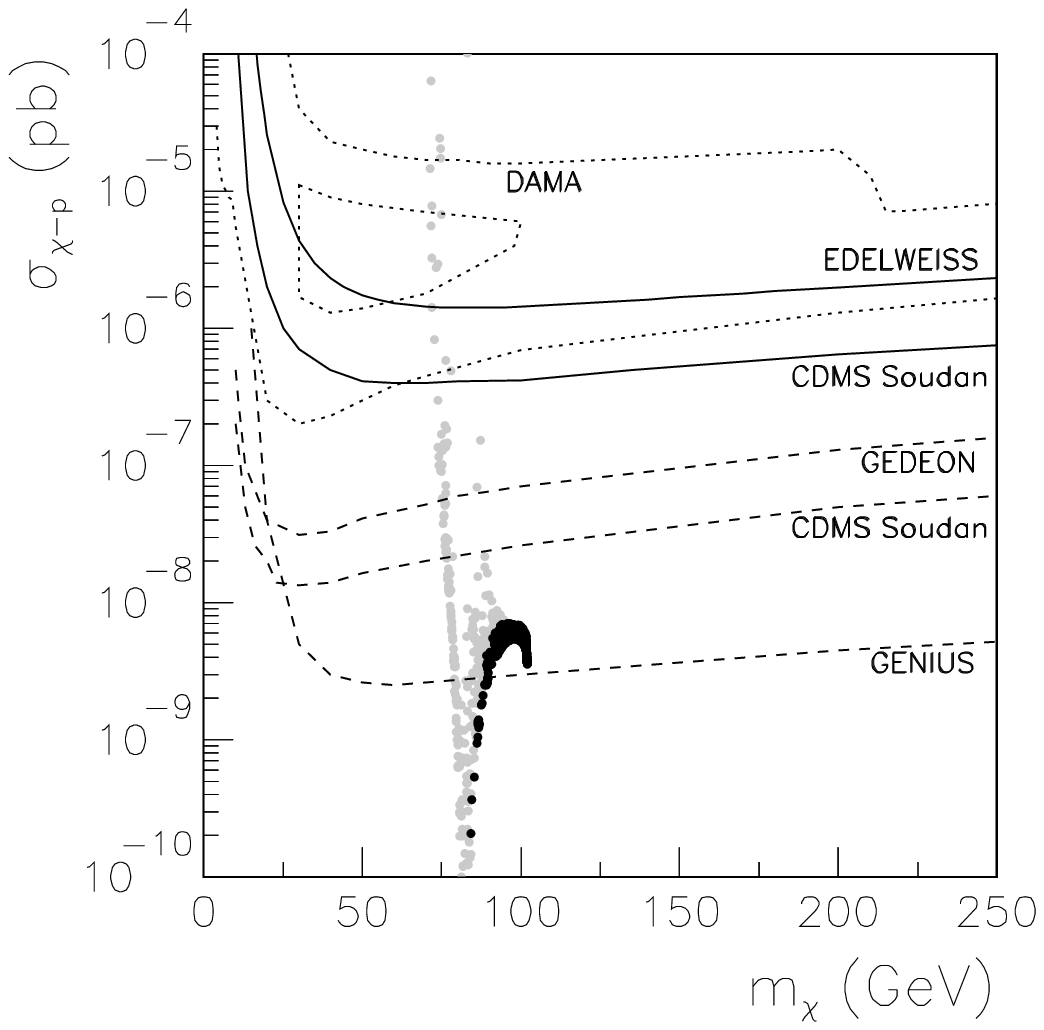,height=8cm}
  \captions{The same as in Fig.\,\ref{++-crossa}a but for the cases
    $A_\lambda=200$ GeV, $A_\kappa=-200$ GeV, 
    $\mu=110$ GeV, and
    $\tan\beta=2,\,4,\,5$, from left to right and top to bottom.}
  \label{aaaaa}
\end{figure}

Similar examples, but for $A_\kappa=-200$ GeV can be found in 
Fig.\,\ref{aaaaa}, where the predictions for $\crosssec$ are depicted
as a function of the neutralino mass for $\tan\beta=2,\,4,\,5$. As
already mentioned, small
values of $\tan\beta$ favour lighter neutralinos with larger detection
cross section.

\begin{figure}
  \epsfig{file=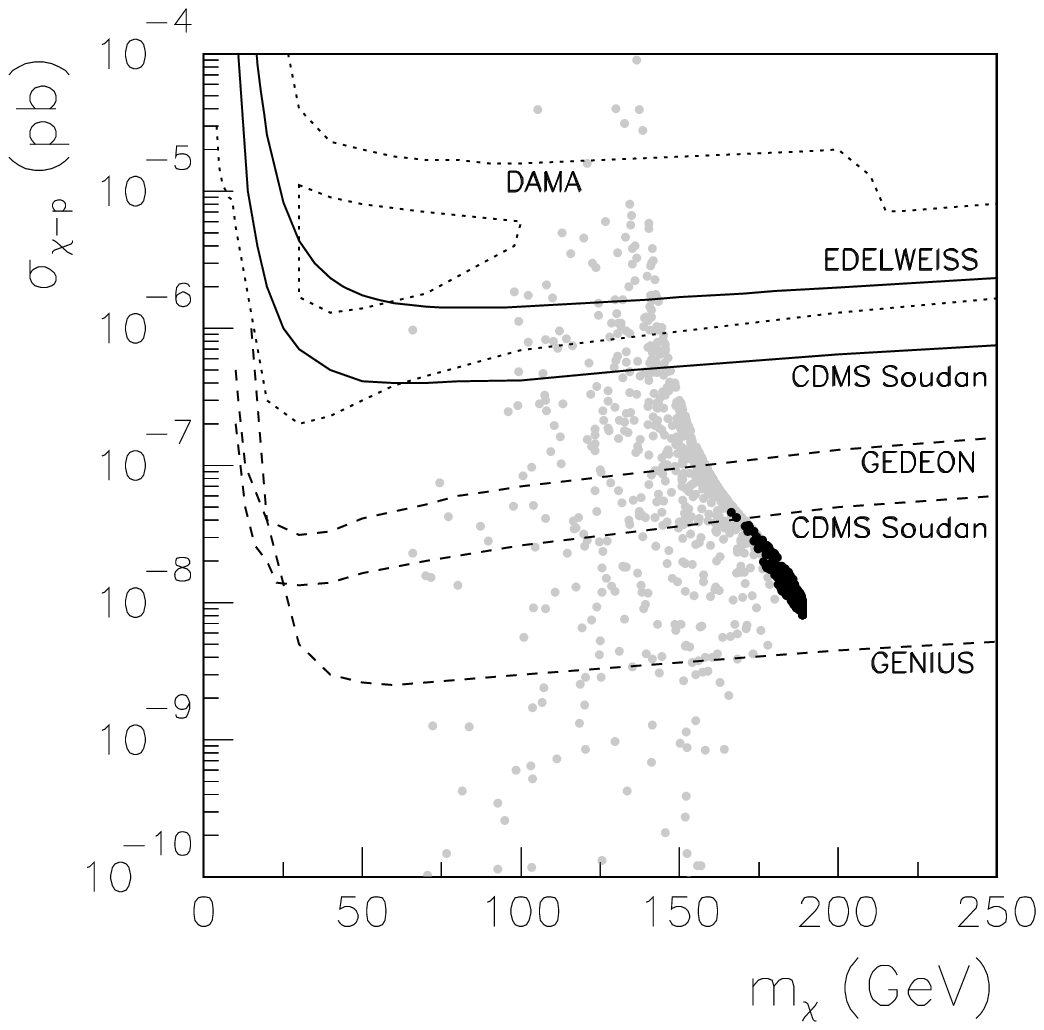,height=8cm}
  \hspace*{-1cm}\epsfig{file=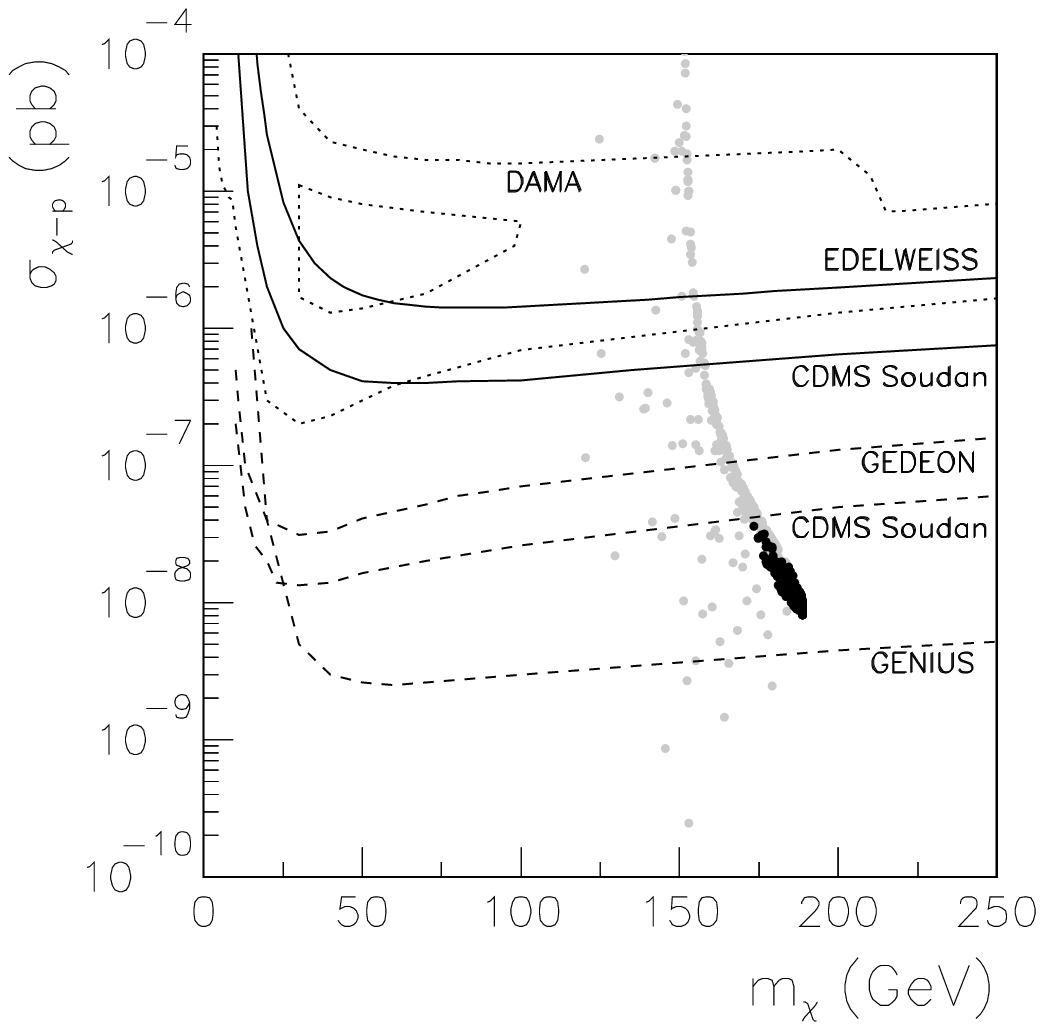,height=8cm}
  \captions{The same as in Fig.\,\ref{++-crossa}a but for the cases
    $A_\lambda=200$ GeV, $\tan\beta=3$, $\mu=200$~GeV, and
    $A_\kappa=-50,\,-200$ GeV, from left to right.} 
  \label{++-mu}
\end{figure}

Heavier neutralinos with a larger singlino composition
can
be obtained if the value of $\mu$ is increased.
For this reason, the regions where direct neutralino production is not
in agreement with experimental bounds become much narrower.
The mass of scalar Higgses also grows in this case, as well as their
doublet character.
Constraints on the Higgs
sector are still strong enough to forbid those points where the
neutralino is mostly singlino, and in the end $\neut$ preserves its
mixed singlino-Higgsino character.
In the remaining allowed area the predictions for $\crosssec$ can
vary, being typically smaller than in cases with
low $\mu$.
This is shown in Fig.\,\ref{++-mu} for two examples with
$\mu=200$~GeV, $A_\lambda=200$~GeV, $\tan\beta=3$, and
$A_\kappa=-50,\,-200$~GeV. In particular, in the case with
$A_\kappa=-200$ GeV, the 
detection cross section is much smaller than in the analogous example 
with $\mu=110$ GeV presented in
Fig.\,\ref{++-ka}. 

\begin{figure}[!t]
  \epsfig{file=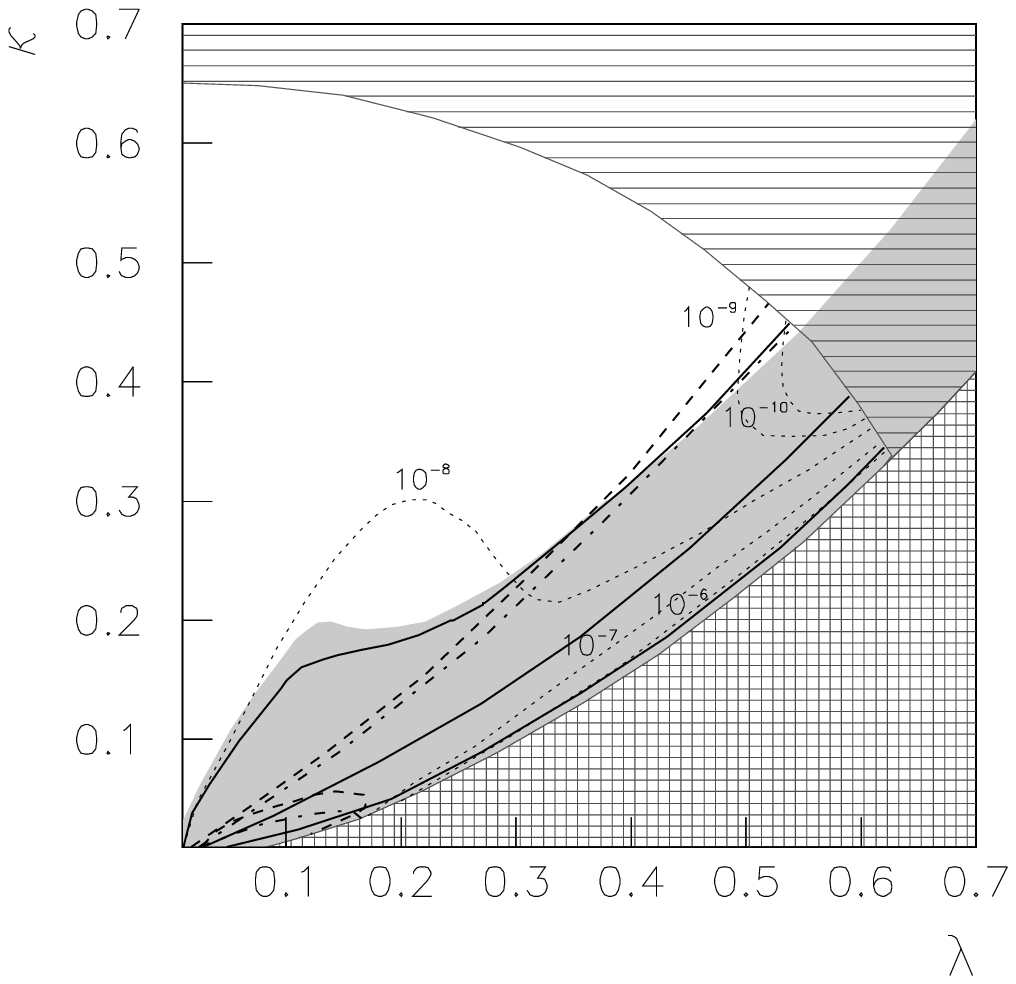,height=8cm}
  \hspace*{-1cm}\epsfig{file=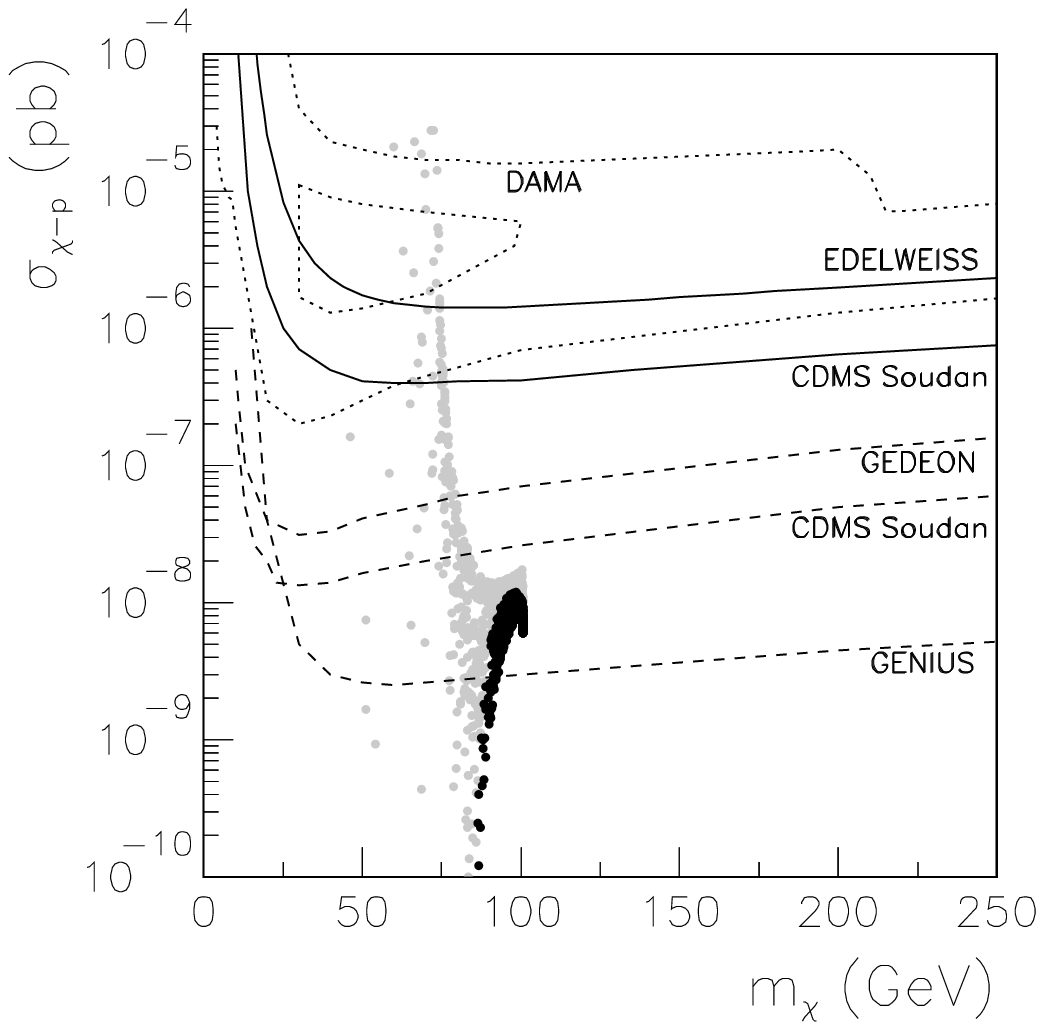,height=8cm}
\\[-3ex]
    \epsfig{file=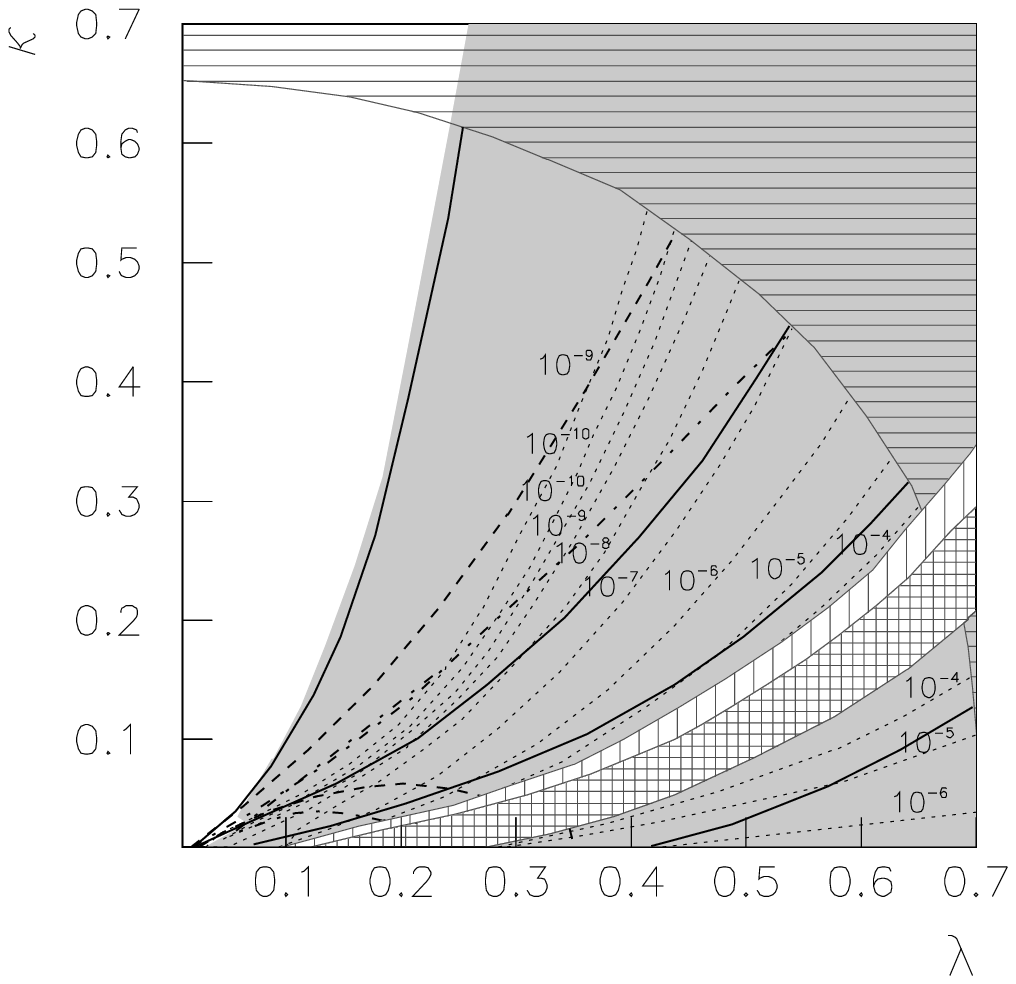,height=8cm}
  \hspace*{-1cm}\epsfig{file=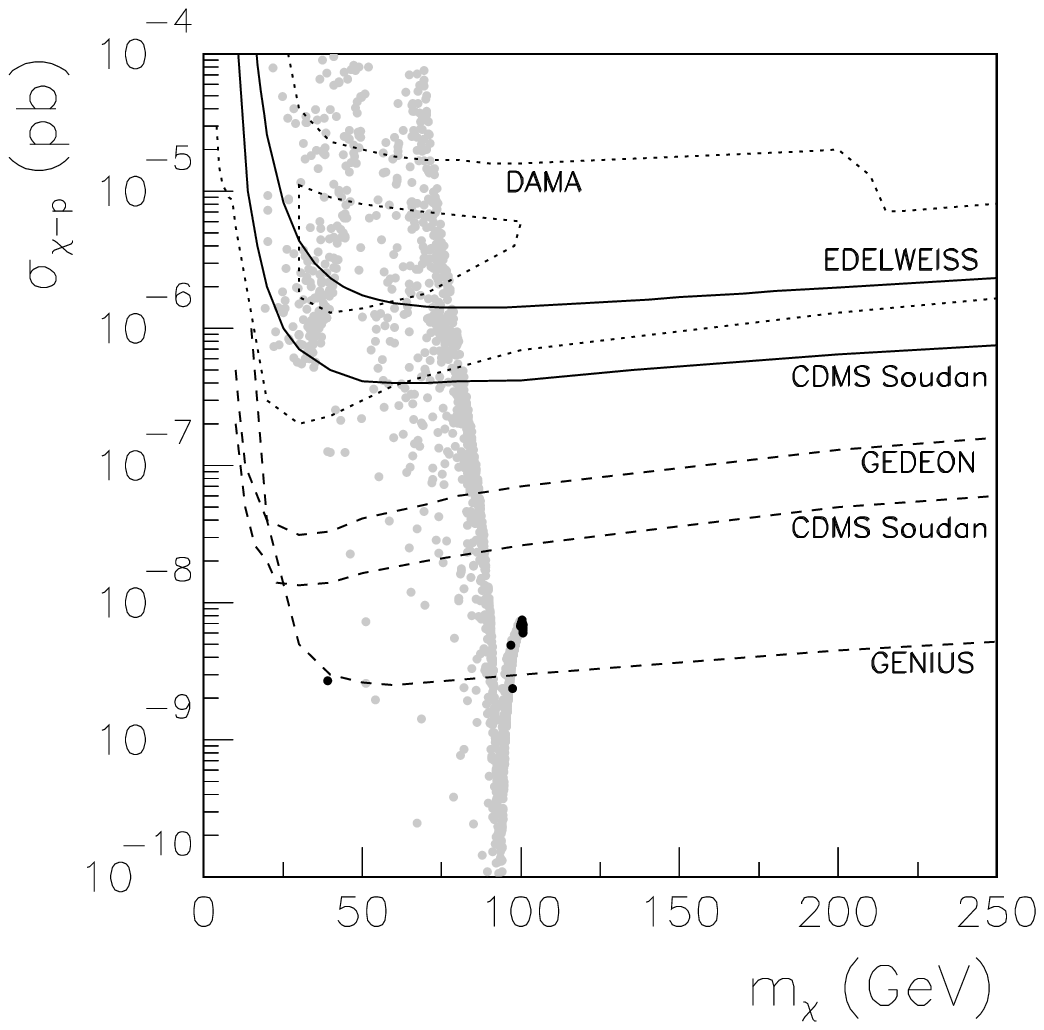,height=8cm}
  \captions{The same as in Fig.\,\ref{++-ka} but for the cases
    $A_\kappa=-50$ GeV,  $\mu=110$ GeV, $\tan\beta=3$, and
    $A_\lambda=50,450$ GeV, from top to bottom.}
  \label{++-alam}
\end{figure}

Finally, variations in the value of $A_\lambda$ also influence the
theoretical  predictions on $\crosssec$.
There is a range of
$A_\lambda$ for which
the eigenvalues of the CP-even Higgs mass matrix
are positive.
However, for smaller or larger $A_\lambda$,
off-diagonal terms may become
large enough to ease the appearance of tachyons in the large $\lambda$
regime.
For instance, in Fig.\,\ref{++-alam} we have represented
the $(\lambda,\kappa)$ plane
and the corresponding predictions for
$\crosssec$ in two cases with
$A_\kappa=-50$ GeV,  $\mu=110$ GeV, $\tan\beta=3$, and
$A_\lambda=50,\,450$~GeV.
We find that, in agreement with
the discussion above, the tachyonic regions are larger
than those for $A_\lambda=200$ GeV in both cases.
Also, the areas excluded by experimental constraints associated to
IHDM and DHDM are more extensive, and in the case of $A_\lambda=450$
GeV 
they forbid most of the parameter space.
The neutralino is mostly Higgsino in the remaining allowed points,
with $N_{15}^2\lsim0.1$($0.2$) and
$N_{13}^2+N_{14}^2\gsim0.9$($0.8$) 
for $A_\lambda=50$($450$)~GeV, and
there is a slight decrease in the predictions for $\crosssec$.

\begin{figure}
  \epsfig{file=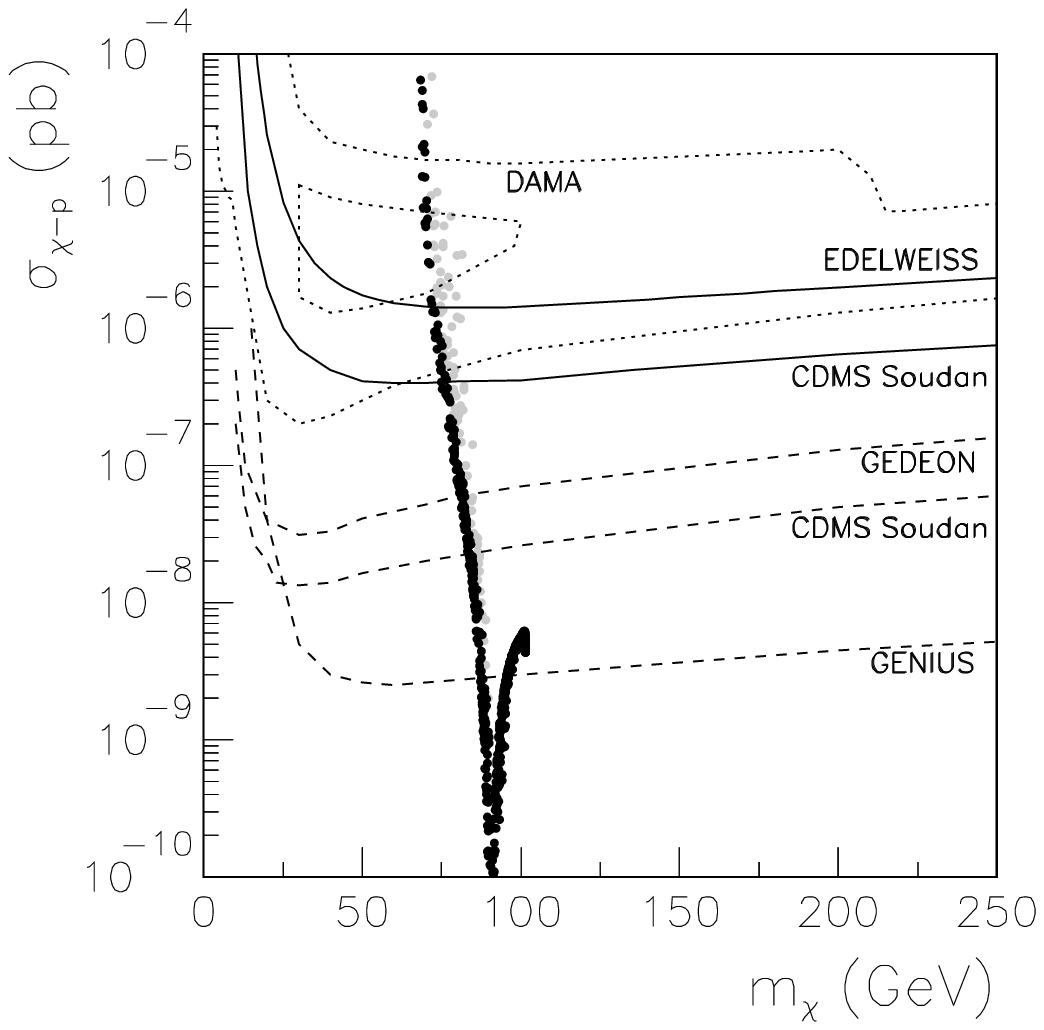,height=8cm}
  \hspace*{-1cm}\epsfig{file=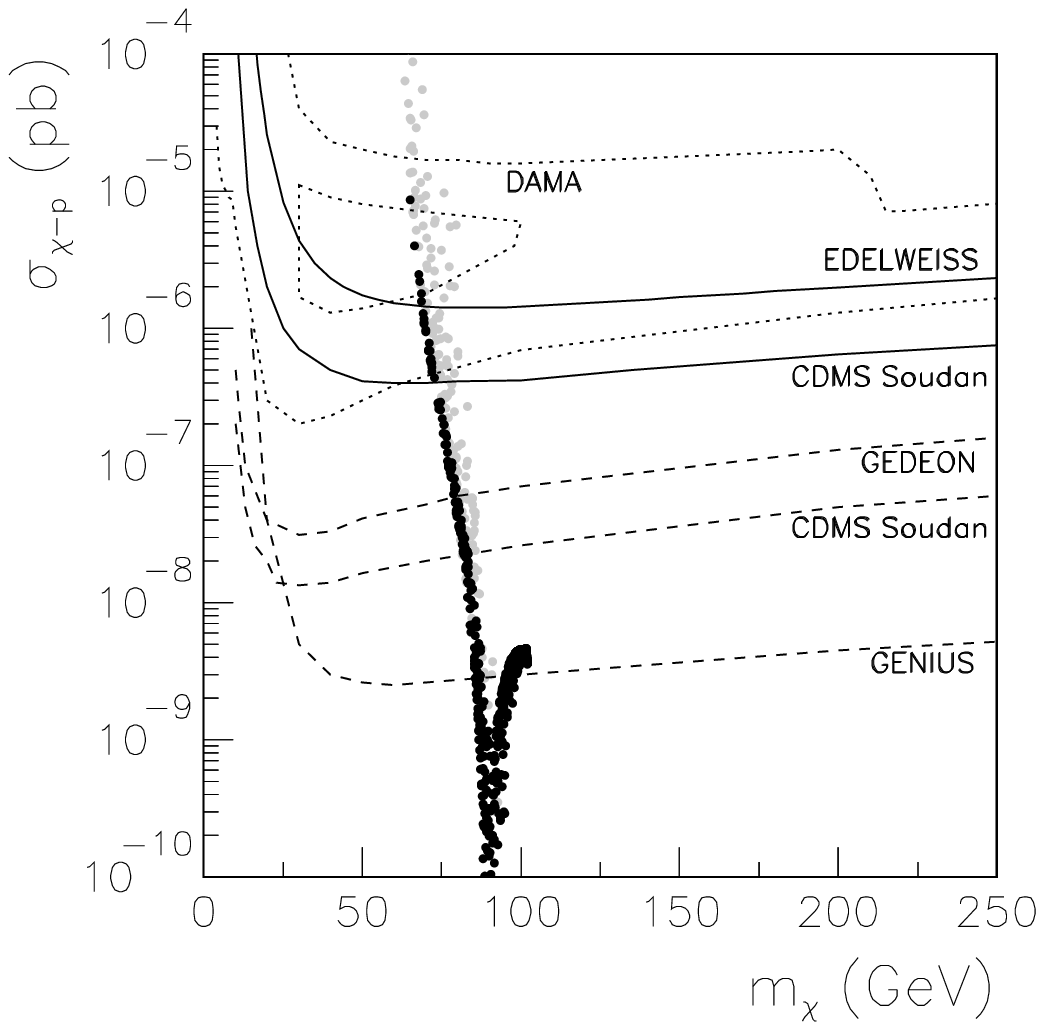,height=8cm}
  \captions{The same as in Fig.\,\ref{++-crossa}a but for the cases
    $A_\kappa=-200$ GeV,  $\mu=110$ GeV, $\tan\beta=4$ and
    $A_\lambda=300$ GeV, on the left, and $A_\kappa=-200$ GeV,
    $\mu=110$ GeV, $\tan\beta=5$ and 
    $A_\lambda=450$ GeV on the right.}
  \label{bbbbb}
\end{figure}

The range of values of $A_\lambda$ for which the
allowed area is more extensive is very dependent on the rest
of the inputs. In particular, since large $\tan\beta$ and
$|A_\kappa|$ increase the diagonal term, ${\cal M}_{P,22}^2$, in the
CP-odd Higgs mass matrix, larger values of $A_\lambda$ can be taken
before ${\cal M}_{P,12}^2$ gets too big. 
For example, in the case with
$A_\kappa=-200$ GeV one can still obtain large accepted regions for
$A_\lambda=300-450$ GeV and $\tan\beta=4-5$, as evidenced in 
Fig.\,\ref{bbbbb}, where points entering the sensitivities of the
present dark matter detectors are obtained with $\neumass\lsim75$ GeV.

\begin{figure}
  \epsfig{file=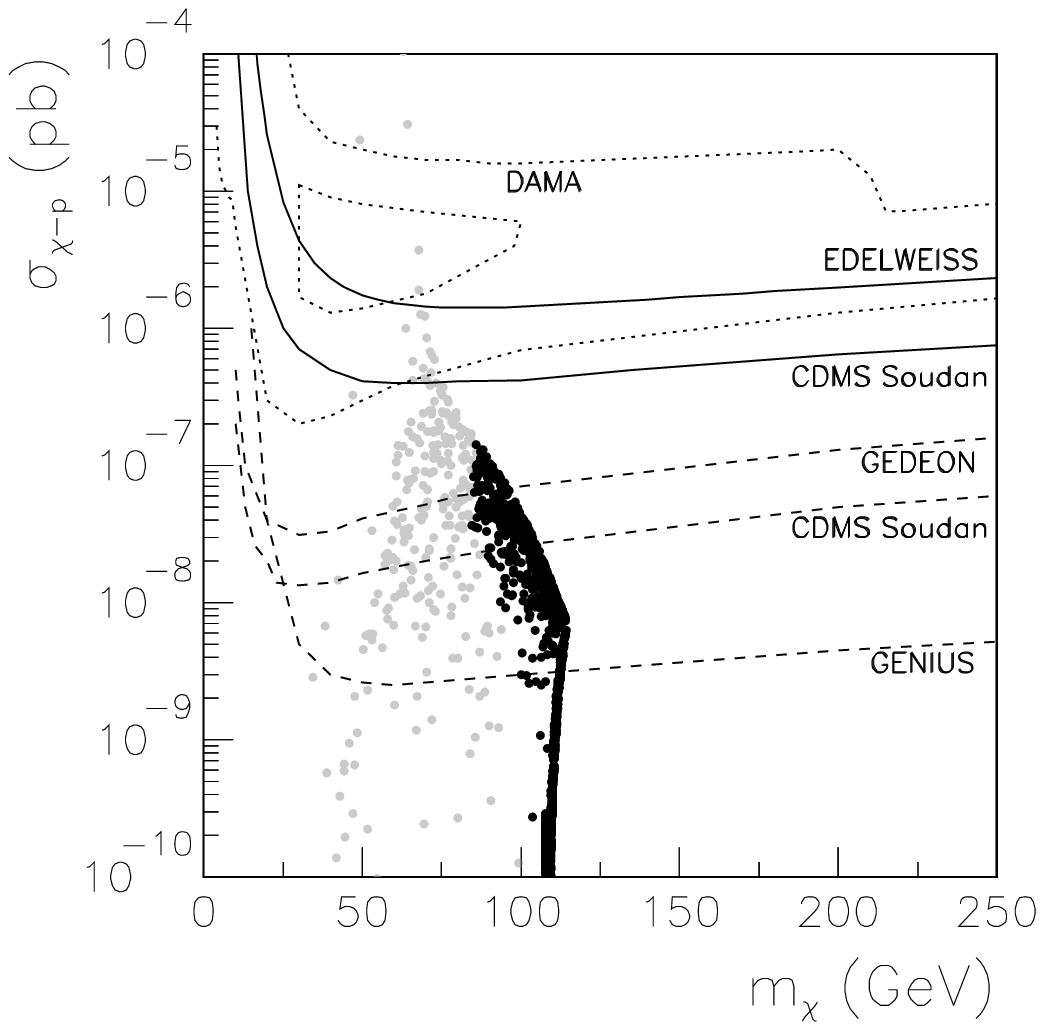,height=8cm}
  \hspace*{-1cm}\epsfig{file=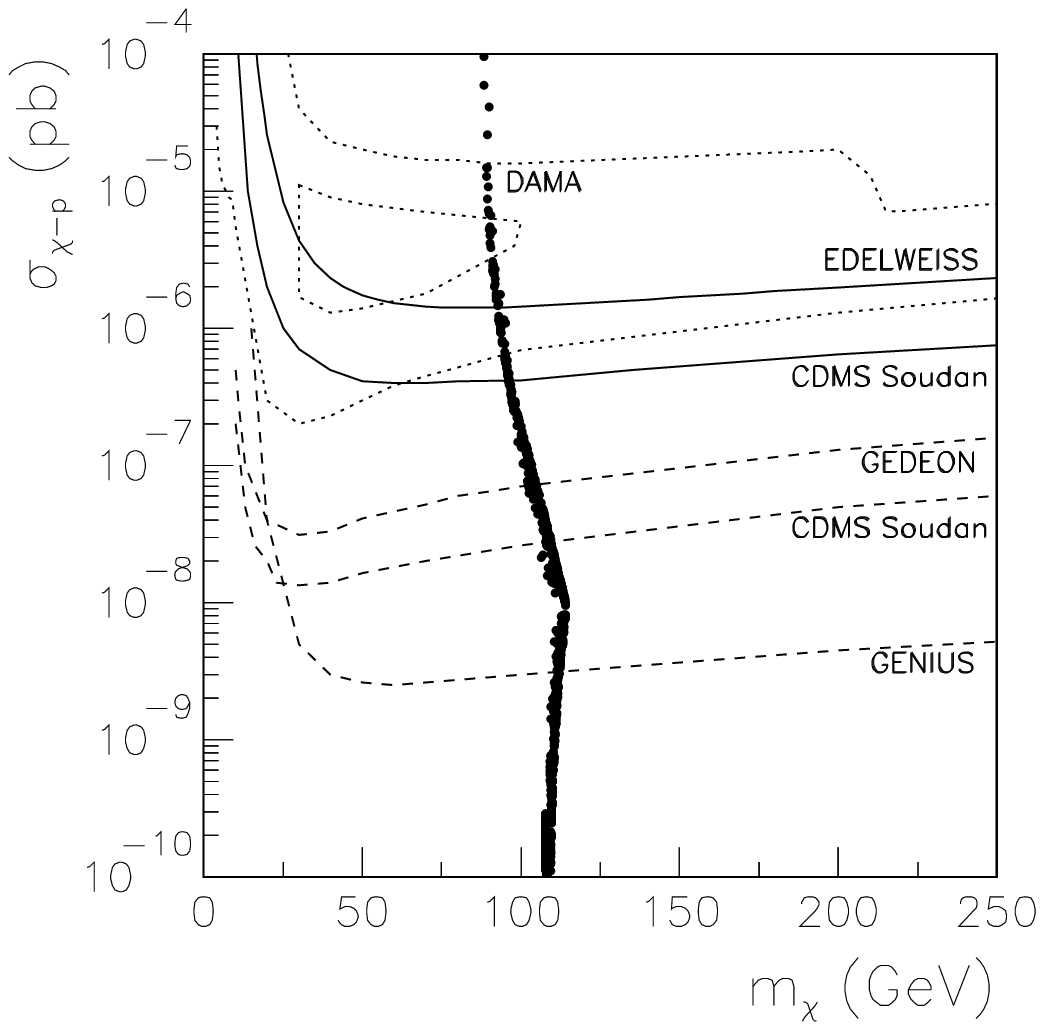,height=8cm}
  \captions{The same as in Fig.\,\ref{++-crossa}a but for the cases
    $A_\lambda=-200$ GeV, $\mu=-110$ GeV, $\tan\beta=3$, and
    $A_\kappa=50,\,200$ GeV,  from left to right.}
  \label{++-sign}
\end{figure}

To complete the analysis of the cases with $\mu A_\lambda>0$ and  $\mu
A_\kappa<0$, we
must address the possibility of having $\mu,\,A_\lambda,\,-A_\kappa<0$.
Note from (\ref{2:pot}) that the tree-level potential, $V_{\rm
  neutral}^{\rm Higgs}$, and therefore the Higgs mass matrices, are
invariant under the exchange of the signs of $\mu$, $A_\lambda$ and
$A_\kappa$, provided that the signs of $\mu A_\lambda$ and $\mu
A_\kappa$ do not change.
This implies that the above analysis regarding the Higgs sector
is identical in this case.
Differences arise, however, in the
neutralino sector
since the signs of $M_{1,2}$
were not altered. Therefore, the neutralino mass spectrum differs, as
well as the lightest neutralino composition. Also the
experimental constraints exhibit a slight variation.
This case presents the same qualitative behaviour as the one
formerly discussed in what the minimization of the Higgs potential is
concerned.
Nevertheless, differences arise regarding the theoretical predictions for
$\crosssec$ due
to the experimental constraints and the different
position of the accidental suppressions in the Higgs-exchange diagrams.
These differences can be sizable for large
$\tan\beta$.
For instance, we have represented in Fig.\,\ref{++-sign} two examples
with $A_\lambda=-200$~GeV, $\mu=-110$~GeV, $\tan\beta=3$, and
$A_\kappa~=~50,\,200$~GeV, where the suppression in $\crosssec$ is
found to occur for $\neumass\approx110$~GeV.

To sum up, we have found that large values of $\crosssec$, even within
the reach of dark matter detectors, can be obtained in the scenarios
analysed in this Subsection. The NMSSM nature is evidenced in these
examples
by the
compositions of the lightest neutralino (which is a singlino-Higgsino
mixed state) and the scalar Higgs (which can be mostly singlet and as
light as $m_{h^0_1}\gsim20$ GeV).

\subsection{$\mu A_\kappa<0$ and $\mu A_\lambda<0$ $(\kappa>0)$}
\label{+--}

This choice comprises the cases $\mu,\,-A_\lambda,\,-A_\kappa>0$ and
$\mu,\,-A_\lambda,\,-A_\kappa<0$.

We first address the possibility
$\mu,\,-A_\lambda,\,-A_\kappa>0$.
When compared with the cases discussed in the previous Subsection,
the occurrence of tachyons in the Higgs sector gives rise to stronger
constraints in this case, both in the CP-even and CP-odd Higgses.

For CP-even Higgses tachyons are now more likely to occur, due
to the negative contributions in ${\cal M}^2_{S,33}$, induced by the
terms proportional to $\mu A_\kappa$ and $A_\lambda/\mu$. Similarly, ${\cal
  M}^2_{S,11}$
receives a sizeable negative contribution from the term proportional
to $\mu A_\lambda$ which is particularly dangerous
for large values of $\tan\beta$.
In the CP-odd sector, an analogous study of the mass matrix shows that
tachyons are more restrictive for large
values of $\lambda$ and small values of $\kappa$. Actually,
from the naive requirement
${\cal M}^2_{P,11}\geq0$ the following constraint is obtained
$\kappa\geq-\lambda A_\lambda/\mu$.
In fact, this ensures the positiveness of the denominator in
condition (ii)
derived from the minimization
of the Higgs potential in Section\,\ref{Sec:minimization}, and gives a
qualitative idea on the dependence of the tachyonic region on the
parameters $A_\lambda$ and $\mu$.
When compared with the cases treated in the
former Subsection, larger regions of the parameter space are now
excluded. We found that tachyons in the CP-odd sector typically
give rise to stronger constraints than those from CP-even sector,
although the corresponding excluded regions practically coincide.

The experimental constraints from the neutralino sector
are not very stringent in these examples, owing to the fact that
the regions where the neutralino would have a small mass
are typically
excluded by the occurrence of tachyons.

\begin{figure}
  \epsfig{file=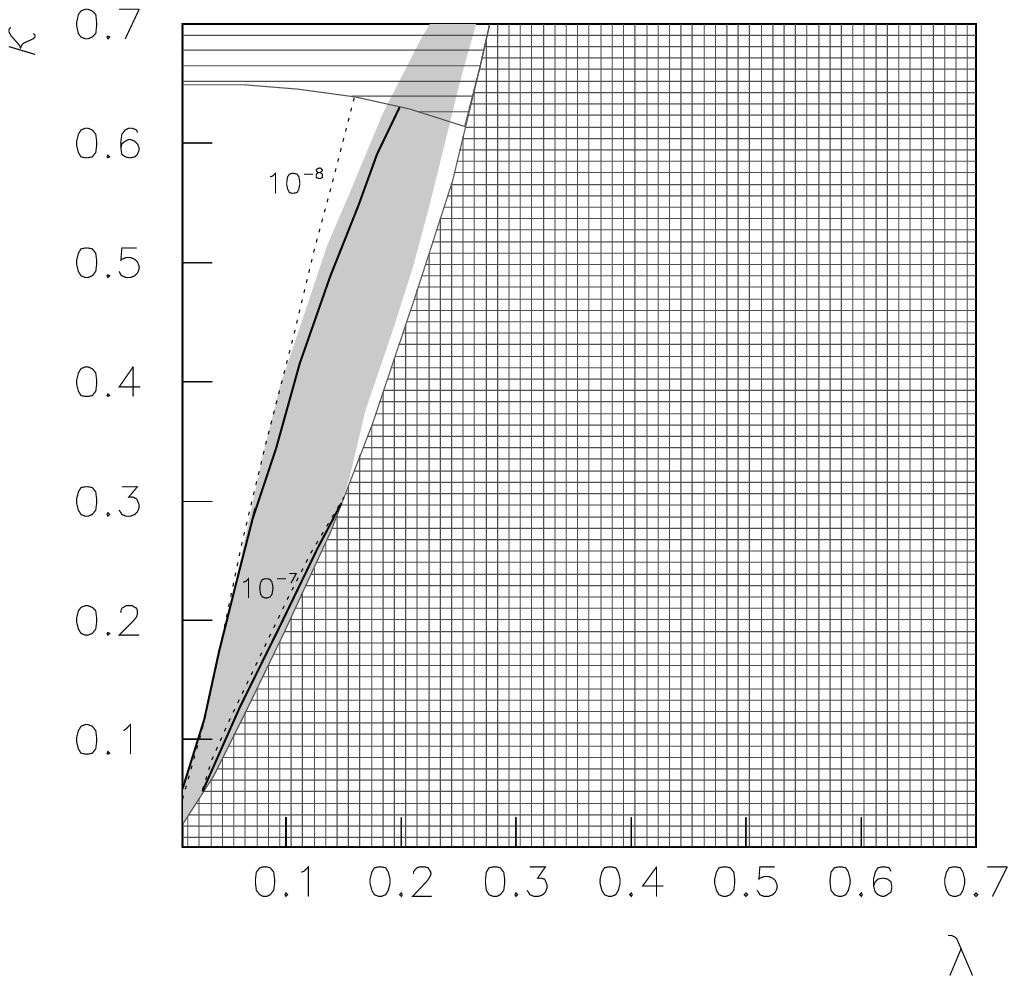,height=8cm}
  \hspace*{-1cm}\epsfig{file=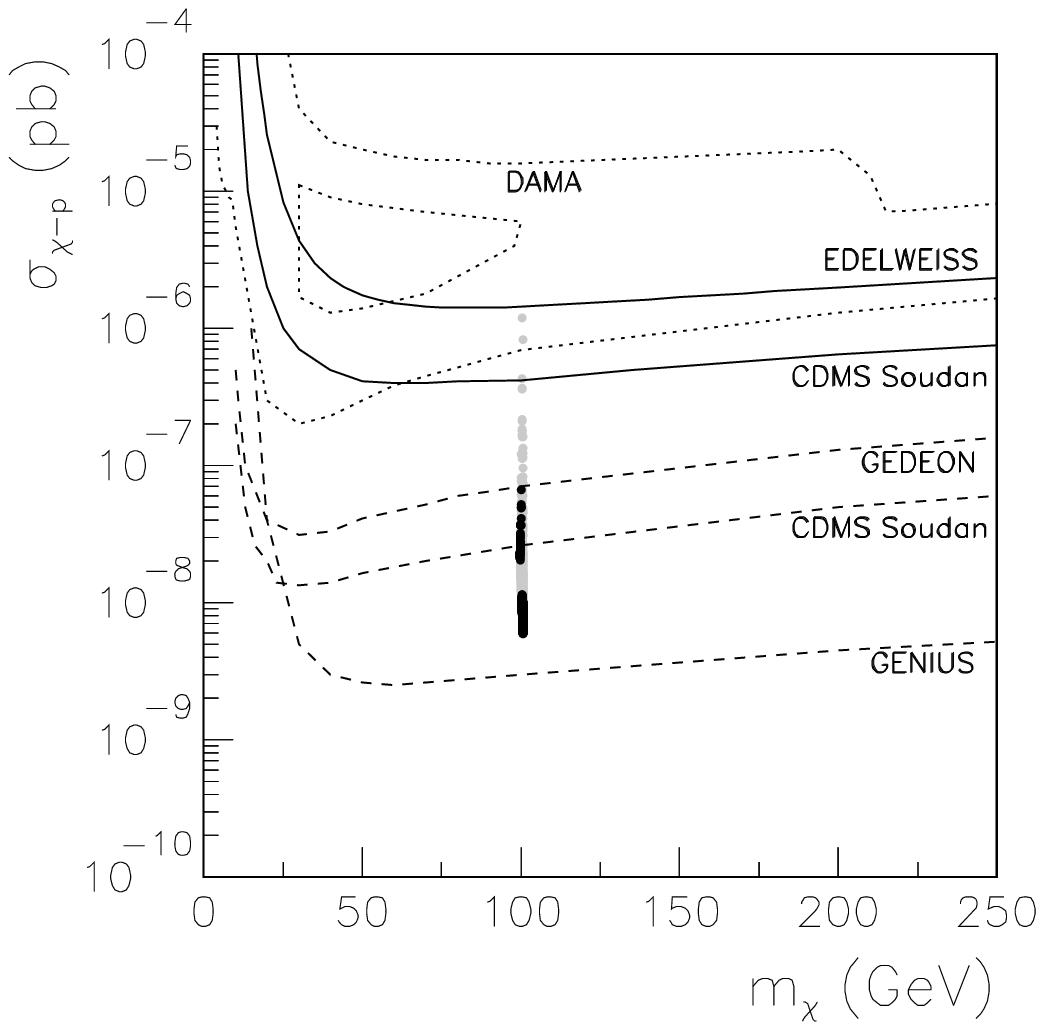,height=8cm}
  \captions{The same as in Fig.\,\ref{++-ka} but for
    $\tan\beta=3$, $A_\lambda=-200$ GeV, $A_\kappa=-50$ GeV and
    $\mu=110$ GeV. Only the lines with $m_{h_1^0}=114,\,75$~GeV are
    represented, and none of the 
    lines showing the lightest scalar Higgs and
    neutralino composition is depicted, since $S_{13}^{\,2}<0.1$ and 
    $N_{15}^2<0.1$ in all the plane.}
  \label{+--a}
\end{figure}

As an example, the $(\lambda,\kappa)$ plane is represented 
in Fig.\,\ref{+--a} for $\tan\beta=3$, \linebreak $A_\lambda=-200$ GeV,
$A_\kappa=-50$ GeV and  $\mu=110$ GeV. In this case, and in contrast
with what was displayed in Fig.\,\ref{++-a}, there exists a very large
region where one cannot find minima of $V_{\rm neutral}^{\rm
  Higgs}$.
In particular, $\lambda\gsim0.25$ is now excluded for
this reason.
In the rest of the parameter space, experimental constraints become
very important in those regions with small values of the CP-even and
CP-odd masses.
Although the most important exclusion is
due to
DHDM constraints ($h^0\to b\bar b$ and $h^0\to \tau^+\tau^-$),
some regions not
fulfilling the bounds on APM ($h^0a^0\to 4b$'s)
also appear.
It is worth emphasizing that in the remaining allowed regions the
lightest Higgs is doublet-like ($S_{13}^{\,2}\lsim0.003$)
and its mass is never too small,
$m_{h^0_1}\gsim85$ GeV.

Regarding the composition of the lightest neutralino, it
turns out to be Higgsino-like in all the allowed
parameter
space ($N_{13}^2+N_{14}^2\gsim0.98$).
As we already mentioned, those regions with small $\lambda$ and
$\kappa$ that would lead to a singlino-like neutralino are
excluded by the absence of physical minima in the potential.
For this reason the mass of the neutralino is
dictated by the value of the $\mu$ term and we found
$\neumass\approx\mu$ throughout the allowed parameter space.

This is shown in
Fig.\,\ref{+--a}, which represents the corresponding values of the
neutralino-nucleon cross section as a function of 
the neutralino mass. The
cross section ranges from
$6\times10^{-9}$ pb $\lesssim\crosssec\lesssim7\times10^{-8}$ pb in
this case. Once more, although higher values could be obtained, these
are typically excluded due to the constraints on the CP-even Higgses.

Let us now address the relevance of variations in $A_\kappa$ in the
allowed regions of the parameter space and thus on the predictions
for $\crosssec$.
It can be seen that the increase in $|A_\kappa|$ (i.e., making it more
negative) translates into an almost negligible enlargement in the
allowed area,
while
the experimental constraints on CP-even Higgses become more
restrictive.
On the other hand,
a decrease in the value of $|A_\kappa|$ leads to a lighter CP-odd
Higgs and the tachyonic region increases, as can be easily understood
from the mass matrix (\ref{2:Amass}). For instance,
in the particular case of
$A_\kappa=0$, and unless
$|A_\lambda|$ is also very small, the entire parameter space can
be excluded.

\begin{figure}
  \epsfig{file=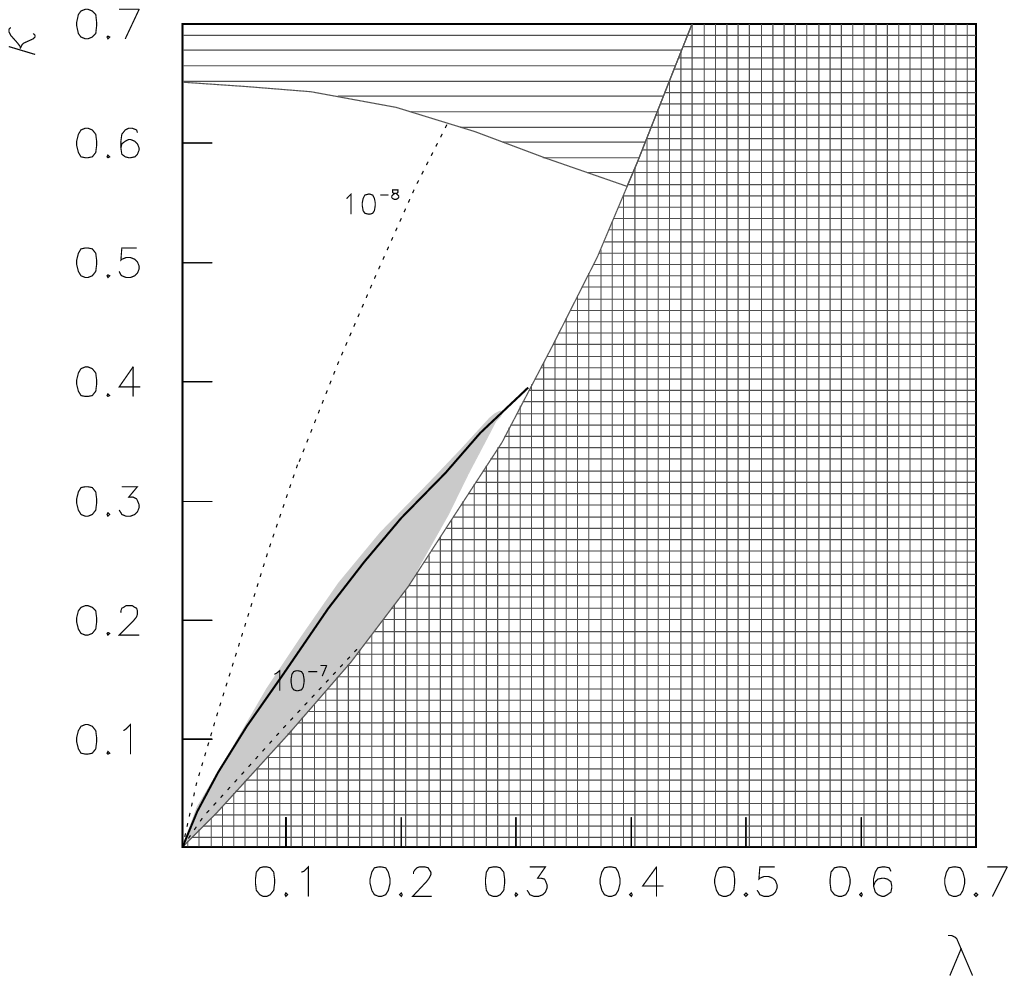,height=8cm}
  \hspace*{-1cm}\epsfig{file=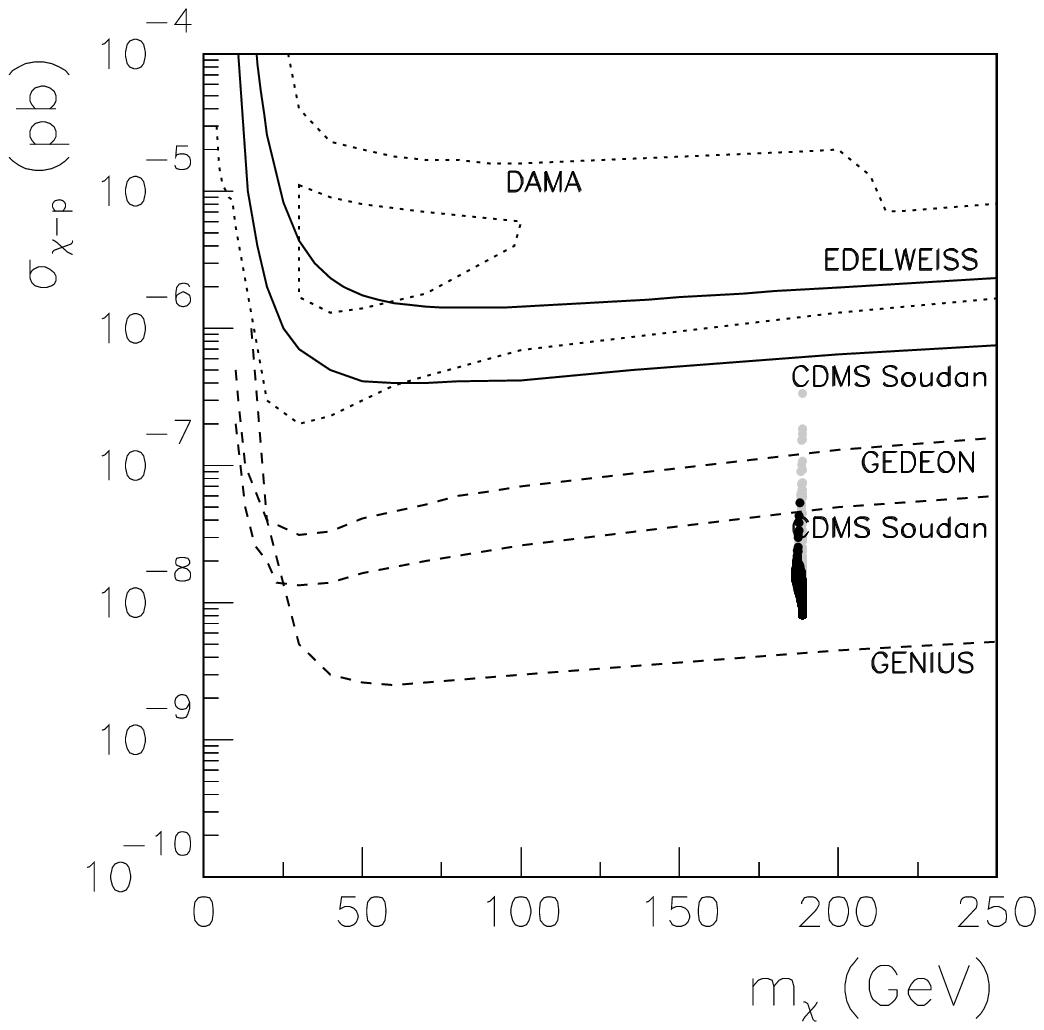,height=8cm}
  \captions{The same as in Fig.\,\ref{++-ka} but for
    $\tan\beta=3$, $A_\lambda=-200$ GeV, $A_\kappa=-50$ GeV and
    $\mu=200$ GeV. Only the line with $m_{h_1^0}=114$~GeV is
    represented, and none of the 
    lines showing the lightest scalar Higgs and
    neutralino composition is depicted, since $S_{13}^{\,2}<0.1$ and 
    $N_{15}^2<0.1$ in all the plane.}
  \label{+--mu}
\end{figure}

Regarding changes in $A_\lambda$ and $\mu$, these clearly affect the
regions excluded by tachyons. Large values of $\mu$ and small
$|A_\lambda|$
allow an increase in the accepted regions, in agreement with the
condition
on $\kappa$ derived above, $\kappa\geq-\lambda A_\lambda/\mu$.
Also, note that, since the masses of
the Higgses increase, the associated experimental constraints become less
restrictive and the allowed area is larger.
Nevertheless, the region where the neutralino would have an
important singlino composition is still typically excluded, and therefore
in the allowed region $\neut$ is still Higgsino-like, with
$\neumass\approx\mu$.
Despite the increase of $\neumass$, the predictions for the cross
section are essentially unaltered. An example with $\tan\beta=3$,
$A_\lambda=-200$ GeV,
$A_\kappa=-50$ GeV and $\mu=200$ GeV is represented in
Fig.\,\ref{+--mu}, displaying both the $(\lambda,\kappa)$ plane
and the neutralino-nucleon cross section versus the neutralino
mass.
We find $\crosssec\lsim5\times10^{-8}$ pb, similar to what was found
in Fig.\,\ref{+--a}, but now
with $\neumass\approx190$ GeV. The singlino component of $\neut$ is
negligible ($N_{15}^2\lsim0.006$) and the scalar Higgs is doublet-like
($S_{13}^{\,2}\lsim0.001$).

\begin{figure}
  \epsfig{file=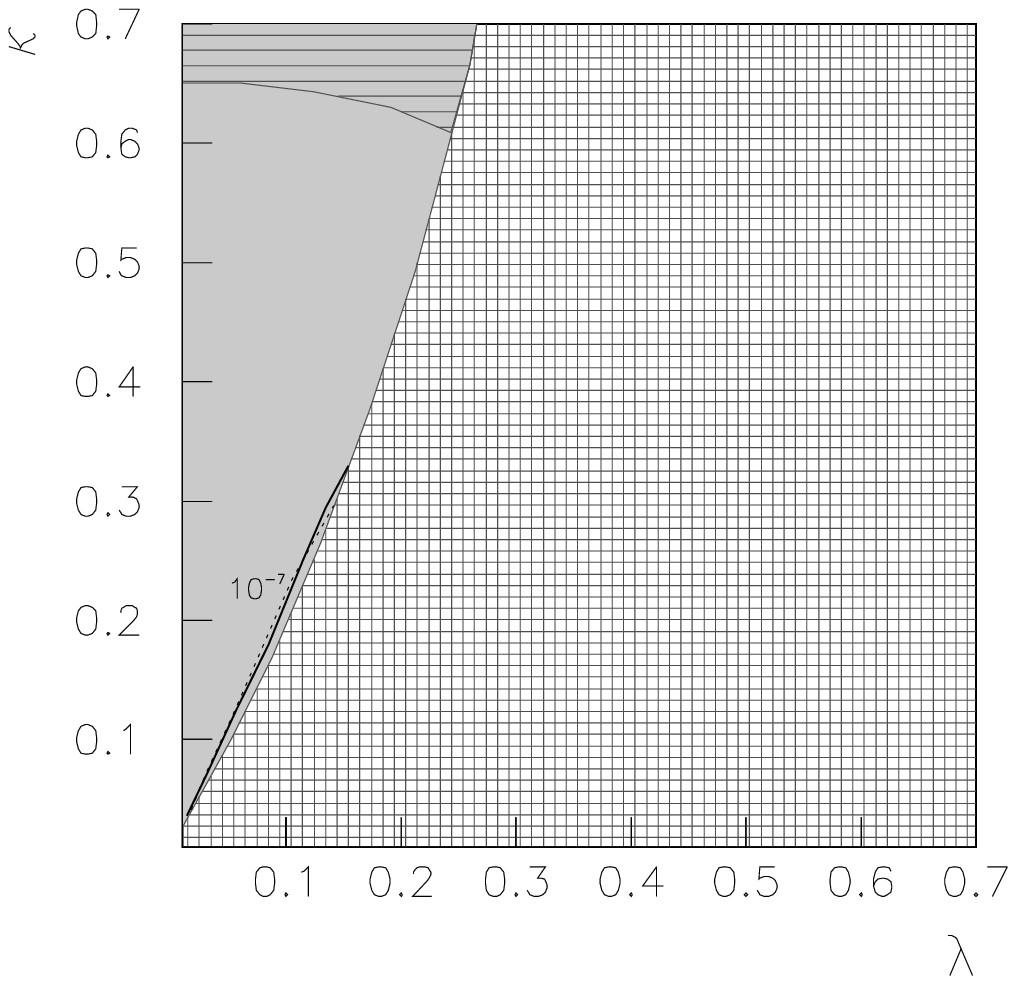,height=8cm}
  \hspace*{-1cm}\epsfig{file=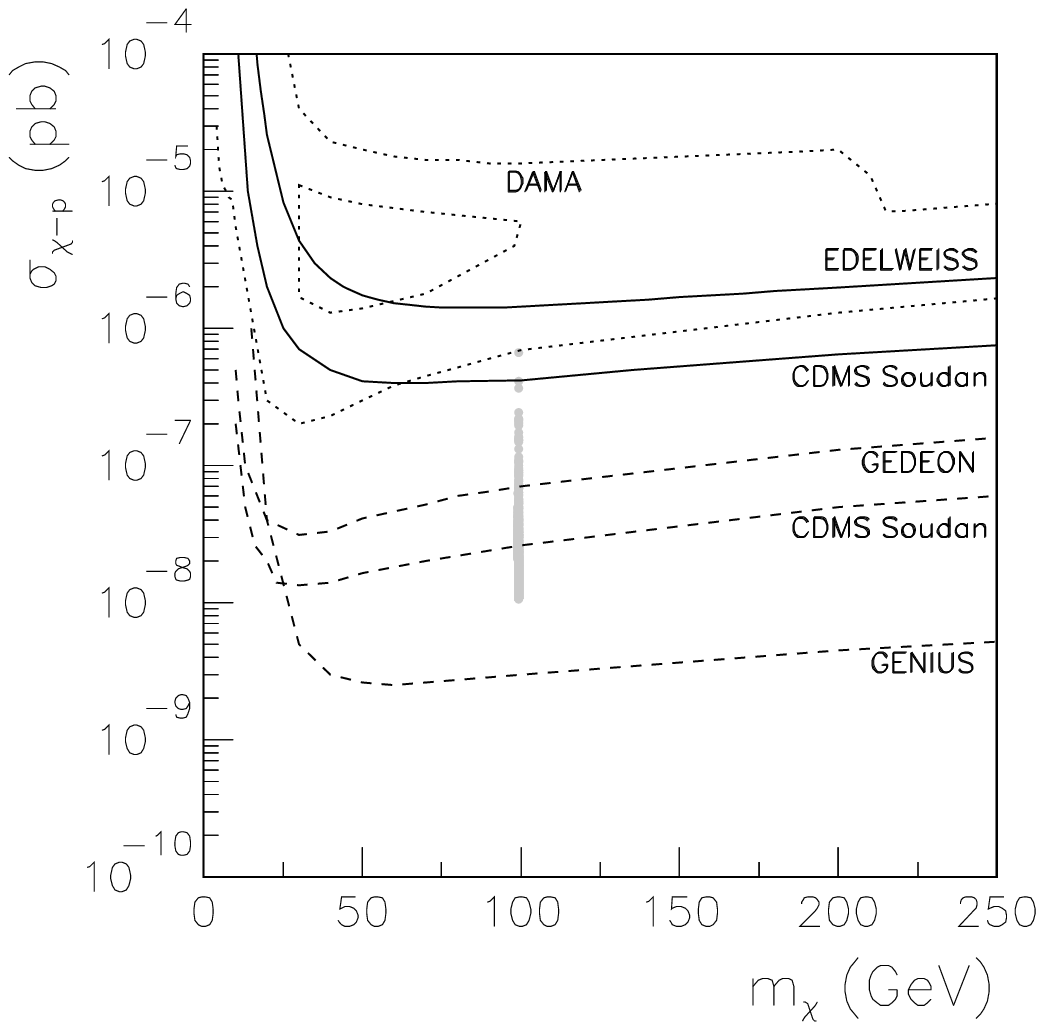,height=8cm}
  \\[-3ex]
  \epsfig{file=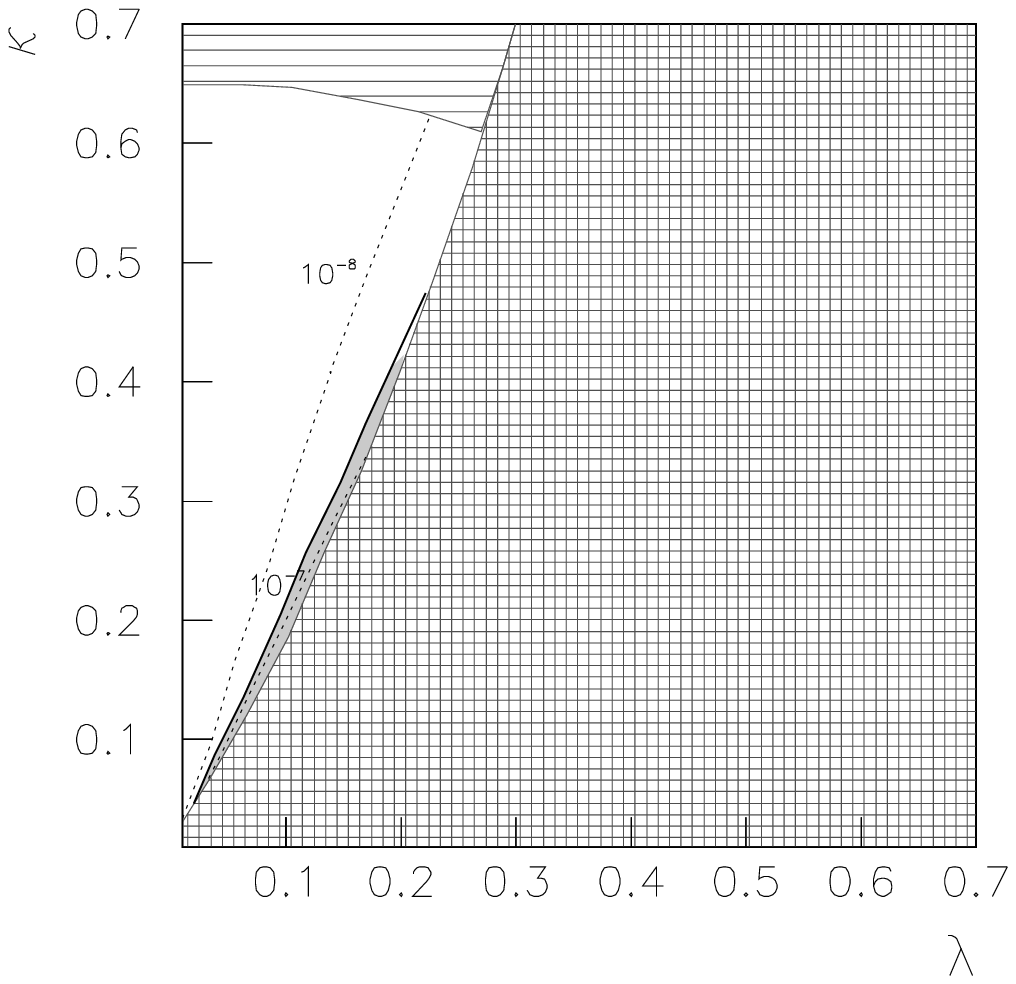,height=8cm}
  \hspace*{-1cm}\epsfig{file=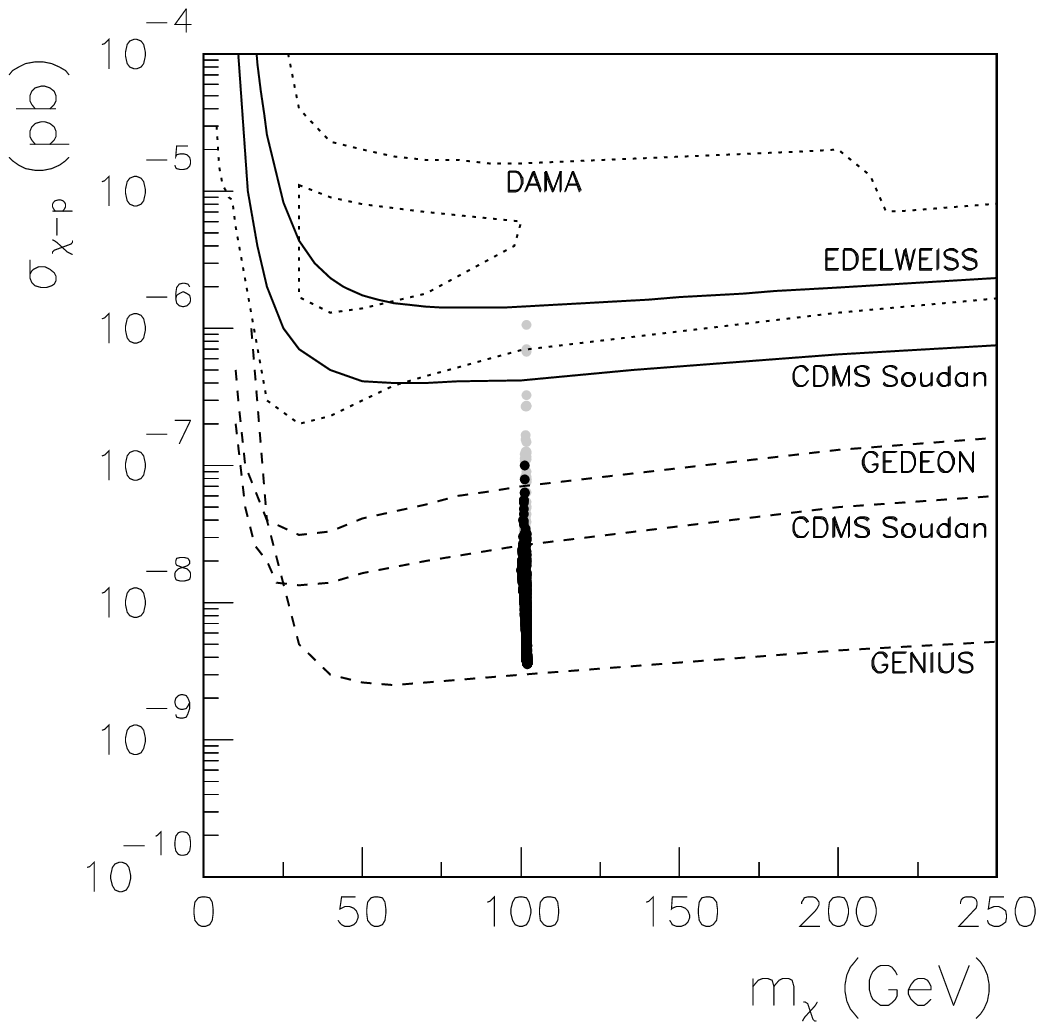,height=8cm}
  \captions{The same as in Fig.\,\ref{++-ka} but for the cases
    $A_\lambda=-200$ GeV, $A_\kappa=-50$ GeV,  $\mu=110$ GeV, and
    $\tan\beta=2,\,5$, from top to bottom. In the left frames,
    only the line with
    $m_{h_1^0}=114$~GeV is 
    represented, and none of the 
    lines showing the lightest scalar Higgs and
    neutralino composition is depicted, since $S_{13}^{\,2}<0.1$ and 
    $N_{15}^2<0.1$ in all the plane.}
  \label{+--crossa}
\end{figure}

Finally, regarding variations in the value of $\tan\beta$, these
have little effect on the shape of the tachyonic region, whereas
experimental constraints are more sensitive to them.
As in the former scenario, for low values
of $\tan\beta$ light scalar Higgses are obtained. Since these are
predominantly doublet-like,
experimental constraints
(especially those associated with DHDM, namely, $h^0\to b\bar b$ and
$h^0\to \tau^+\tau^-$)
become very important and forbid, for
instance, the whole parameter space in the case $\tan\beta=2$. On the
other hand, larger values of $\tan\beta$ are welcome in order to
increase the value of $m_{h_1^0}$, obtaining also
a moderate enhancement of the
cross section. In order to illustrate this discussion, we
represent in Fig.\,\ref{+--crossa} two cases with $\tan\beta=2,5$, for
$A_\lambda=-200$ GeV,
$A_\kappa=-50$~GeV and  $\mu=110$ GeV. We find that the cross section
can reach $\crosssec\approx10^{-7}$ pb in the case where
$\tan\beta=5$.
None of these examples displays any qualitative change regarding the
neutralino and Higgs compositions.

To complete the analysis of the cases with $\mu A_\lambda<0$ we still
have to consider the case $\mu,\,-A_\lambda,\,-A_\kappa<0$. As we
explained in the former Subsection, the analysis of the
Higgs sector will be
analogous to that of the case we have just studied.
Despite the differences in the neutralino
sector, its mass, composition and detection cross section will also be
qualitatively equal to those previously discussed.

Summarizing, all the cases we have analysed in this Subsection present
as common features the appearance of Higgsino-like neutralinos with a
detection cross section which can be as large as
$\crosssec\lsim10^{-7}$ pb, and doublet-like Higgses.

\subsection{$\mu A_\kappa>0$ and $\mu A_\lambda>0$ $(\kappa>0)$}
\label{+++}

We consider now those cases where $\mu A_\kappa>0$, and $\mu
A_\lambda>0$, conditions which are fulfilled in the cases
$\mu,\,A_\lambda,\,A_\kappa>0$ and
$\mu,\,A_\lambda,\,A_\kappa<0$.

We will begin
with all $\mu$,
$A_\lambda$, and $A_\kappa$ positive. As in the previous
cases, a simple analysis
of the tree-level Higgs mass matrices gives a qualitative
understanding on the nature and extension of the tachyonic regions in
the parameter space.

In this particular case tachyons in the CP-odd sector arise
through the negative contribution
$-\frac{3\kappa\mu}{\lambda}A_\kappa$ in ${\cal M}^2_{P,22}$. Since this
is mainly compensated by the positive term
$\frac{\lambda^2v^2}{\mu}A_\lambda\sin2\beta$, the tachyonic region
occurs for small values of $\lambda$. The excluded region is obviously
more important for small values of
$A_\lambda$ and large $\mu$, $A_\kappa$,
and $\tan\beta$.
The occurrence of tachyons in the CP-even Higgses is analogous to the
case $\mu A_\lambda>0$, $\mu A_\kappa<0$ discussed in
Subsection\,\ref{++-},
due to the increase of the
off-diagonal terms in the mass matrix. As in that case, tachyons
appear for large values of $\lambda$ and small $\kappa$
and become more stringent
as $\tan\beta$ grows.

Experimental constraints play also a very relevant role in this
case.
Close to the tachyonic regions the experimental bounds on the
Higgs sector are very severe. In particular IHDM, DHDM ($h^0\to b\bar
b$, $h^0\to 2\,{\rm jets}$) are responsible for the most important
exclusions, although APM (mainly $h^0a^0\to4\,b$'s)
may also be violated.
Finally, excluding those regions where the direct neutralino
production is in disagreement with the experimental bounds leads to
important
constraints in the region with light $\neut$.

\begin{figure}
  \epsfig{file=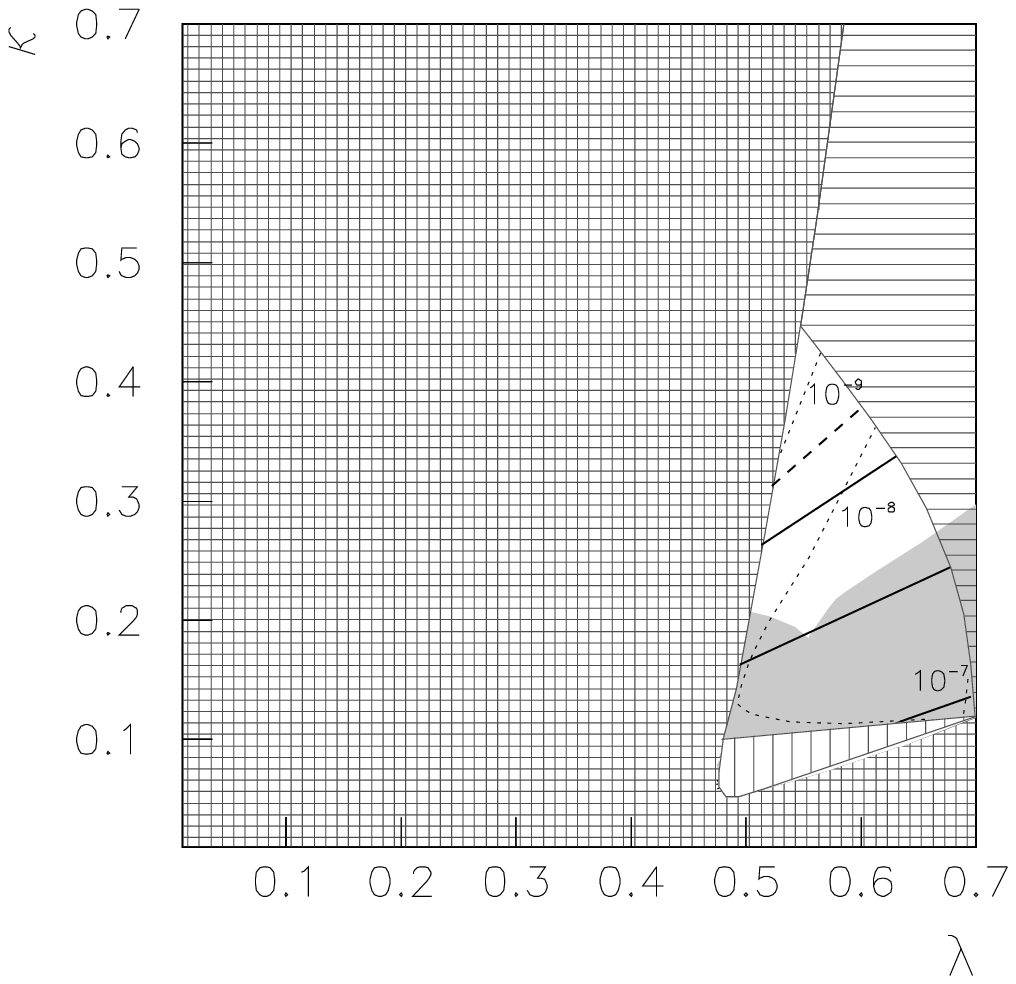,height=8cm}
  \hspace*{-1cm}\epsfig{file=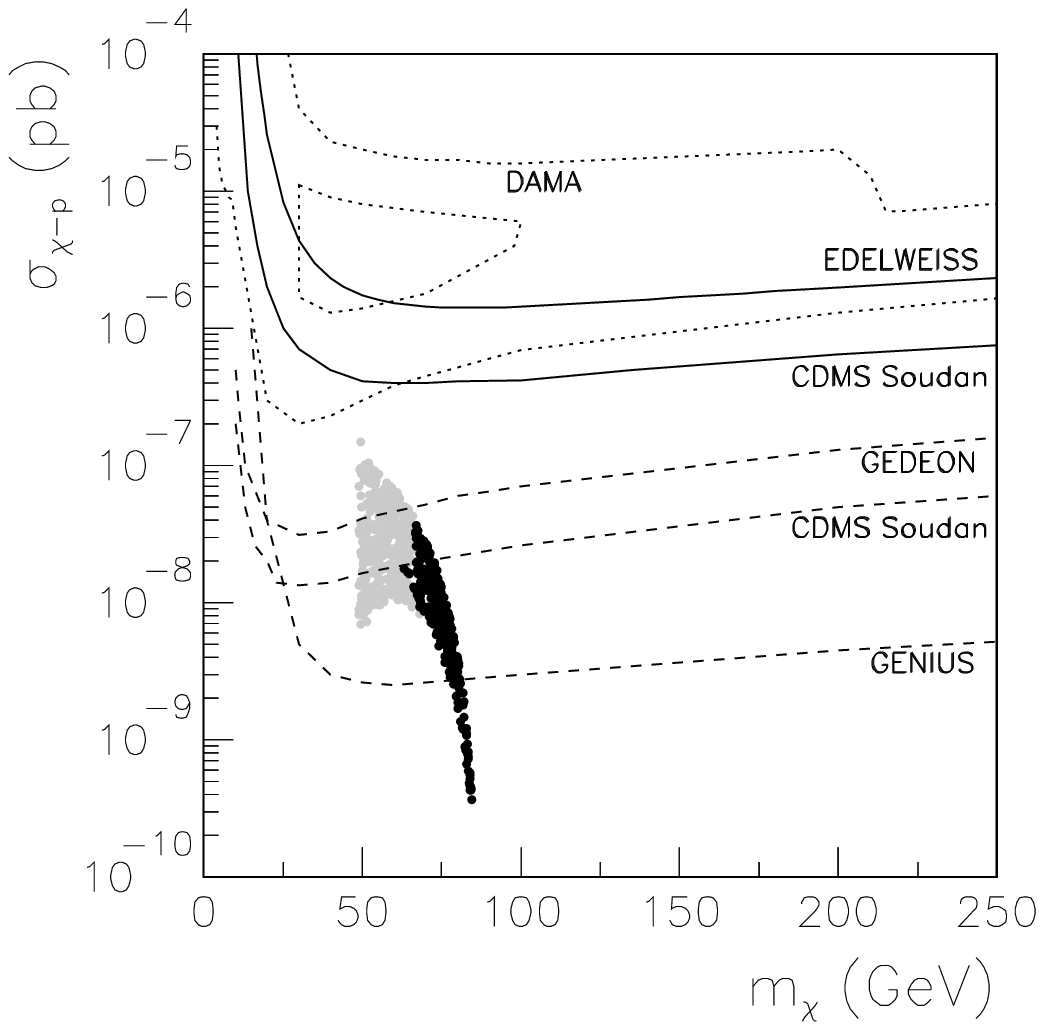,height=8cm}
  \captions{The same as in Fig.\,\ref{++-ka} but for
    $A_\lambda=200$ GeV, $\mu=110$
    GeV, $A_\kappa=50$ GeV,
    and $\tan\beta=3$. Only the line with $S_{13}^{\,2}=0.1$ is
    represented, and none of the lines showing the neutralino
    composition is depicted 
    since $0.5>N_{15}^2>0.1$ in all the plane.}
  \label{+++a}
\end{figure}

An example with $A_\lambda=200$ GeV, $\mu=110$ GeV, $A_\kappa=50$ GeV
and $\tan\beta=3$ is represented in Fig.\,\ref{+++a}, depicting the
constraints on the
$(\lambda,\kappa)$ plane and the corresponding predictions
for $\crosssec$ versus the neutralino mass.
In the small experimentally allowed region the lightest neutralino is
a mixed singlino-Higgsino state, with $N_{15}^2\lsim0.4$ and
$N_{13}^2+N_{14}^2\gsim0.6$,
and the lightest scalar Higgs can have an important
singlet component ($S_{13}^{\,2}\lsim0.8$). The experimental constraints impose
$\neumass\gsim70$ GeV and  $m_{h_1^0}\gsim85$ GeV, which set a limit on the
theoretical predictions for the neutralino-nucleon cross section at
$\crosssec\lsim 4\times10^{-8}$ pb.

Variations of $A_\kappa$ have an important impact on the allowed
parameter space. As already commented, the region excluded due to
tachyons in the CP-odd Higgs sector 
increases for larger values of $A_\kappa$. For instance,
in the example with
$A_\lambda=200$ GeV, $\mu=110$ GeV,
and $\tan\beta=3$, the allowed region completely
disappears for $A_\kappa\gsim110$~GeV.
On the other hand, decreasing the value of $A_\kappa$ the parameter
space is enlarged. Recall that the minimal value $A_\kappa=0$ has
already been analysed in Section \ref{++-} in the context of a
scenario with $\mu,\,A_\lambda,-A_\kappa>0$.

\begin{figure}
  \epsfig{file=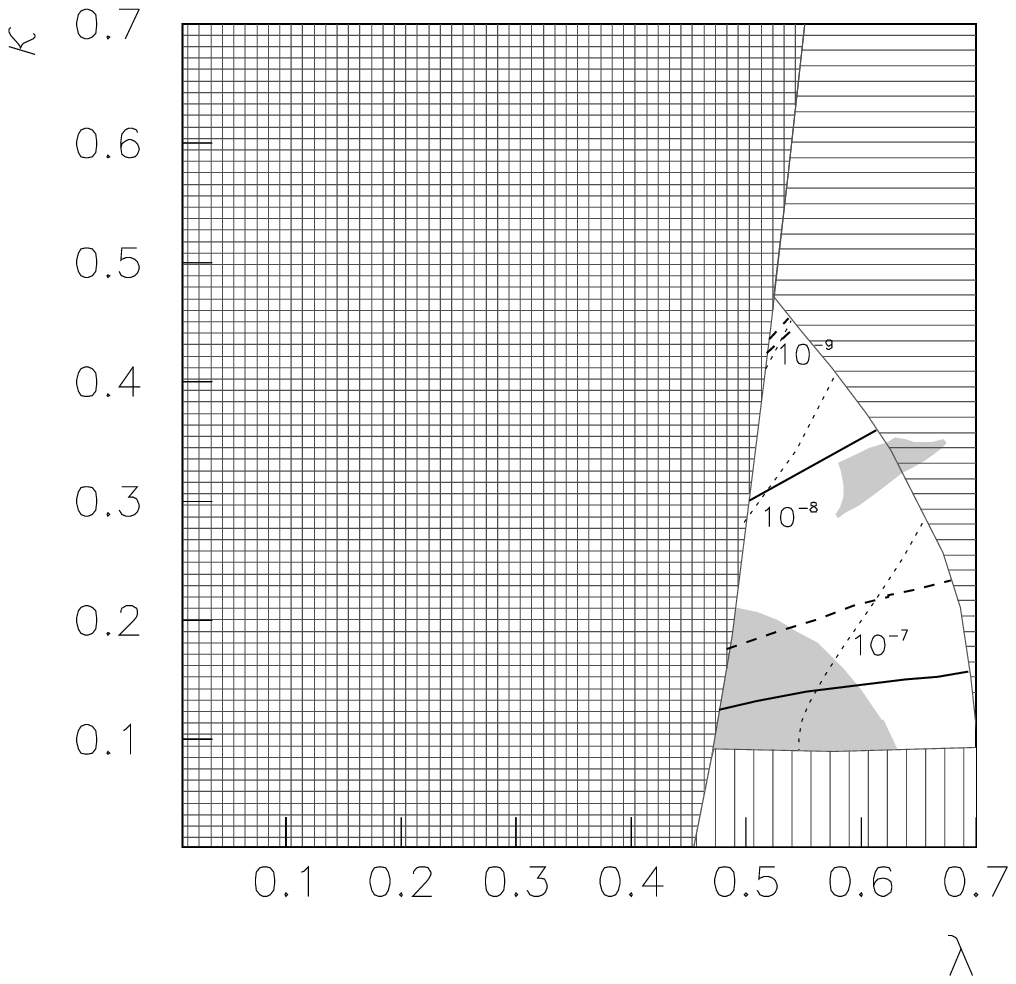,height=8cm}
  \hspace*{-1cm}\epsfig{file=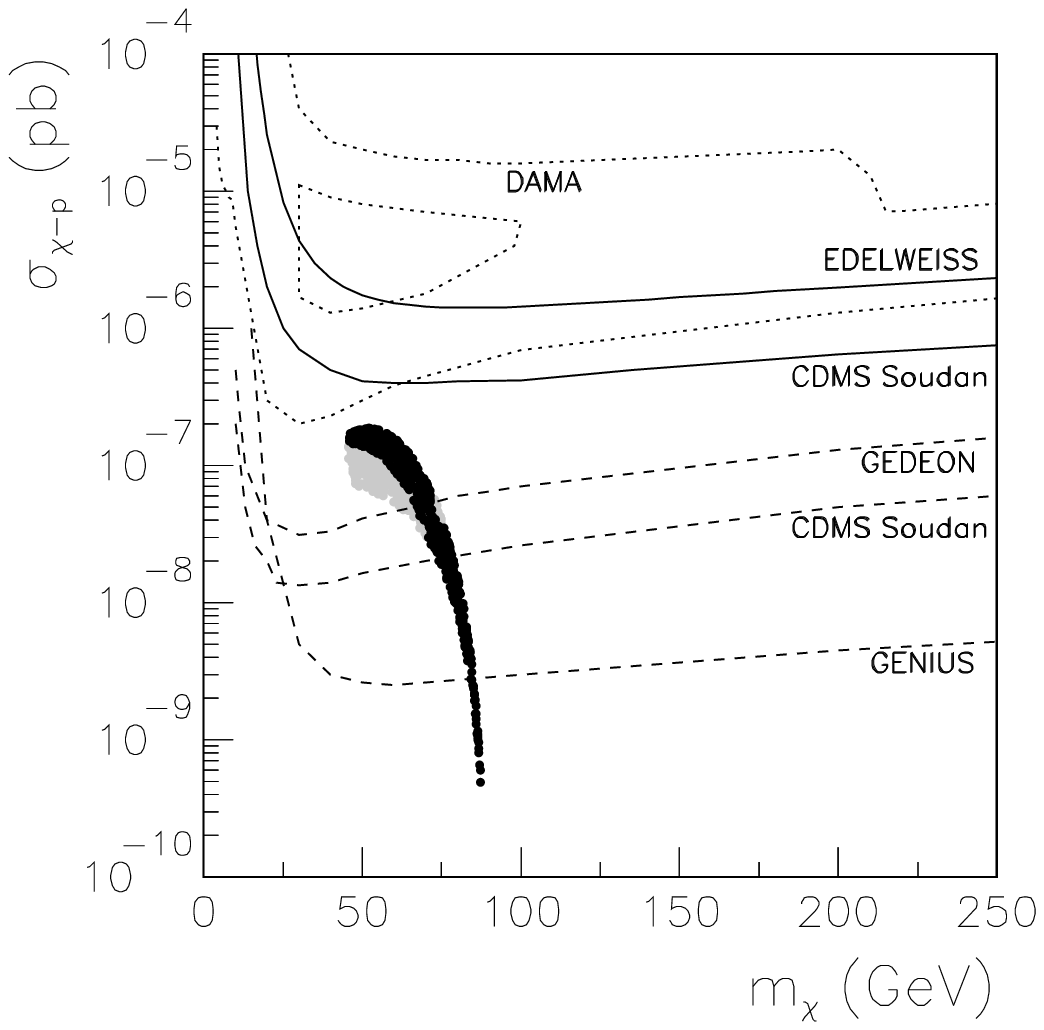,height=8cm}
  \captions{The same as in Fig.\,\ref{++-ka} but for
    $A_\lambda=300$ GeV, $\mu=110$
    GeV, $A_\kappa=50$ GeV,
    and $\tan\beta=3$. Only the lines with $m_{h_1^0}=114,\,75$~GeV are
    represented. Regarding the neutralino composition, only the line
    with $N_{15}^2=0.1$ is shown in the upper corner of the allowed
    region, since in the rest of the parameter space
    $0.5>N_{15}^2>0.1$.} 
  \label{+++al}
\end{figure}

Decreasing the value of $A_\lambda$ also leads to an increase of
regions with a tachyonic pseudoscalar. If  $\mu=110$ GeV, $A_\kappa=50$
and $\tan\beta=3$, the whole parameter space is excluded for
$A_\lambda\lsim50$ GeV. On the other hand, a moderate increase of
$A_\lambda$ helps avoiding tachyons, especially in the CP-even
sector.
An example with $A_\lambda=300$ GeV can be found in
Fig.\,\ref{+++al}, where the
$(\lambda,\kappa)$ plane and the theoretical predictions for
$\crosssec$ are represented.
Since in this case the experimental constraints from Higgs decays
are less severe, we find that the regions
with very light Higgs and $\neut$ are now experimentally viable.
In particular, neutralinos with an important singlino composition,
$N_{15}^2\lsim0.45$, can be obtained with $\neumass\gsim45$ GeV,
whereas the lightest Higgses ($m_{h^0_1}\approx65-90$ GeV) are all
singlet-like.
This in turn
favours larger values of the cross section
($\crosssec\lsim2\times10^{-7}$ pb),
and compatibility with present experiments
is almost obtained.
Should we further increase the value of $A_\lambda$, the experimental
constraints associated with the scalar Higgs would become again more
important.

In order to prevent the occurrence of tachyons in the CP-odd Higgses,
the value of $\mu$ has to be small. For instance, taking $\mu=200$
GeV in the example with $A_\lambda=200$~GeV,
$A_\kappa=50$ GeV and $\tan\beta=3$, all the $(\lambda,\kappa)$
plane would be excluded.

\begin{figure}
  \epsfig{file=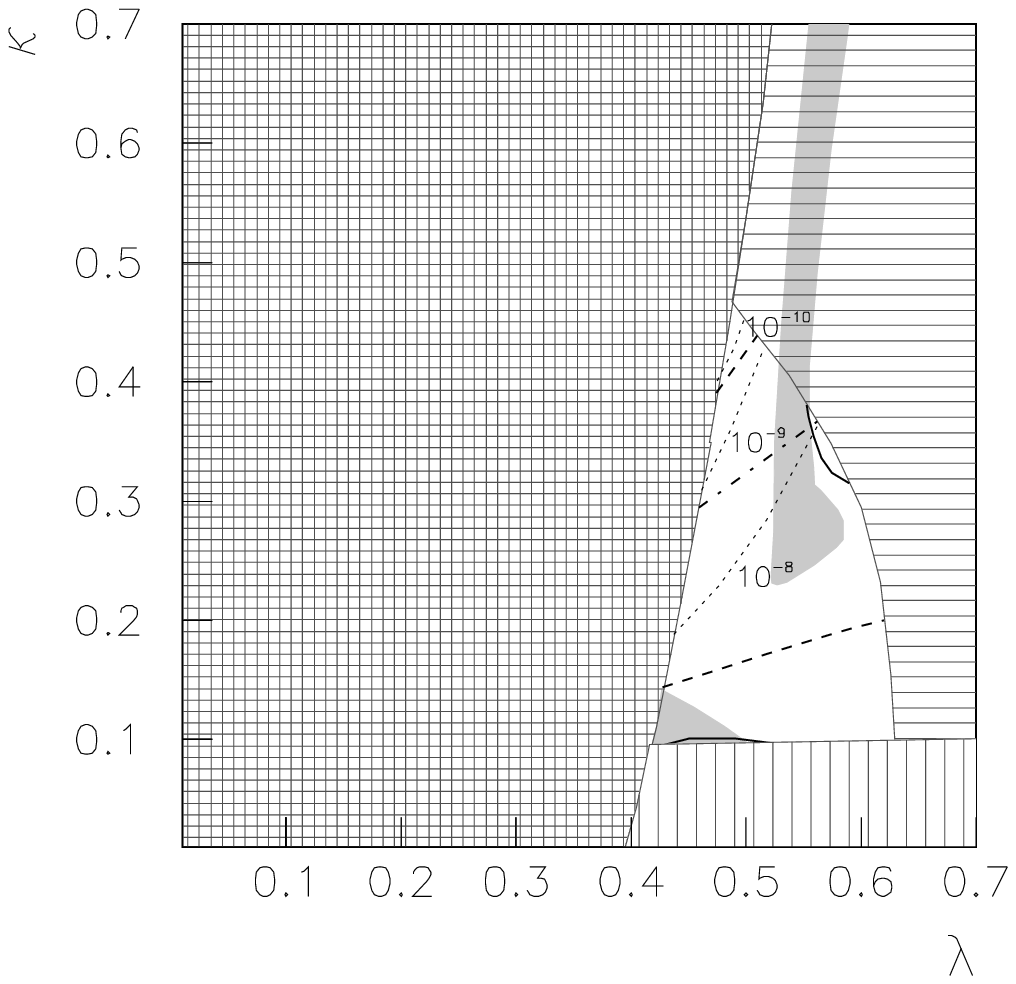,height=8cm}
  \hspace*{-1cm}\epsfig{file=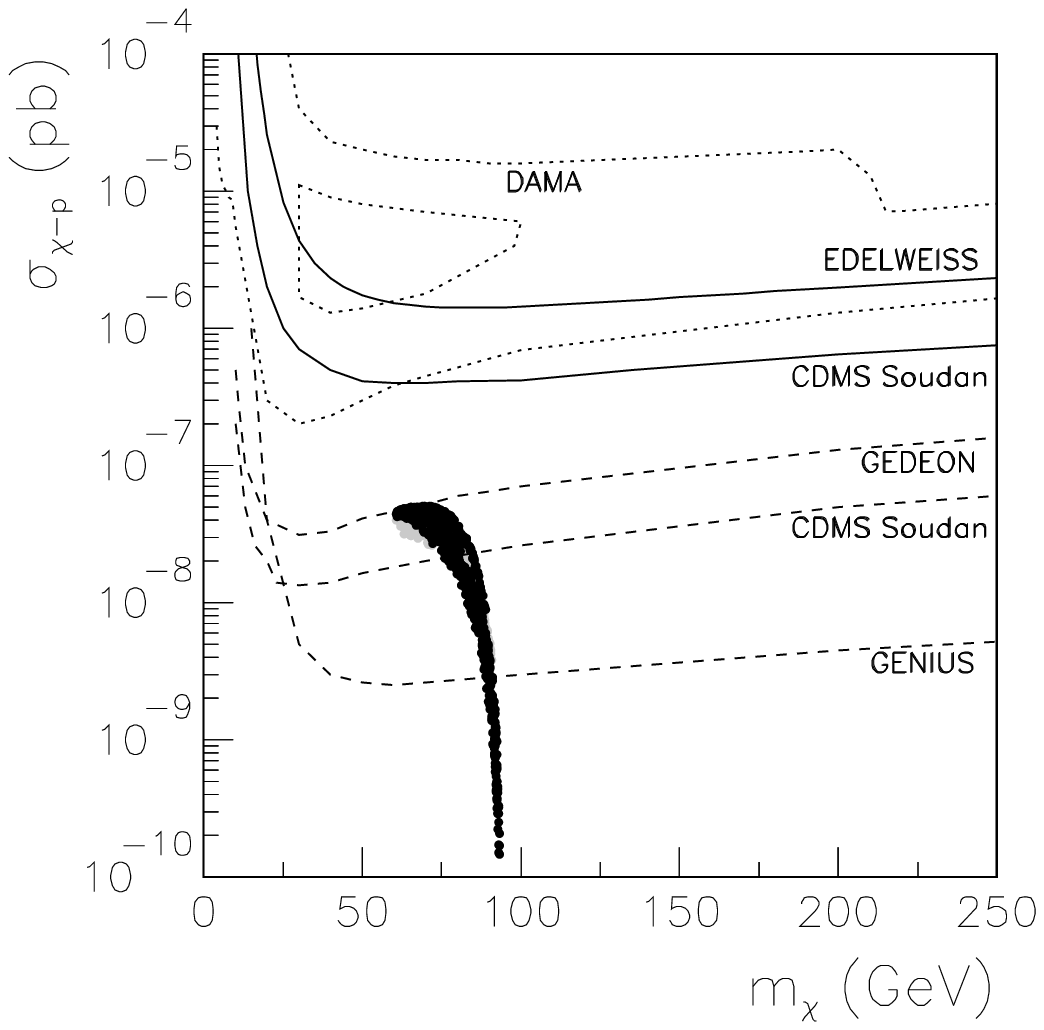,height=8cm}
  \\[-3ex]
  \epsfig{file=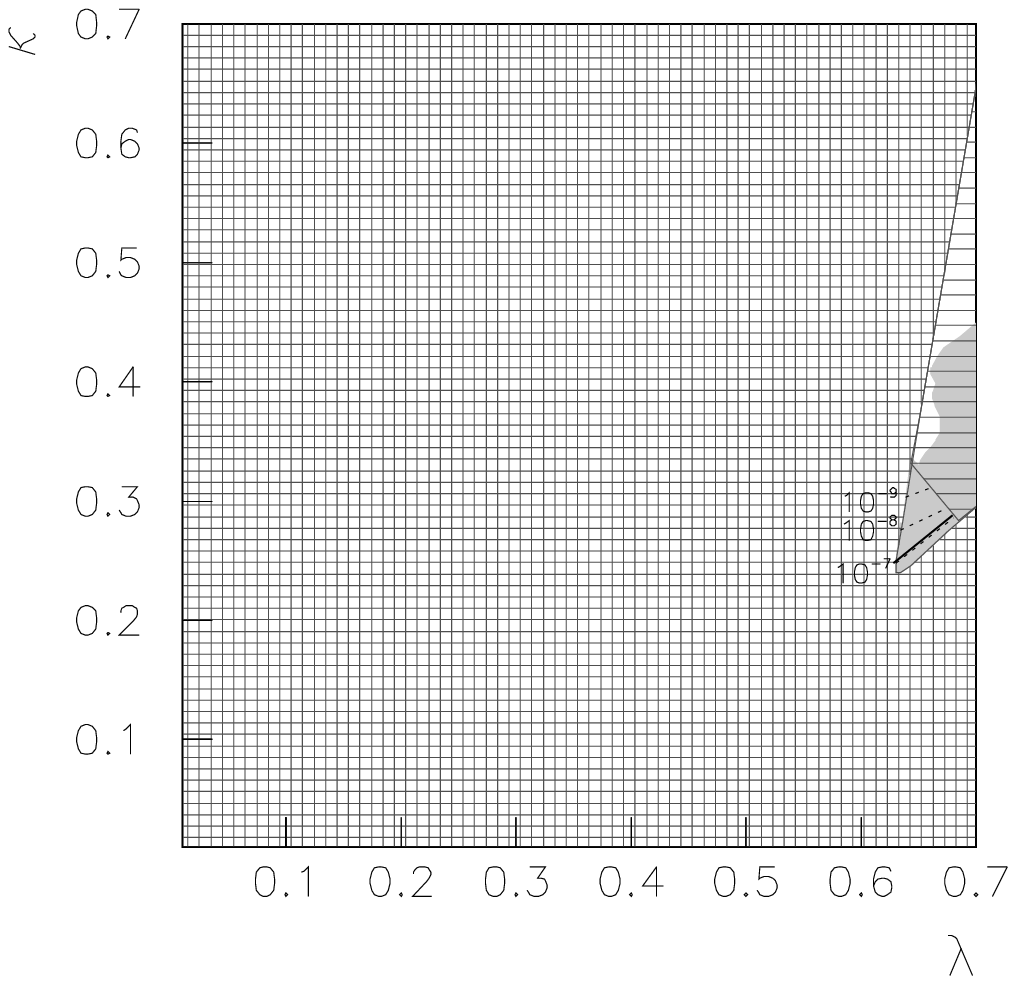,height=8cm}
  \hspace*{-1cm}\epsfig{file=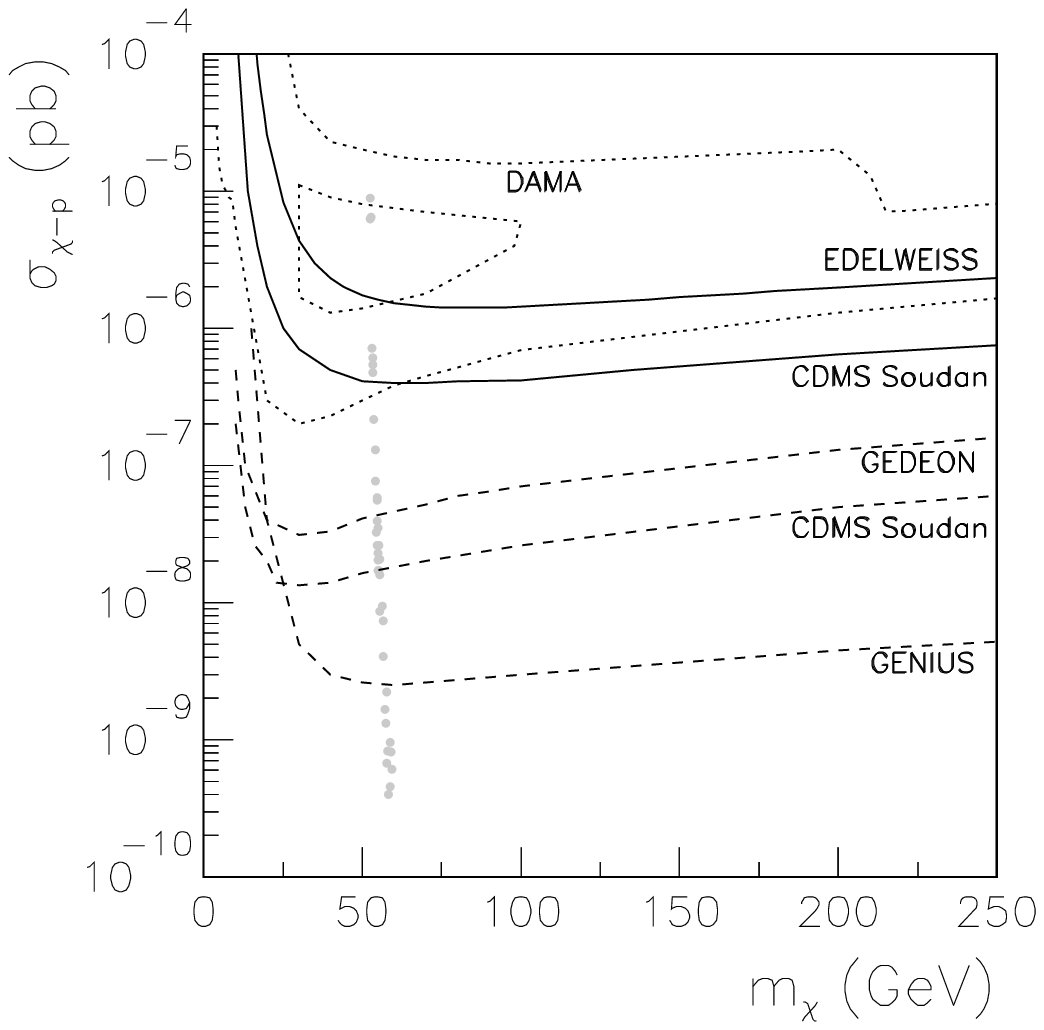,height=8cm}
  \captions{The same as in Fig.\,\ref{++-ka} but for the cases
    $A_\lambda=200$ GeV, $\mu=110$
    GeV, $A_\kappa=50$~GeV,
    and $\tan\beta=2,\,5$, from top to bottom. In the case with
    $\tan\beta=2$ only the lines with  
    $m_{h_1^0}=114,\,75$~GeV, and those with $S_{13}^{\,2}=0.1$ and
    $N_{15}^2=0.1$  are 
    represented. Similarly, in the case with $\tan\beta=5$ only the
    line with $m_{h_1^0}=25$~GeV is drawn, and none of the lines
    showing the compositions of the lightest scalar Higgs and neutralino 
    is depicted, since $0.7>S_{13}^{\,2}>0.1$ and $0.4>N_{15}^2>0.1$
    in all the plane.}
  \label{+++tgb}
\end{figure}

Regarding the value of $\tan\beta$, as already mentioned,
the larger it is, the
more extensive the regions excluded by $m^2_{h^0_1}<0$
become.
In Fig.\,\ref{+++tgb} we represent two examples with
$A_\lambda=200$ GeV,
$A_\kappa=50$ GeV, $\mu=110$ GeV and $\tan\beta=2,\,5$. We
find that low values of $\tan\beta$ still allow physical minima of the
potential.
Moreover, both the singlino component of $\neut$ and the singlet
component of the scalar Higgs can be slightly enhanced.
For example, for $\tan\beta=2$ light neutralinos ($\neut\gsim60$ GeV)
can be obtained with $N_{15}^2\lsim0.55$, while Higgses in the mass
range $m_{h^0_1}\approx75-100$ GeV are singlet-like.
However, the predictions for the detection cross section suffer a
moderate decrease and only $\crosssec\lsim5\times 10^{-8}$ pb is obtained.
On the other
hand, for $\tan\beta\gsim5$ the whole parameter space is in general
excluded.

We should now address the complementary choice of the sign of the
parameters, namely $\mu,\,A_\lambda,\,A_\kappa<0$, for which we
already know that
the analysis of the Higgs sector
still holds. Once more,
differences arise in the theoretical
predictions for $\crosssec$.

To sum up, the choices for the signs of the parameters which have
been considered in this section also permit obtaining large
values for the theoretical prediction of the neutralino-nucleon cross
section, despite the fact
that the parameter space is very constrained both experimentally and
by the occurrence of tachyons. In particular, values of
$\crosssec$ close to the sensitivities of the present detectors can be
found in some regions of the parameter space.
The lightest neutralino displays a mixed singlino-Higgsino character,
and the scalar Higgs is singlet-like and light in those regions with
larger $\crosssec$.

\subsection{$\mu A_\kappa>0$ and $\mu A_\lambda>0$ $(\kappa<0)$}
\label{k-+++}

We will now focus our attention on those cases with a negative value
for $\kappa$, namely  $\mu,\,A_\lambda,\,A_\kappa>0$ and
$\mu,\,A_\lambda,\,A_\kappa<0$. 

Let us therefore concentrate on the first of the two 
possibilities, $\mu,\,A_\lambda,\,A_\kappa>0$. The parameter
space is in this case
plagued with tachyons in both the CP-even and CP-odd Higgs sectors.
On the one hand,
regarding the CP-odd Higgses, large values of $|\kappa|$ and $\mu$ and
small values of $\lambda$ and $A_\lambda$ may lead to negative values
in the diagonal terms of the mass matrix, especially in ${\cal
  M}^2_{P,11}$. Moreover, large values of $\lambda$ can also
induce very large off-diagonal terms ${\cal M}^2_{P,12}$ if
$A_\lambda$ is large and a negative eigenvalue can be obtained in that
case.
Similar arguments lead to analogous conclusions
concerning tachyons in the scalar sector,
being the region with small $\lambda$ the one facing the most severe
restrictions.

Note that experimental constraints will play a very important
role in the vicinity of these regions. Although the largest exclusions
typically arise from the bounds on IHDM and DHDM (mainly in $h^0\to b\bar
b$, $h^0\to 2\,{\rm jets}$),
APM can also exclude some regions with a small $m_{a^0_1}$.
All these become particularly restrictive in the low $\tan\beta$
regime. In fact, in most of the cases with $\tan\beta\lsim3$ all of
the parameter space is excluded.
Experimental constraints in the neutralino sector can
also be very stringent, especially for small values of $\mu$, where
$\neut$ is light and Higgsino-like.

In the remaining allowed regions of the parameter space,
the lightest CP-even Higgs is mostly dominated by the
doublet component. Concerning
the lightest neutralino, it turns out to be
Higgsino-like. Owing to this, the predictions for
$\crosssec$ are very similar to those
obtained in Section\,\ref{+--}.

\begin{figure}
  \epsfig{file=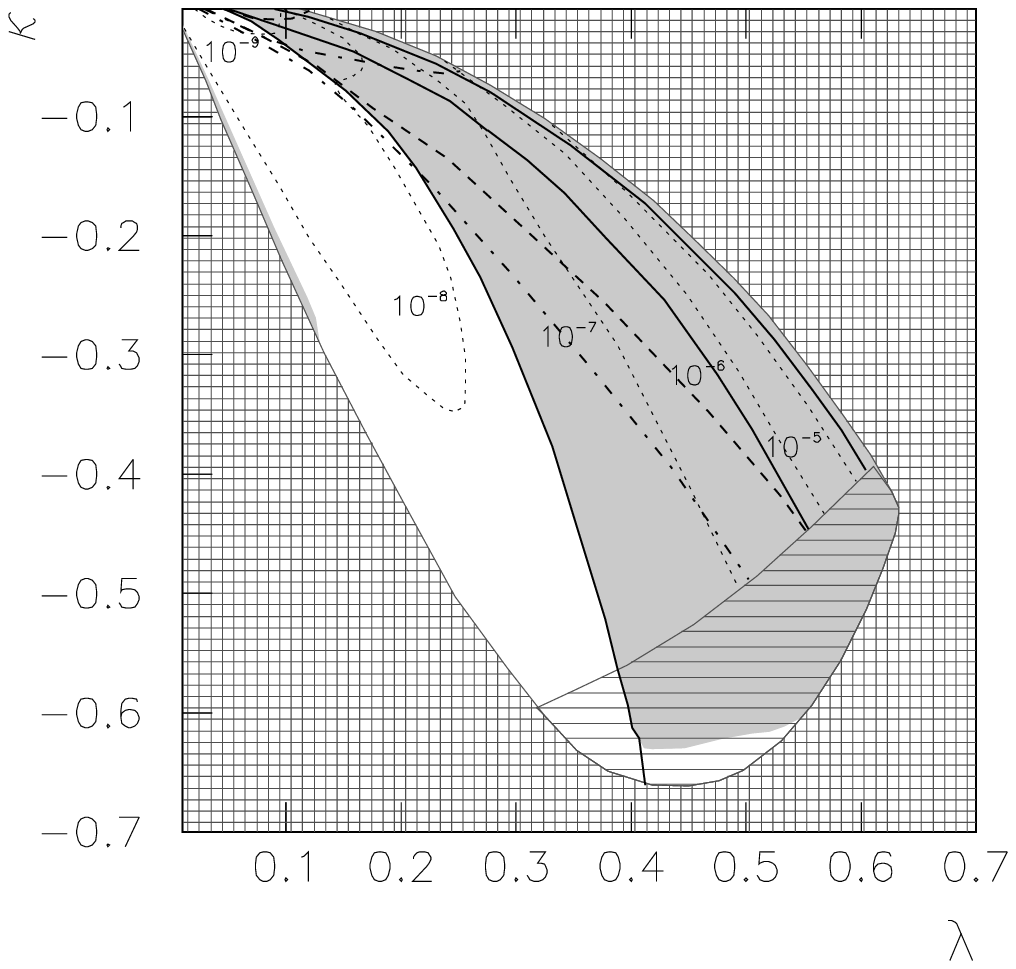,height=8cm}
  \hspace*{-1cm}\epsfig{file=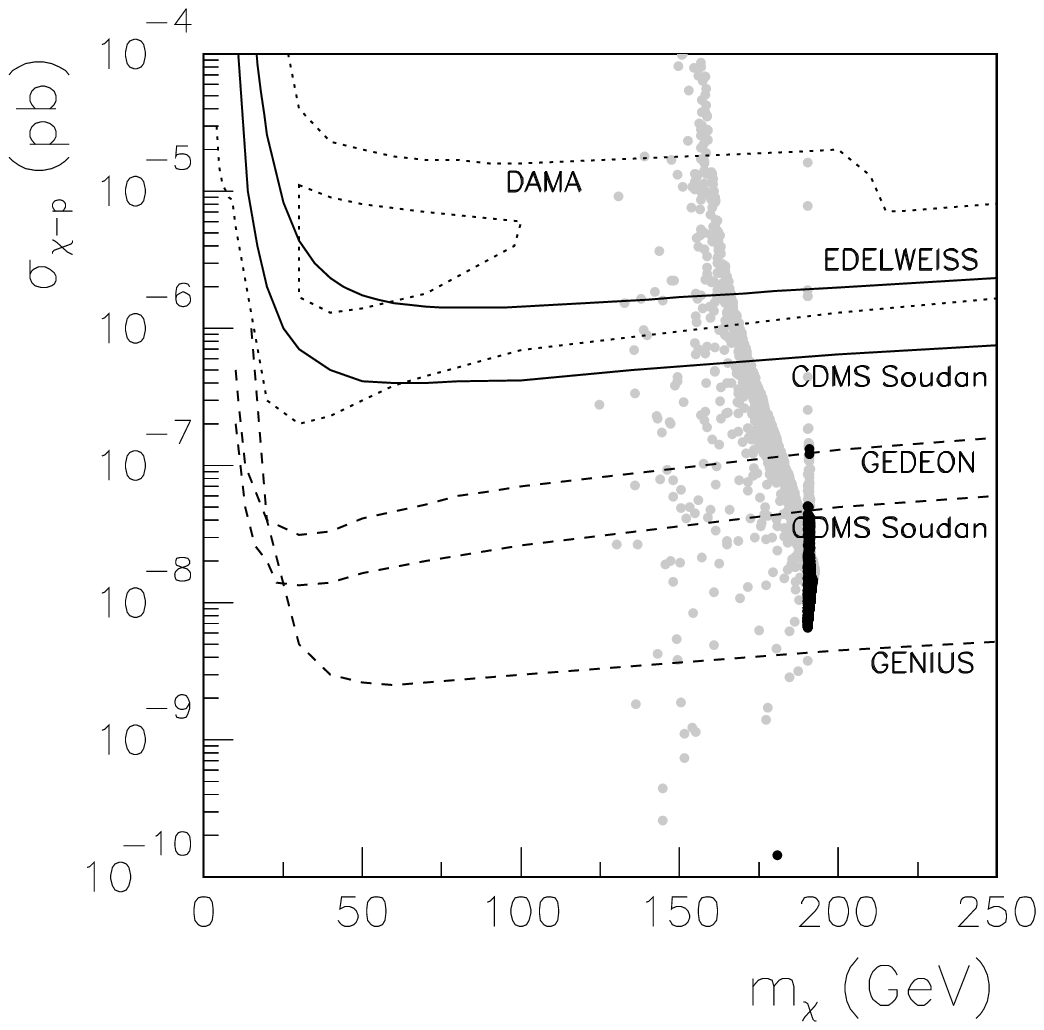,height=8cm}
  \captions{The same as in Fig.\,\ref{++-ka}
    but for negative $\kappa$, with $A_\lambda=450$ GeV, $\mu=200$
    GeV, $A_\kappa=50$ GeV
    and $\tan\beta=5$.}
  \label{K+++a}
\end{figure}

The above discussion can be illustrated with Fig.\,\ref{K+++a}, where
the $(\lambda,\kappa$) plane and the predictions for $\crosssec$ are
presented for a case with $A_\lambda=450$~GeV, $\mu=200$ GeV,
$A_\kappa=50$ GeV, and $\tan\beta=5$.
Light singlino-like neutralinos can only be obtained in the very small
area with $|\kappa|\lsim0.06$, where the scalar Higgs may be as
light as $m_{h^0_1}\gsim50$ GeV with a large singlet component.
In this particular region the predicted values for the detection cross
section are not large, $\crosssec\lsim10^{-9}$ pb. In the rest of the
parameter space $\neut$ is Higgsino-like, which implies
$\neumass\approx\mu$, and
the lightest scalar Higgs is a doublet with $m_{h^0_1}\gsim112$ GeV.
Slightly higher values for the cross section are obtained, which are
bounded by the experimental constraints on the scalar Higgs at
$\crosssec\lsim6\times10^{-8}$ pb.
This prediction can
be slightly increased with larger values of $\tan\beta$. For instance,
with $\tan\beta=10$ one finds $\crosssec\lsim2\times10^{-7}$ pb. None
of the above remarks concerning the masses and compositions of the
lightest neutralino and scalar Higgs would change in this case.

\begin{figure}
  \hspace*{3.5cm}\epsfig{file=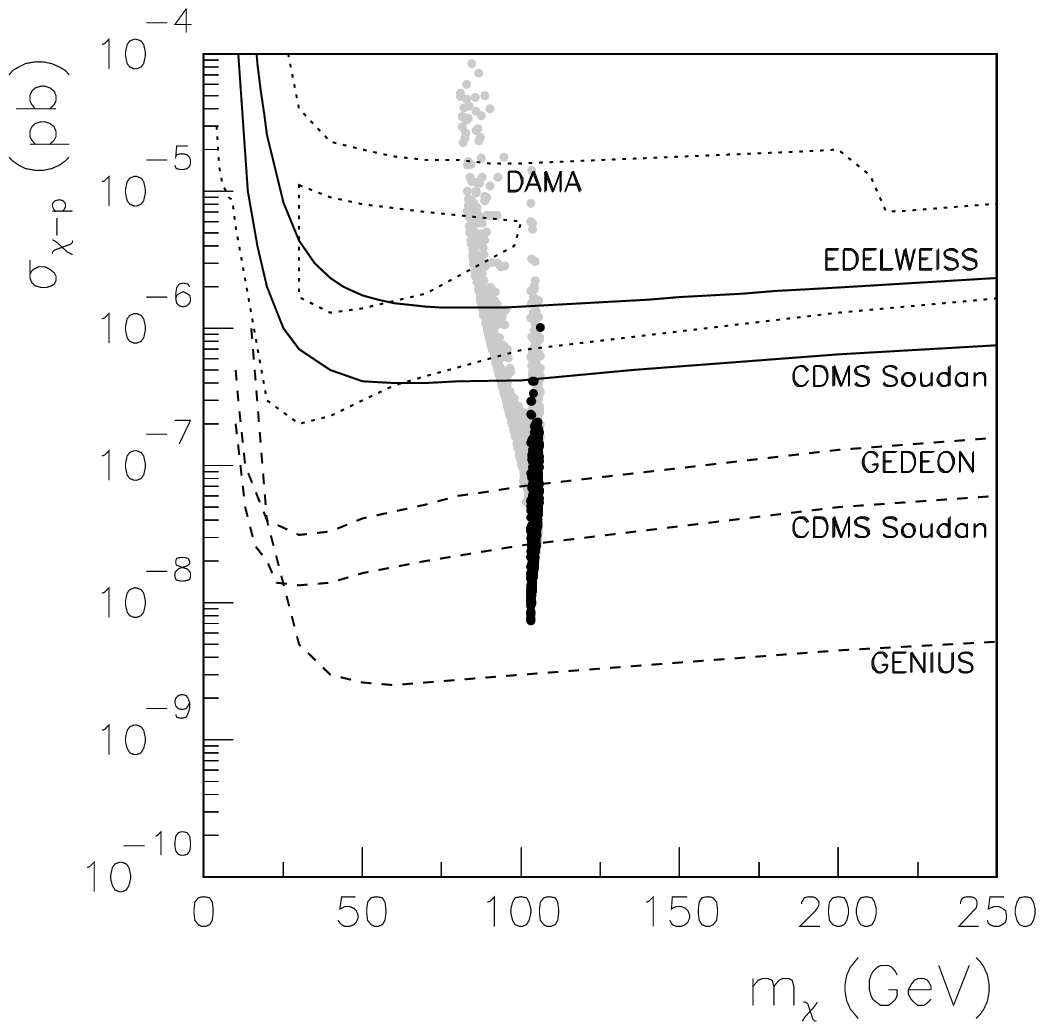,height=8cm}
  \captions{The same as in Fig.\,\ref{K+++a}
    but for $A_\lambda=200$ GeV, $\mu=110$
    GeV, $A_\kappa=200$ GeV
    and $\tan\beta=10$.}
  \label{K+++b}
\end{figure}

Variations in the rest of the parameters are very constrained due to
the extensive tachyonic regions and the
strong experimental bounds, especially those associated to the
bounds on IHDM and DHDM.
This is, for instance, what happens when the value of $A_\lambda$
decreases.
The very narrow region for small $|\kappa|$ where singlino-like
neutralinos can be obtained is usually ruled out and the only
surviving areas are those featuring heavy Higgsino-like neutralinos
and doublet-like scalar Higgses. For this reason, the predictions for
the detection cross section are always similar to those presented in 
Fig.\,\ref{K+++a}.
As an example of some of the most favourable results for the cross section 
that can be obtained, we show in Fig.\,\ref{K+++b} a
case with $A_\lambda=200$~GeV, $\mu=110$ GeV,
$A_\kappa=200$~GeV, and $\tan\beta=10$, where the sensitivity of the
CDMS Soudan experiment is almost reached.

Finally, concerning the case $\mu,\,A_\lambda,\,A_\kappa<0$, nothing
changes in the analysis of the Higgs sector.
Once again, the
differences in the neutralino sector and the slight changes in the
experimental constraints can induce variations in the predicted
$\crosssec$. Nevertheless, a similar global upper bound of
$\crosssec\lsim10^{-7}$ pb is obtained.

The examples analysed in this Subsection feature a lightest neutralino
which is Higgsino-like in most of the parameter space, together with a
doublet-like lightest scalar Higgs. The neutralino-nucleon cross
section is bounded by experimental constraints on the Higgs sector at
$\crosssec~\lsim~2~\times~10^{-7}$~pb.
Singlino-like neutralinos can only be obtained in extremely small
regions of the parameter space and predict smaller cross sections,
$\crosssec~\lsim~2~\times~10^{-9}$~pb.

\subsection{Variations in the gaugino mass parameters}
\label{hierarchyg}

In order to complete our analysis we must now address 
variations in the gaugino mass parameters. 
These clearly affect the neutralino sector, altering both the mass and
composition of the lightest neutralino. In the former analysis we
always assumed the relation $\mu\,<\,M_1\,<M_2$, which lead to
neutralinos with important Higgsino compositions. We will now
generalize our results for different hierarchies among these
parameters. Namely, we will investigate the consequences of having
$M_1\,<\,\mu,\,M_2$ or $M_2\,<\,\mu,\,M_1$.

The Higgs sector is not so sensitive to variations in the gaugino
masses, 
since they
only enter through loop corrections. For this reason,
all the analysis regarding tachyons remains qualitatively
valid, and the experimental constraints associated to Higgses
exclude similar areas in all these cases. However, it is important to
note
that increasing the gaugino masses, especially the gluino
mass, generally implies also an increase in the mass of the lightest 
scalar Higgs and for this reason, a heavy gaugino spectrum would
typically lead to low values of $\crosssec$.

\begin{figure}[!t]
  \epsfig{file=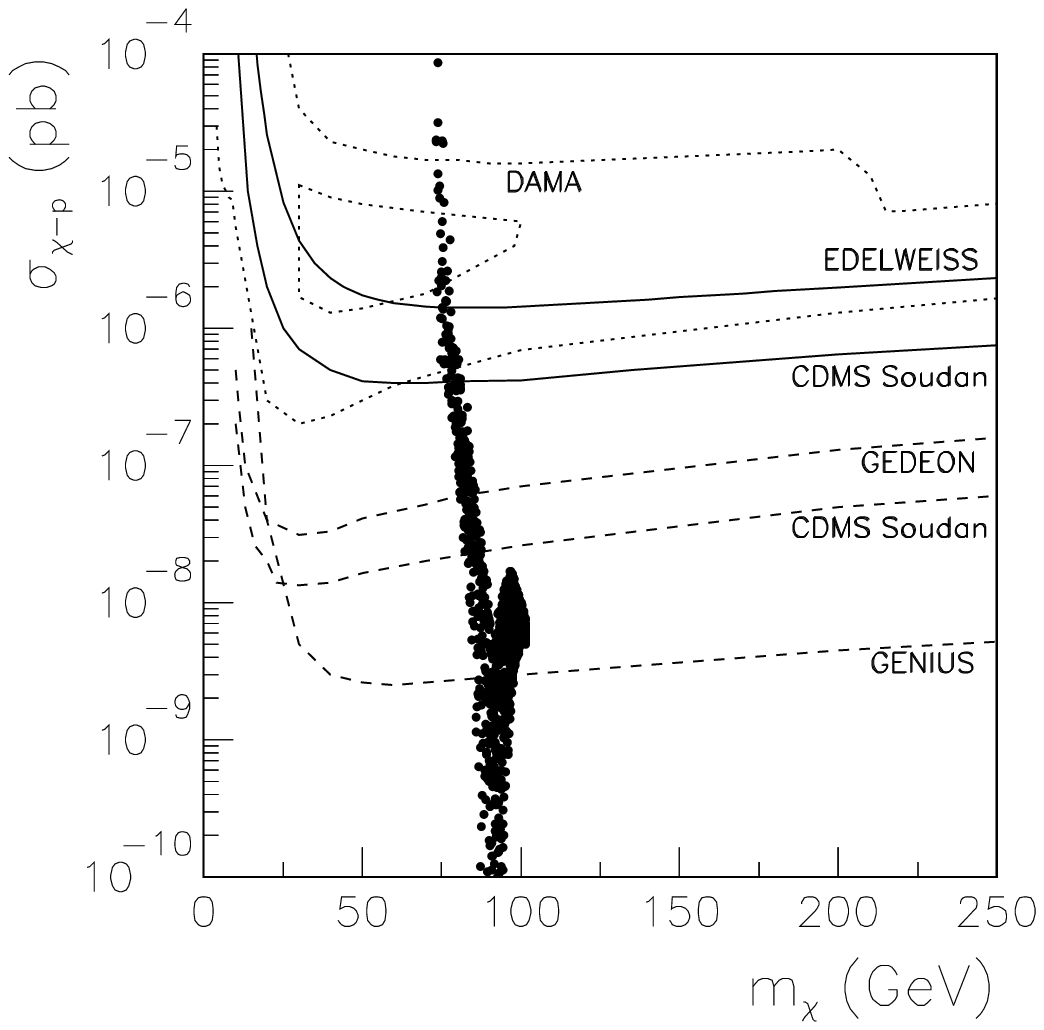,height=8cm}
  \hspace*{-1cm}\epsfig{file=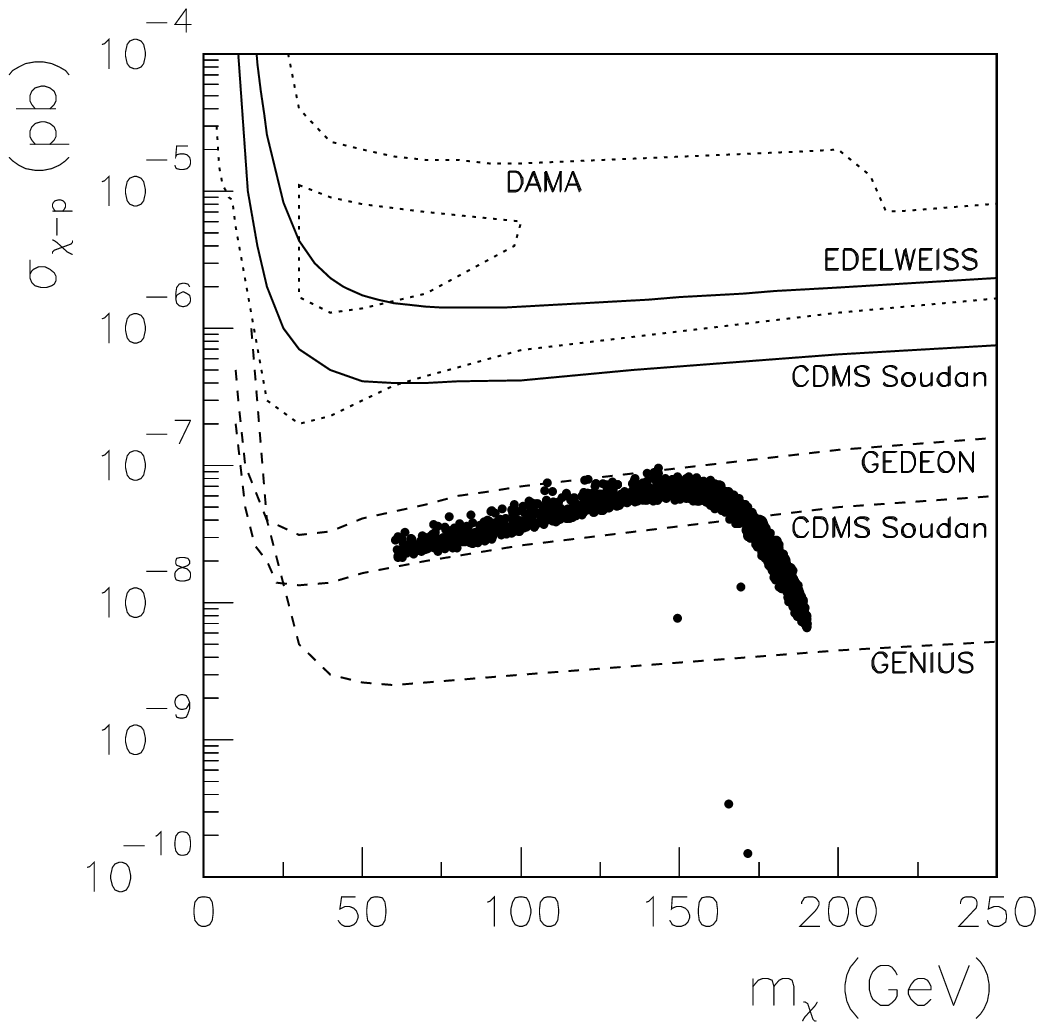,height=8cm}
  \\[-3ex]
  \hspace*{3.5cm}\epsfig{file=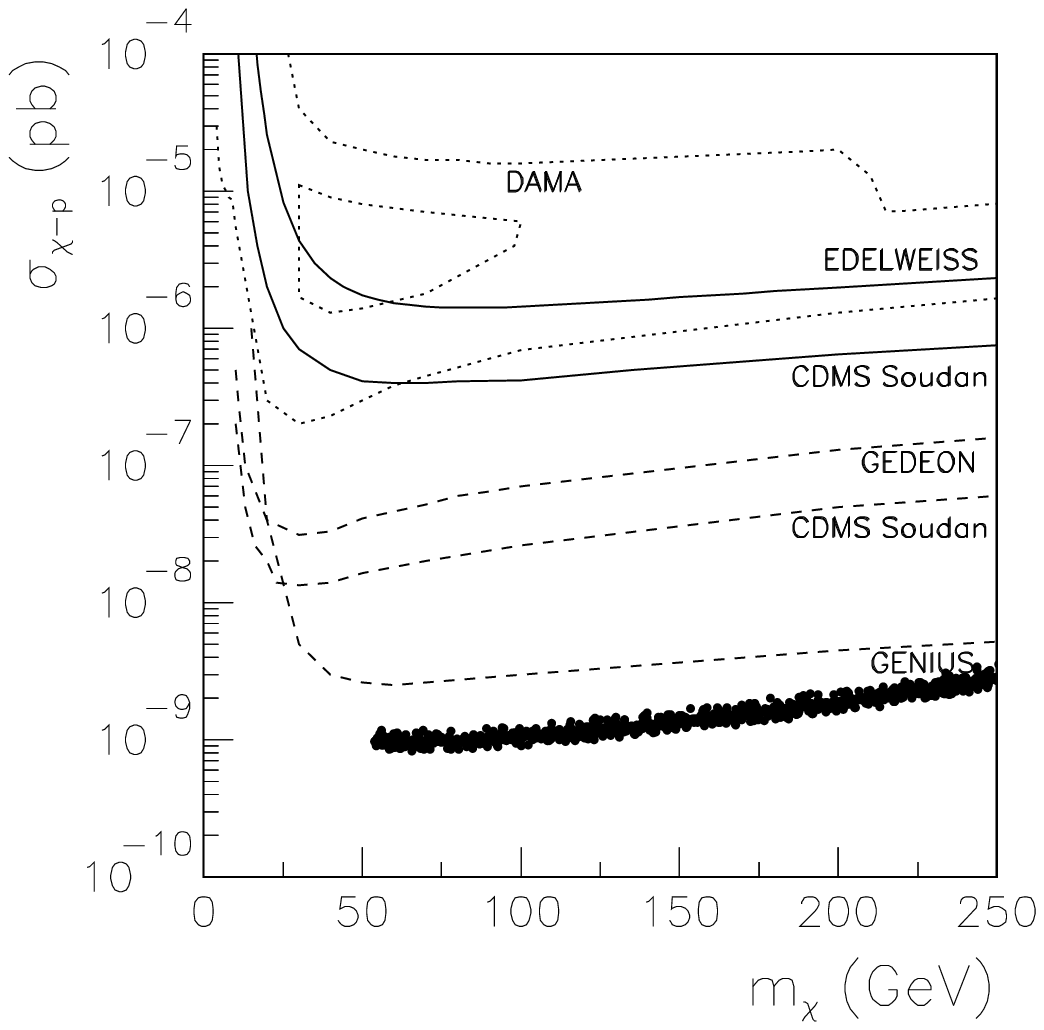,height=8cm}
  \captions{Scatter plot of the scalar neutralino-nucleon cross
    section as a function of the neutralino mass for $A_\lambda=200$
    GeV, $A_\kappa=-200$ GeV, $\tan\beta=3$, $\mu=110,\,200,\,500$
    GeV from left to right and from top to bottom. 
    The gaugino masses satisfy the GUT
    relation and the gaugino mass parameter is varied in the range
    $50$ GeV $\leq M_1\leq 500$ GeV. Only those points fulfilling all the
    constraints are represented.}
  \label{gutxx}
\end{figure}

Let us first vary the values of the gaugino masses preserving their 
GUT relation, but allowing also changes in the $\mu$ parameter so that
the relation $M_1< \mu,\,M_2$ can be achieved, thus increasing the
gaugino character of $\neut$.
For this purpose we choose the example with $A_\lambda=200$ GeV,
$A_\kappa=-200$ GeV, and $\tan\beta=3$ that was represented in
Fig.\,\ref{++-ka}, where now three different values for the $\mu$
parameter are taken, $\mu=110,\,200,\,500$~GeV. Regarding the gaugino
masses, we consider variations in the Bino mass as $50$ GeV $\leq
M_1\leq 500$ GeV, and the GUT relation $M_1=\frac{1}{2}\,M_2=\frac{1}{7}\,M_3$.
The results are shown in
Fig.\,\ref{gutxx}, where only those points fulfilling all the
constraints are represented.
Since the experimental constraint on the chargino mass imposes a lower
bound on $M_2$, the value of the Bino mass is also constrained.
For this reason, for low values of $\mu$ the lightest neutralino is
still a singlino-Higgsino state ($N_{15}^2\lsim0.3$ and
$N_{13}^2+N_{14}^2\gsim0.7$). However, as the value of $\mu$
increases, so does the gaugino composition of $\neut$ (note that both
the Higgsino and singlino compositions decrease when increasing
$\mu$). 
For instance,
in the example with $\mu=500$ GeV neutralinos lighter than
$\neumass\lsim 375$~GeV are Bino-like.

The appearance of Bino-like neutralinos has as a consequence the
enormous decrease in the neutralino-nucleon cross section. In
particular, the Higgs mediated interaction is now negligible and
detection would only take place through the squark mediated
interaction. In contrast to the huge predictions for $\crosssec$ in
the case of a singlino-Higgsino neutralino (see, e.g., the plot with
$\mu=110$ GeV in Fig.\,\ref{gutxx}), Bino-like neutralinos would have
$\crosssec\lsim10^{-9}$ pb, and thus would be beyond the sensitivities
of even the largest projected dark matter detectors.

\begin{figure}[t]
  \epsfig{file=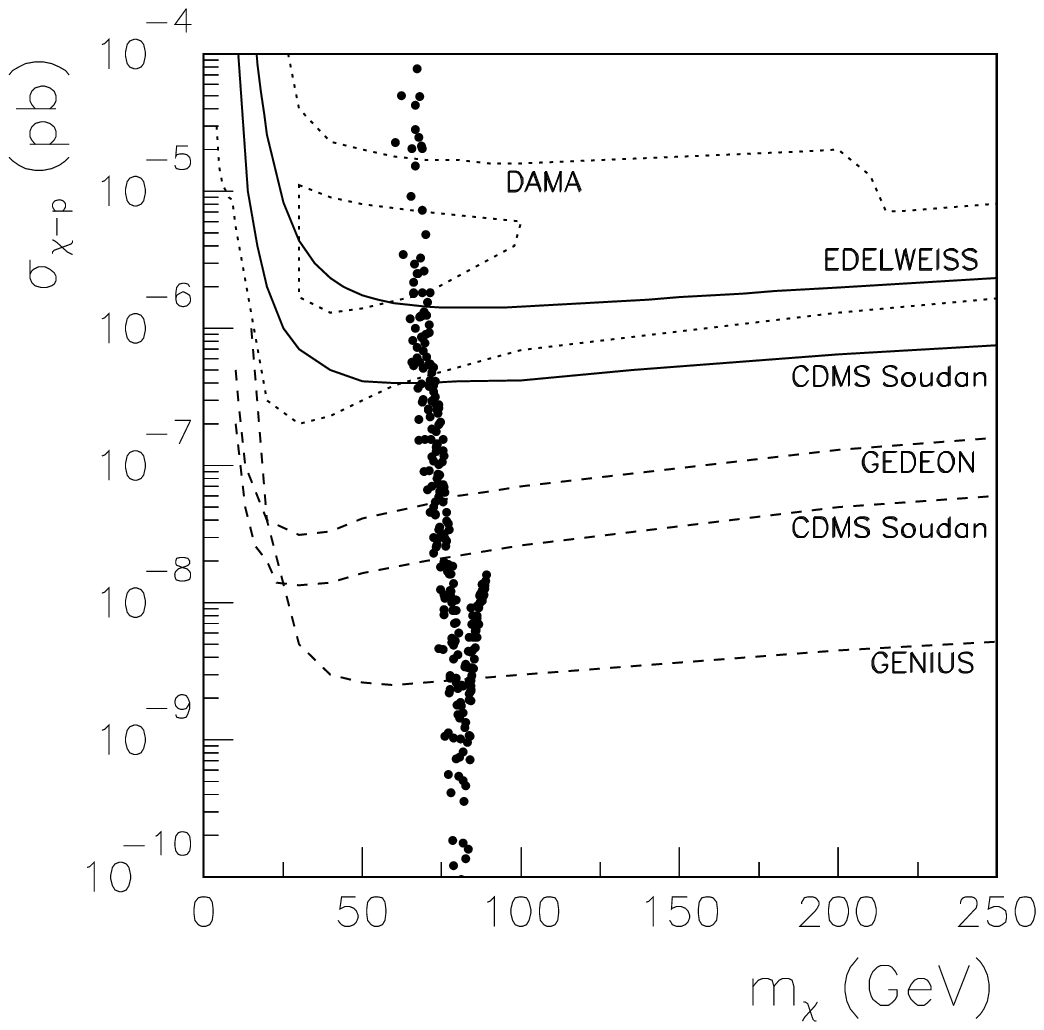,height=8cm}
  \hspace*{-1cm}\epsfig{file=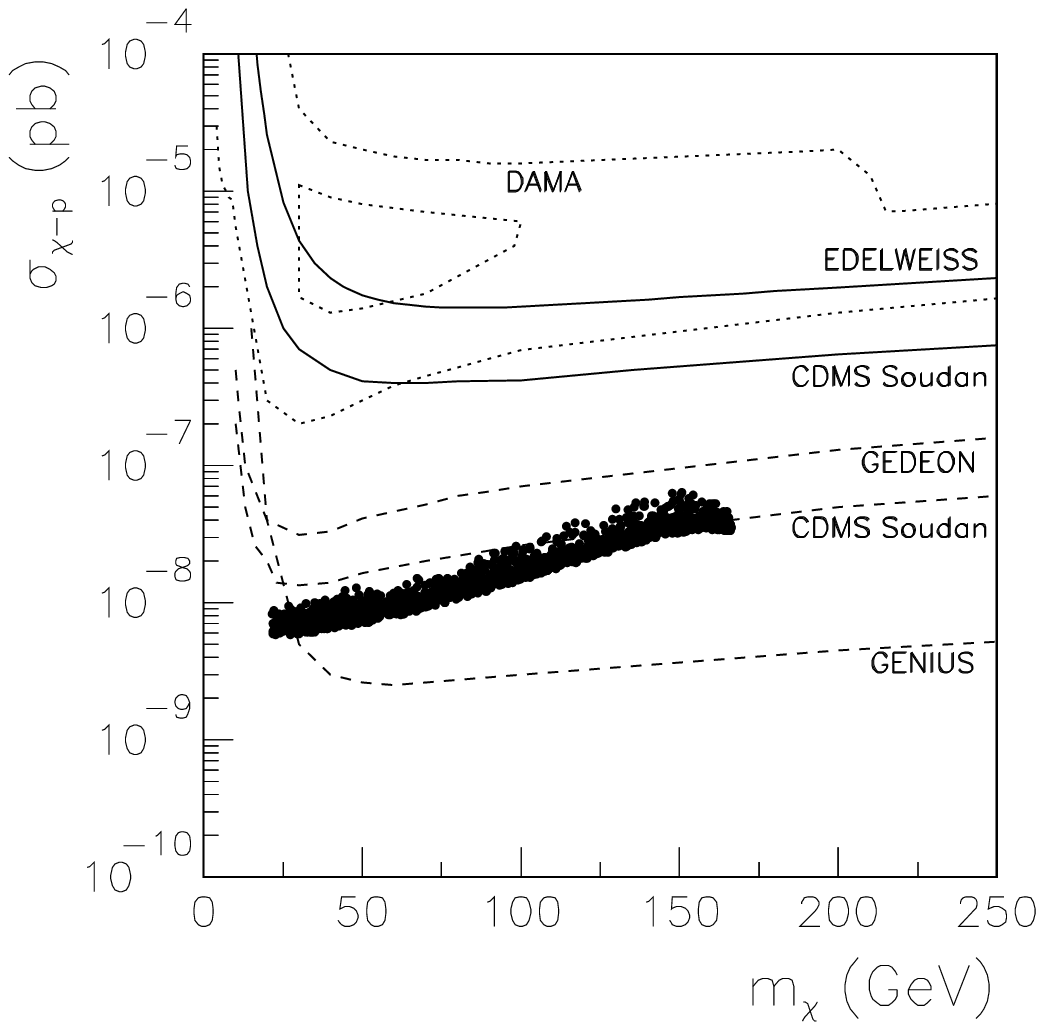,height=8cm}
  \\[-3ex]
  \hspace*{3.5cm}\epsfig{file=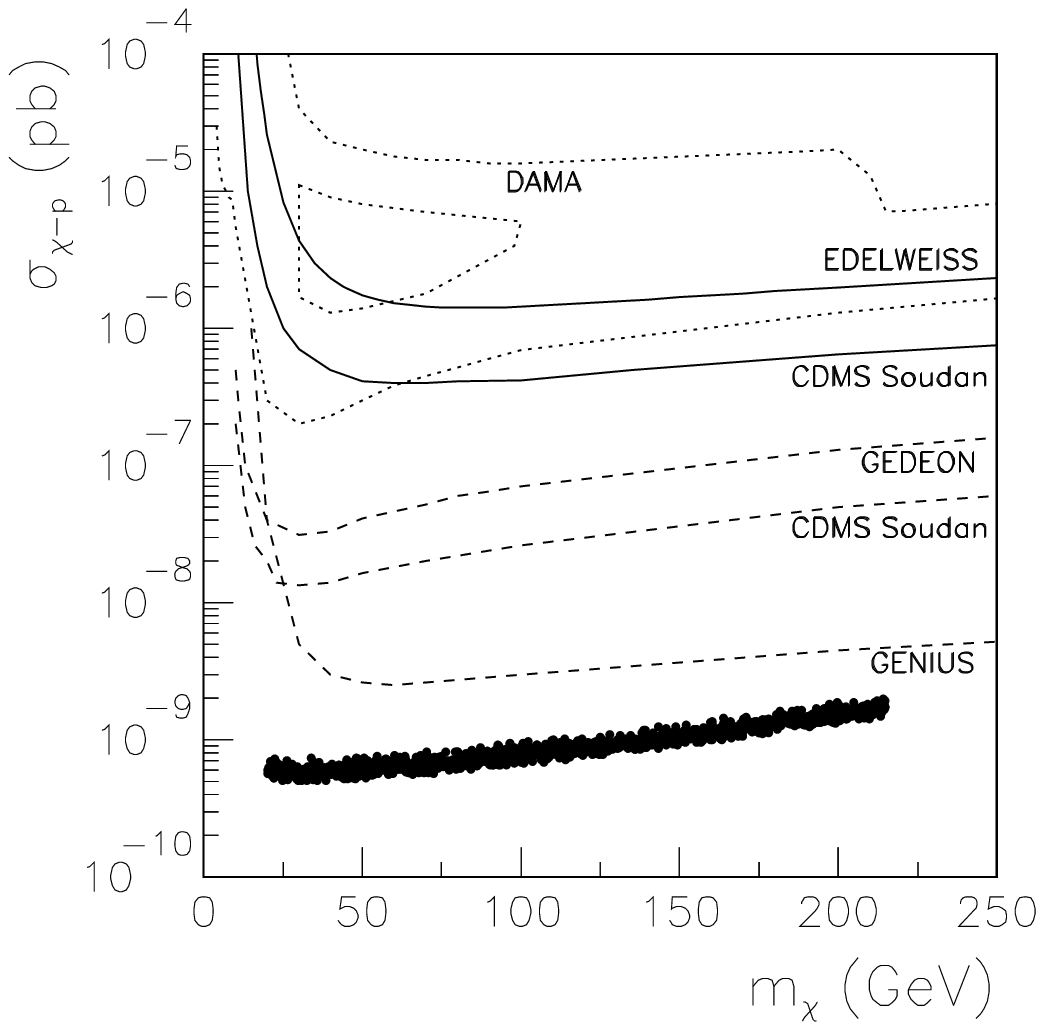,height=8cm}
  \captions{The same as Fig.\,\ref{gutxx} but for the relation
    $M_1=0.2\,M_2$.}
  \label{binoxx}
\end{figure}

Very
light Bino-like neutralinos can also appear when the GUT relation is
relaxed.
This is due to
the freedom to choose very small values of $M_1$, unrelated to
the more constrained (from the bound on the chargino mass)
$M_2$. Furthermore, Bino-like neutralinos are not subject to such
strong bounds on direct neutralino production as Higgsino-like
neutralinos. 
In order to exemplify this possibility, we have chosen the relation
$M_1=0.2\,M_2$ (the
GUT relation among the Wino and gluino masses,
$M_2=\frac27M_3$, is preserved)
and applied it to the same examples as before. The
results are
represented in Fig.\,\ref{binoxx}.
Neutralinos as light as $\neumass\gsim25$ GeV are attainable, although
the predicted values for the detection cross section lie beyond the
sensitivities of all present and projected experiments.

\begin{figure}[t]
  \epsfig{file=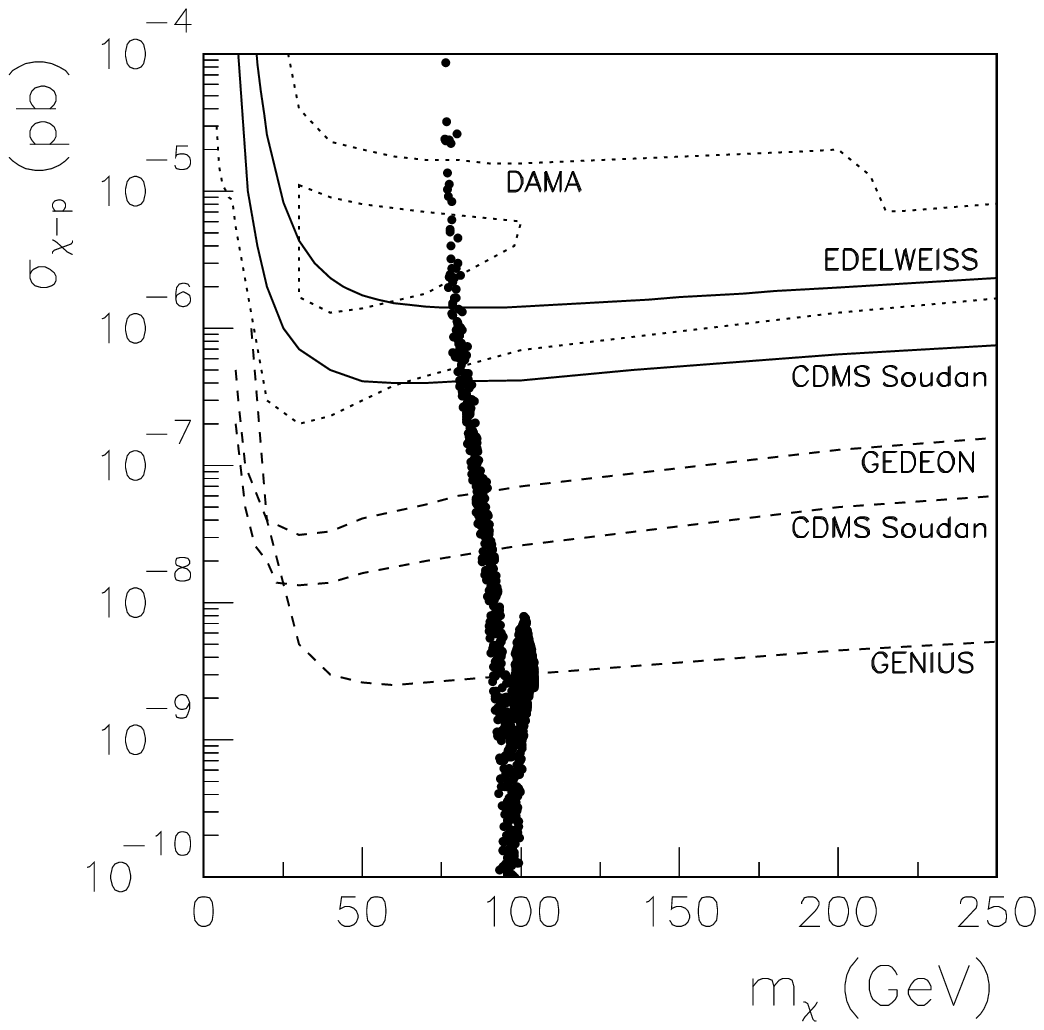,height=8cm}
  \hspace*{-1cm}\epsfig{file=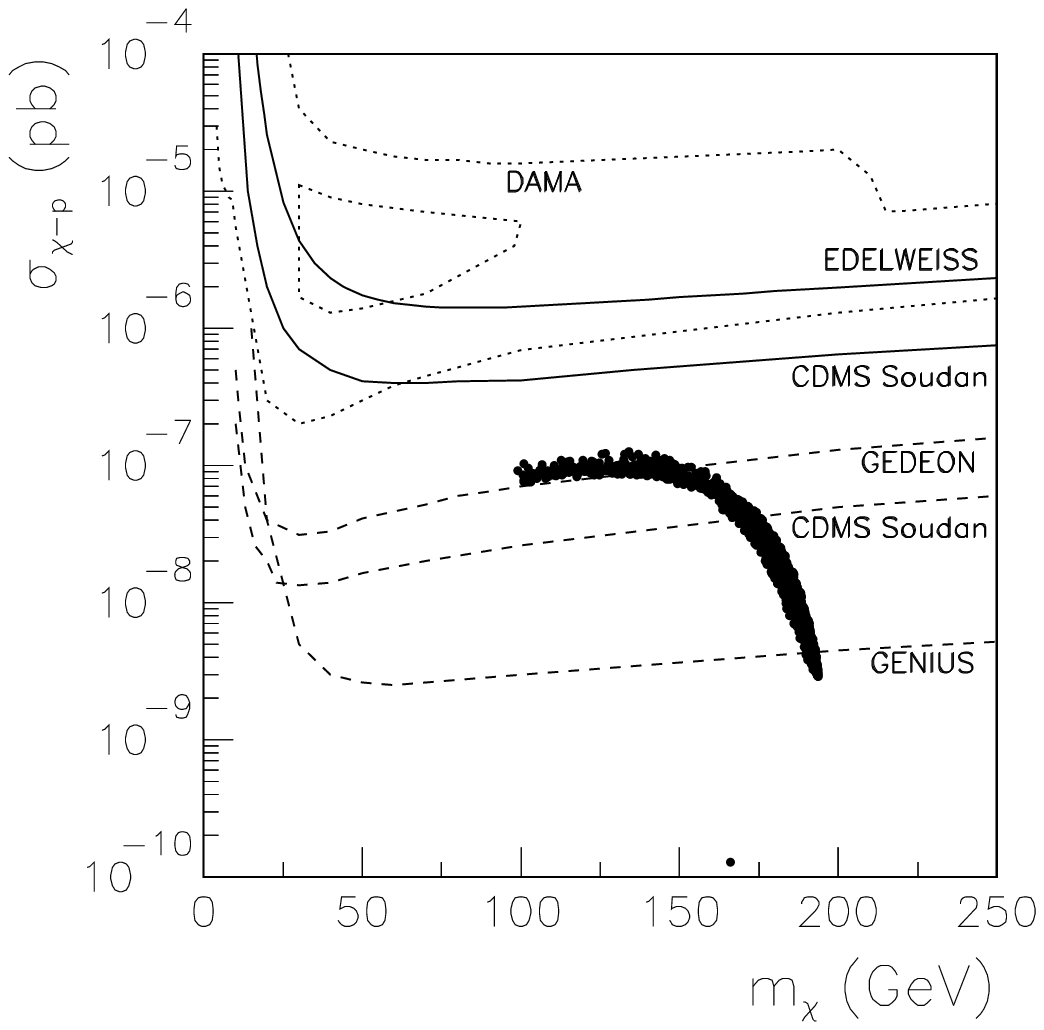,height=8cm}
  \\[-3ex]
  \hspace*{3.5cm}\epsfig{file=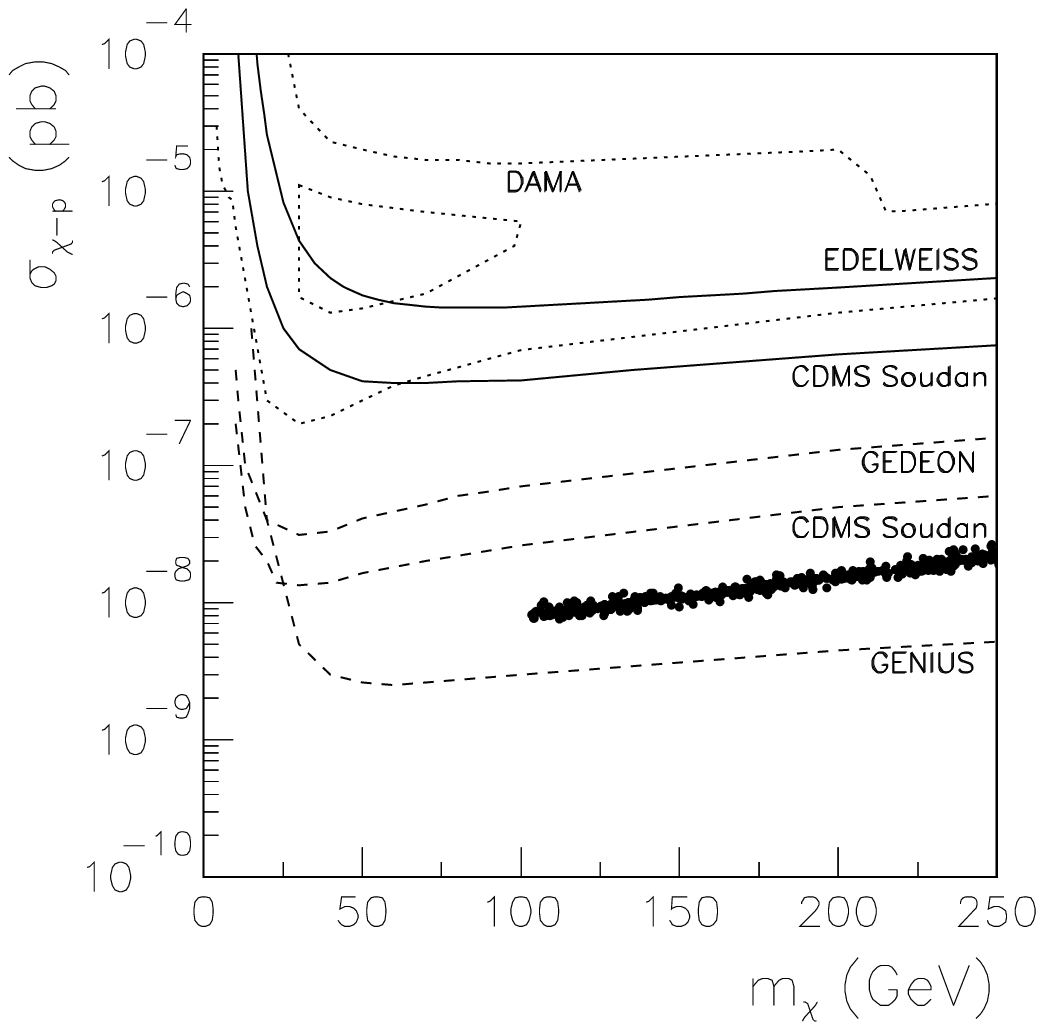,height=8cm}
  \captions{The same as Fig.\,\ref{gutxx} but for the relation
    $M_1=2\,M_2=\frac{1}{7}\,M_3$.}
  \label{winoxx}
\end{figure}

Finally, 
more general 
compositions for the neutralino can also be found. In particular,
Wino-like 
neutralinos are attainable if $M_2<M_1,\,\mu$. This is an interesting
possibility, since Wino-like neutralinos predict in general larger
values for the cross section than Bino-like $\neut$.
In order to explore this possibility we
have repeated the same scan in the parameter space as above, but now
with the relation $M_1=2\,M_2=1/7\,M_3$ for the gaugino masses 
at the EW scale. The results
are shown in Fig.\,\ref{winoxx} for $\mu=110,\,200,\,500$~GeV.
Obviously, the Wino component of $\neut$ becomes more important for
large values of $\mu$. Thus, for instance, whereas in the example with
$\mu=200$ GeV $\neut$ is a mixed Higgsino-Wino state (with
$N_{12}^2\lsim0.6$), Wino-like neutralinos populate the example with
$\mu=500$~GeV.
Because of the experimental constraint on the chargino mass,
Wino-like neutralinos cannot be obtained below $\neumass\approx100$
GeV. 
Note that although the theoretical predictions for the detection cross
section are larger than in the case of Bino-like neutralinos of
Fig.\,\ref{gutxx}, there is an important decrease with respect to
those in the case of Higgsino-singlino neutralinos. In particular, in
this case $\crosssec\lsim2\times10^{-8}$~pb. 

All the variations discussed in this Subsection imply a variation in
the mass and composition of the lightest neutralino and scalar
Higgs. We have seen how Bino- and Wino-like neutralinos are attainable
by decreasing $M_1$ and $M_2$, respectively, and increasing $\mu$. 
The singlino-Higgsino 
character of the neutralino is therefore lost in these
cases. 
Furthermore, the increase in $\mu$ also leads to doublet-like heavier
Higgses. For these reasons the Higgs mediated interaction is no longer
effective and the theoretical predictions for the
neutralino-nucleon cross section can have a huge decrease.
Therefore, the optimal situation is the one we have analysed in all
the examples of
Subsections.\,\ref{++-} to \ref{k-+++}, where $\mu$ is close to its
lower accepted value and smaller than the gaugino masses.

\subsection{Overview}
\label{overview}

In the previous Subsections we have presented a separate 
analysis of the distinct
regions of the $A_\lambda$, $A_\kappa$, and $\tan \beta$
parameter space. 
This kind of approach was useful in order to comprehend the
implications of individual variations of these parameters. 
Still, in order to obtain a global view on the theoretical
predictions for $\crosssec$, and their compatibility with present and
projected dark matter detectors, it is useful to conduct a more general
survey of the parameter space.
Such an analysis constitutes a good overview of the 
properties of the cases studied in the previous Subsections.

Since we are interested in regions predicting large $\crosssec$,
according to the conclusions of Subsection\,\ref{hierarchyg} we will
focus our attention on the case $\mu=110$ GeV, with heavy gaugino
masses, $M_1=\frac12\,M_2=500$ GeV. The rest of the input
parameters are allowed to vary in the ranges,
$-600$~GeV~$\leq~A_\lambda~\leq~600$~GeV,
$-400$~GeV~$\leq~A_\kappa~\leq~400$~GeV, and
$\tan\beta=2,\,3,\,4,\,5,\,10$, with
$\lambda,\,|\kappa|\in[0.01,0.8]$.

\begin{figure}
  \epsfig{file=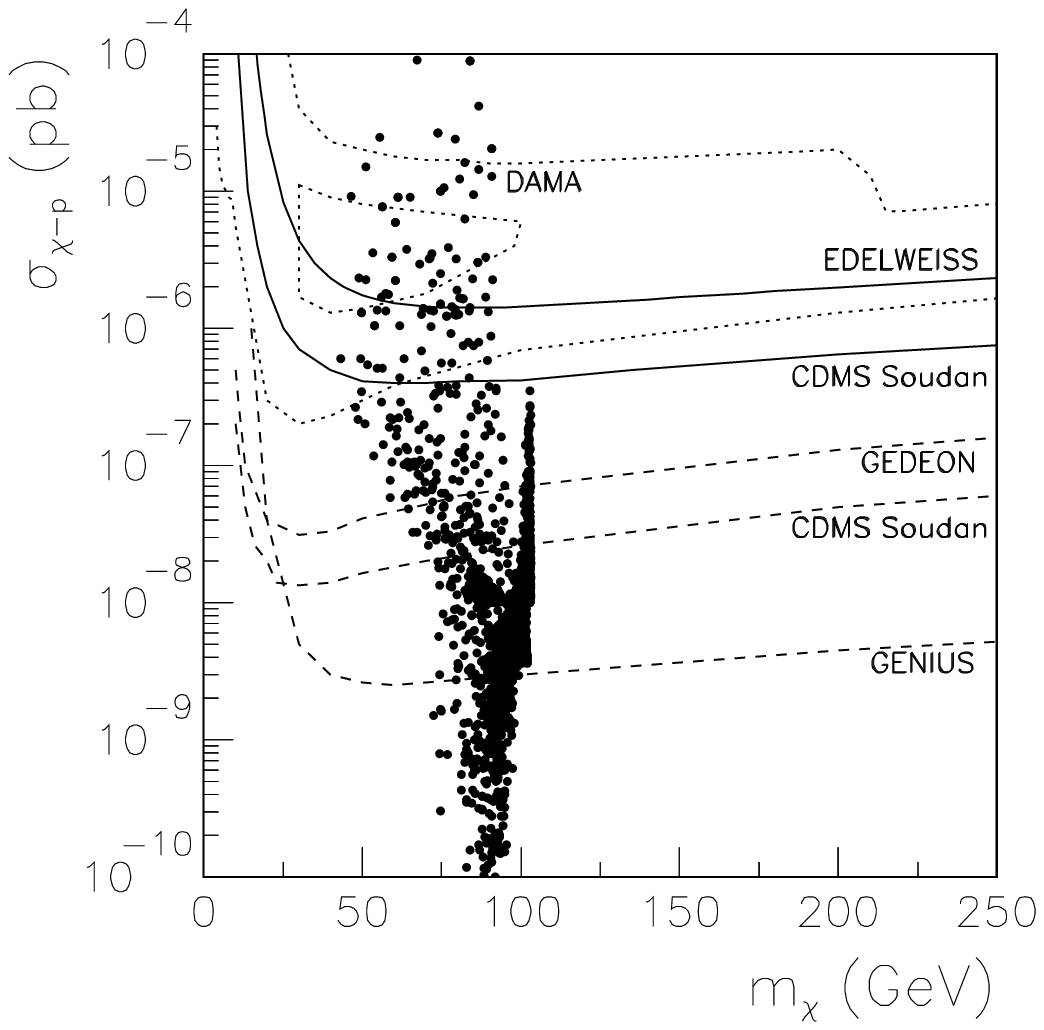,height=8cm}
  \hspace*{-1cm}\epsfig{file=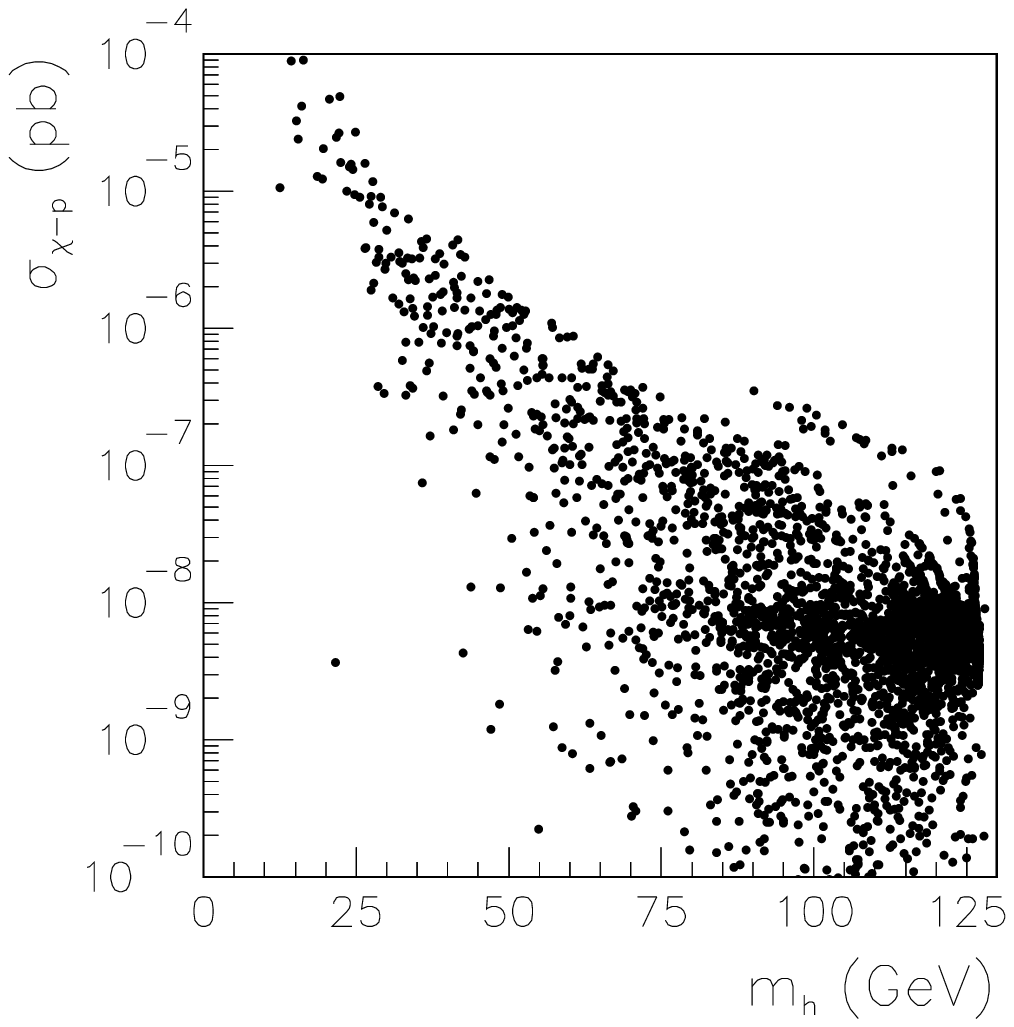,,height=8cm}
  \captions{The same as in Fig.\,\ref{++-crossa}  but for $\mu=110$ GeV 
    and the rest of the parameters in the ranges
    $-600$ GeV $\leq A_\lambda\leq600$ GeV,
    $-400$ GeV $\leq A_\kappa\leq400$ GeV, and $\tan\beta~=~2,\,3,\,4,
    \,5,\,10$, and with positive values of $\kappa$. 
    Only those points fulfilling all the constraints are
    represented.}
  \label{summarya}
\end{figure}

\begin{figure}
  \epsfig{file=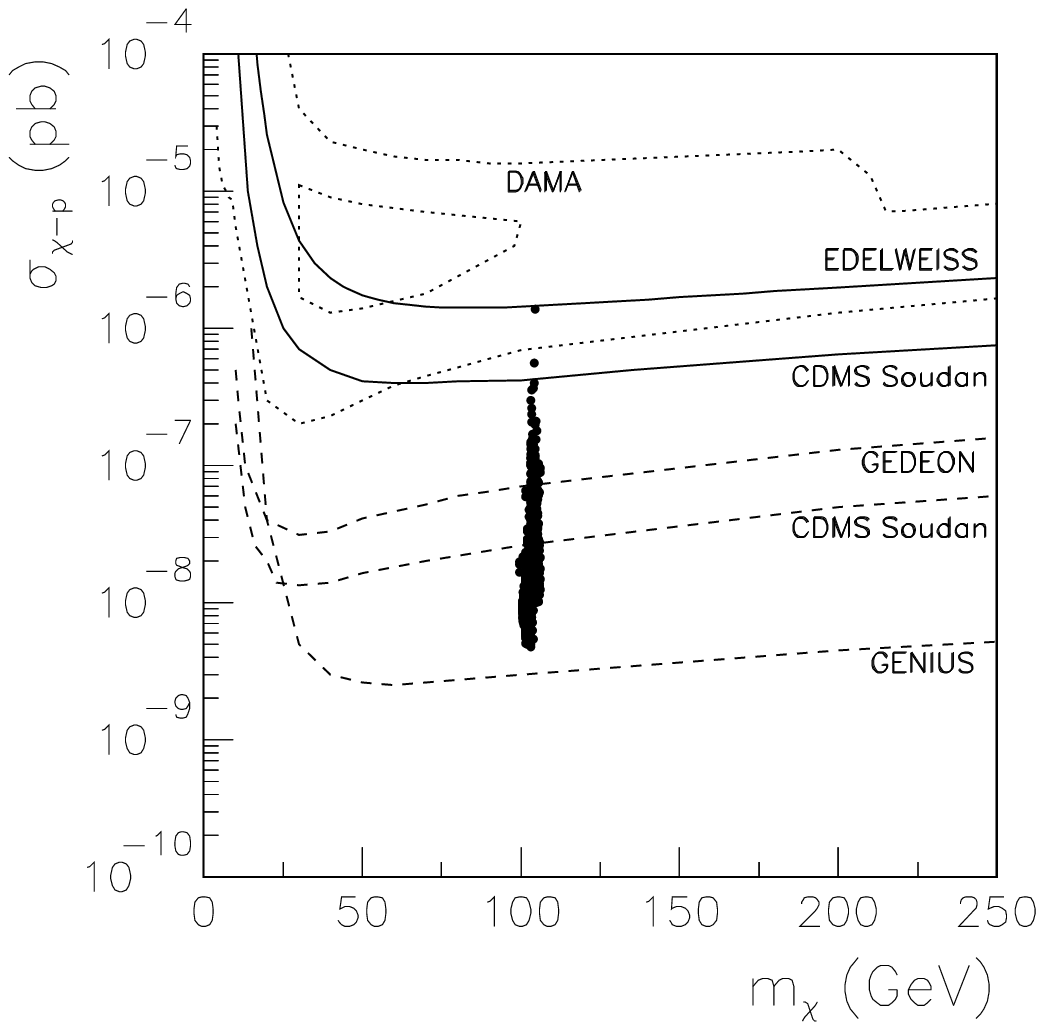,,height=8cm}
  \hspace*{-1cm}\epsfig{file=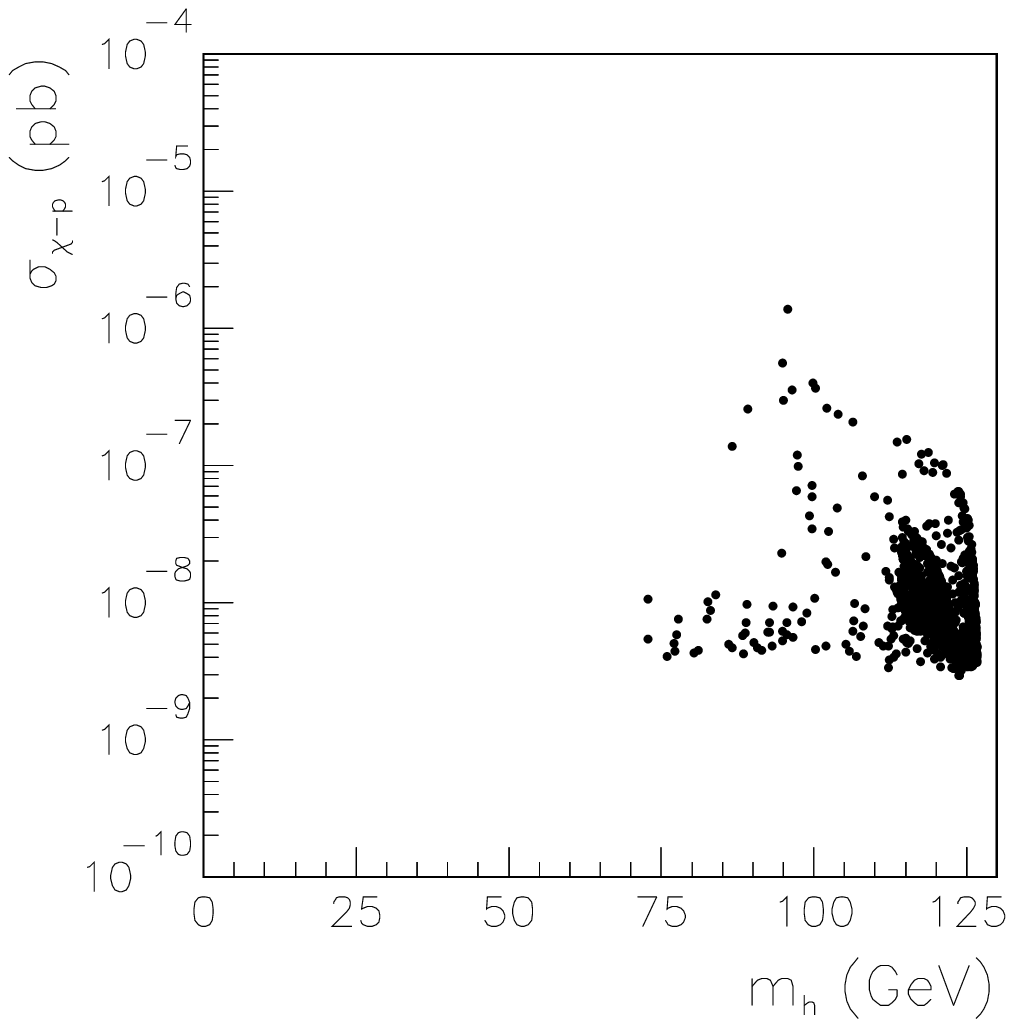,,height=8cm}
  \captions{The same as in Fig.\,\ref{summarya} but for negative
    values of $\kappa$.} 
    \label{summaryb}
\end{figure}

The results of this scan are shown in Figs.\,\ref{summarya} and
\ref{summaryb}, for positive and negative $\kappa$, respectively,
where the theoretical predictions for $\crosssec$ are represented as a
function of the lightest neutralino and scalar Higgs masses. 
Fig.\,\ref{summarya} therefore summarizes the results of
Sections\,\ref{++-}, \ref{+--}, and \ref{+++}. Points with
large predictions for $\crosssec$ are found. 
These correspond to very light
singlet-like Higgses, with even $m_{h_1^0}\gsim15$ GeV, 
which are more easily
obtained for low values of $\tan\beta$ ($\tan\beta\lsim5$). The points
in this region correspond to those with $\mu A_\lambda>0$, 
discussed in Sections\,\ref{++-} 
and \ref{+++}, respectively. The lightest neutralino in those points
has an important singlino composition, $N_{15}^2\lsim0.6$.

Fig.\,\ref{summaryb} generalizes the analysis of Section
\ref{k-+++}. As we had already discussed there, the lightest scalar
Higgs is heavier, $m_{h_1^0}\gsim75$ GeV, and although it can be
singlet-like in some cases, the predictions for the neutralino-nucleon
cross section are typically low. Regarding the neutralino, it is
Higgsino-like in most of the parameter space and therefore
$\neumass\approx\mu$.

\section{Conclusions}\label{conc}

We have performed a systematic analysis of the low-energy parameter space of the
Next-to-Minimal Supersymmetric Standard Model (NMSSM),
studying the implications for the direct detection of neutralino dark
matter. We have thus computed the theoretical predictions for the
scalar neutralino-proton cross section,
$\crosssec$,
and compared it with the sensitivities of present and projected dark
matter experiments.
In the computation we have taken into account 
all available experimental constraints from LEP on the
parameter space.

We have found that large values of $\crosssec$, even within the reach
of present dark matter 
detectors (see e.g. Fig.\,\ref{summarya}), can be obtained in regions of the
parameter space. 
This is essentially due to the exchange of very light Higgses,  
 $m_{h_1^0}\lsim 70$ GeV. 
The NMSSM nature is evidenced in this result, since such Higgses have 
a significant singlet composition, thus escaping detection and being in
agreement with accelerator data. In fact, Higgses as light as $15$~GeV
can be obtained.
The lightest neutralino in those regions exhibits a large
singlino-Higgsino composition, and a mass in the range
 $50\lsim \neumass\lsim 100$~GeV. 

Let us finally mention that our work can be considered as a first step
towards a more complete analysis of the direct detection of 
neutralino dark matter in the NMSSM. 
As discussed in 
Sections \ref{2:parameters} and \ref{res},
other potentially important constraints on the parameter space
should be addressed in forthcoming publications.
This is the case e.g. of the relic density,
the $b\rightarrow s\gamma$ branching ratio,
and the current upper limit on the decay 
$B_s \rightarrow \mu^+ \mu^-$.


\section*{Acknowledgements}\label{ack}
The work of D.G. Cerde\~no was supported in part by the Deutsche
Forschungsgemeinschaft, the DAAD, and the European Union under contract
HPRN-CT-2000-00148.
C. Hugonie was supported by the European Union RTN grant
HPRN-CT-2000-00148.
D.E. L\'opez-Fogliani acknowledges the financial support of the
Spanish DGI through a FPU grant, and also through the
Acci\'on Integrada Hispano-Alemana HA2002-0117.
The work of C. Mu\~noz was supported in part by the Spanish DGI
under Acci\'on Integrada Hispano-Alemana HA2002-0117, and
under contracts BFM2003-01266 and FPA2003-04597, and also
by the European Union under contract HPRN-CT-2000-00148.
A.M. Teixeira acknowledges the support by
Funda\c c\~ao para a Ci\^encia e Tecnologia under
the grant SFRH/BPD/11509/2002,
and also by the Spanish DGI
under Acci\'on Integrada Hispano-Alemana HA2002-0117 and
under contract BFM2003-01266.
D.G. Cerde\~no and A.M. Teixeira are also thankful to the organizers
of the
workshop ``Terrestrial and Cosmic Neutrinos, Leptogenesis and
Cosmology'' 
in the Benasque Center for Science
for their hospitality.

\appendix
\section{Relevant NMSSM interaction vertices}\label{app}

\subsection{Higgs-quark-quark Yukawa coupling}
Parameterising the interaction of neutral CP-even
Higgs fields with quarks as
\begin{equation}
\mathcal{L}_{qqh} = -
\bar q_i \left[ C^i_{YL} P_L + C^i_{YR} P_R \right] q_i \,h_a^0\,,
\end{equation}
where $h_a^0$ denotes the physical (mass) Higgs eigenstates
and $i=1,2$ up- and down-type quarks, one has
\begin{align}
C^i_{YL}= C^i_{YR}= C^i_Y\,; \quad
&C^{1}_{Y} = -\frac{g m_{u}}{2 M_W \sin \beta} S_{a2}\,,
\quad
C^{2}_{Y} = -\frac{g m_{d}}{2 M_W \cos \beta} S_{a1}\,,
\end{align}
where we have omitted the quark generations and
$S$ is the unitary matrix that diagonalises the scalar Higgs mass
matrix, defined in Eq.~(\ref{2:Smatrix}).

\subsection{Neutralino-neutralino-Higgs interaction}
The interaction of scalar Higgs and neutralinos can be parametrised as
\begin{equation}\label{L:HNN}
\mathcal{L}_{h \tilde \chi^0 \tilde \chi^0}
=\frac{1}{4} h_a^0  \bar{\tilde \chi}_{\alpha}^0
\left[ C_{HL}^{a\alpha\beta} P_L + C_{HR}^{a\alpha\beta} P_R \right] \tilde \chi_\beta^0\,,
\end{equation}
where $a=1-3$ refers to the Higgs mass eigenstate, $\alpha,\beta=1-5$
denote the physical neutralino states, and the couplings are defined as
\begin{align}\label{C:HNN}
C_{HL}^{a\alpha\beta} =&
\left\{ -g \left( N_{\alpha2}^*- \tan \theta_W N_{\alpha1}^* \right)
\left( S_{a1} N_{\beta3}^* -S_{a2} N_{\beta4}^*\right) + \right. \nonumber \\
&  \left.
+\sqrt{2} \lambda \left[ S_{a3} N_{\alpha3}^*  N_{\beta4}^* +
N_{\beta5}^* \left( S_{a2} N_{\alpha3}^* +S_{a1} N_{\alpha4}^* \right) \right]
+(\alpha\to \beta) \right\}  \nonumber \\
& - 2 \sqrt{2} \kappa S_{a3} N_{\alpha5}^* N_{\beta5}^* \, , \\
C_{HR}^{a\alpha\beta} =&\left(C_{HL}^{a\alpha\beta}\right)^* \,.
\end{align}
In the text, and since we have exclusively analysed interactions involving
the lightest neutralino states (i.e. $\alpha=\beta=1$), we have simplified the above
as $C_{HL}^{a11}=C_{HL}^{a}$ and $C_{HR}^{a11}=C_{HR}^{a}$.

\subsection{Neutralino-squark-quark interaction}
In terms of the mass eigenstates, the Lagrangian reads
\begin{align}
\mathcal{L}_{q \tilde q \tilde \chi^0} =
\bar q_i \left[ C^{\alpha Xi}_{L} P_L + C^{\alpha Xi}_{R} P_R \right]
\tilde \chi^0_\alpha \tilde q_{i}^{X}\,,
\end{align}
where $i=1,2$ denotes an up- or down-type quark and squark,
$X=1,2$ the squark mass eigenstates, and $\alpha=1,\ldots,5$ the neutralino states.
Since we have neglected flavour
violation in the squark sector, only $LR$ mixing occurs, and
squark physical and chiral eigenstates are related as
\begin{equation}
  \left(\begin{array}{c}
      \tilde q_1 \\ \tilde q_2
    \end{array}\right)
  \;=\;
  \left(\begin{array}{cc}
      \eta_{11}^{\tilde q}&\eta_{12}^{\tilde q}\\
      \eta_{21}^{\tilde q}&\eta_{22}^{\tilde q}
    \end{array}\right)\;
  \left(\begin{array}{c}
      \tilde q_L \\ \tilde q_R
    \end{array}\right)\,.
  \label{A:sqmatrix}
\end{equation}
One can also make the usual redefinition
$\eta^{\tilde q}_{11}=\eta^{\tilde q}_{22}= \cos \theta_{\tilde q}$ and
$\eta^{\tilde q}_{12}=-\eta^{\tilde q}_{21}= \sin \theta_{\tilde q}$
Therefore, for the up sector, and again omitting quark and squark
generation indices, the coefficients
$C^{\alpha Xi}_{L,R}$ are given by:
\begin{align}
C^{\alpha11}_{L} &=
-\sqrt{2} g \left[\frac{Y_{u}}{2}
\tan \theta_W N^*_{\alpha1} \sin \theta_{\tilde u}
+ \frac{m_{u}}{2 M_W \sin \beta} N^*_{\alpha4} \cos \theta_{\tilde u}
\right] \label{nsqq1l}\,,\\
C^{\alpha11}_{R} &=
-\sqrt{2} g \left\{ \left[
N^*_{\alpha2} T_3^u +\frac{Y_Q}{2} \tan \theta_W N^*_{\alpha1} \right] \cos
\theta_{\tilde u}
+ \frac{m_{u}}{2 M_W \sin \beta} N^*_{\alpha4} \sin \theta_{\tilde u}
\right\} \label{nsqq1r}\,,\\
C^{\alpha21}_{L} &=
-\sqrt{2} g \left[\frac{Y_{u}}{2}
\tan \theta_W N^*_{\alpha1} \cos \theta_{\tilde u}
- \frac{m_{u}}{2 M_W \sin \beta} N^*_{\alpha4} \sin \theta_{\tilde u}
\right] \label{nsqq2l}\,,\\
C^{\alpha21}_{R} &=
-\sqrt{2} g \left\{ \left[-
N^*_{\alpha2} T_3^u +\frac{Y_Q}{2} \tan \theta_W N^*_{\alpha1} \right] \sin
\theta_{\tilde u}
+ \frac{m_{u}}{2 M_W \sin \beta} N^*_{\alpha4} \cos \theta_{\tilde u}
\right\}\,.\label{nsqq2r}
\end{align}
In the above, $Y_{Q({u})}$ denotes the hypercharge of the
$SU(2)_L$ quark doublet (up-singlet) and $T_3^u$ the isospin of the
$u_L$ field.
The analogous for the down sector is trivially obtained by
the appropriate replacements (
$Y_u \to Y_d$, $T_3^u \to T_3^d$,
$m_u \to m_d$, $\theta_{\tilde u} \to \theta_{\tilde d}$,
 $\sin \beta \to \cos \beta$ and
$N_{\alpha4} \to N_{\alpha3}$).
In this paper, and since only $q - \tilde q - \tilde \chi^0_1$
interactions have been considered, we have always used
$C^{1Xi}_{L,R}= C^{Xi}_{L,R}$, i.e., setting $\alpha=1$ in the above.


\begin{thebibliography}{99}

\bibitem{lightreview} For a recent review see, C. Mu\~noz, `Dark matter
detection in the light of recent experimental results', to appear in {\it Int.
J. Mod. Phys.} {\bf A} [arXiv:hep-ph/0309346].

\bibitem{experimento1} DAMA Collaboration, R. Bernabei et al., `Search for WIMP
annual modulation signature: results from DAMA/NaI-3 and DAMA/NaI-4 and the
global combined analysis', {\it Phys. Lett.} {\bf B480} (2000) 23; `Dark matter
search', {\it Riv. N. Cim.} {\bf 26} (2003) 1 [arXiv:astro-ph/0307403].

\bibitem{halo} P. Belli, R. Cerulli, N. Fornengo and S. Scopel, `Effect of the
galactic halo modeling on the DAMA/NaI annual modulation result: an extended
analysis of the data for WIMPs with a purely spin-independent coupling', {\it
Phys. Rev.} {\bf D66} (2002) 043503 [arXiv:hep-ph/0203242].


\bibitem{soudan} CDMS Collaboration, D.S. Akerib et al., `First results from
the cryogenic dark matter search in the Soudan underground lab',
arXiv:hep-ph/0405033.





\bibitem{edelweiss} EDELWEISS Collaboration, A. Benoit et al., `First results
of the EDELWEISS WIMP search using a 320-g heat-and-ionization Ge detector',
{\it Phys. Lett.} {\bf B513} (2001) 15 [arXiv:astro-ph/0106094];
`Improved exclusion limits from the EDELWEISS WIMP
search', {\it Phys. Lett.} {\bf B545} (2002) 43 [arXiv:astro-ph/0206271].

\bibitem{ZEPLINI} ZEPLIN I Collaboration, N.J.T. Smith et al., talk
given at TAUP 2003, Seattle, Washington.


%
\bibitem{conflict} P. Ullio, M. Kamionkowski and P. Vogel,
`Spin-dependent WIMPs in DAMA?',
{\it J. High Energy Phys.} {\bf 07} (2001) 044 [arXiv:hep-ph/0010036];

\noindent D. Smith and N. Weiner, `Inelastic dark matter',
{\it Phys. Rev.} {\bf D64} (2001) 043502 [arXiv:hep-ph/0101138];
`Inelastic dark matter at DAMA, CDMS and future experiments',
talk given at DM 2002 Conference, Marina del Rey, California (2002),
arXiv:astro-ph/0208403;

\noindent R. Bernabei et al.,
`Investigating the DAMA annual modulation data in the framework
of inelastic dark matter',
{\it Eur. Phys. J.} {\bf C23} (2002) 61;

\noindent G. Prezeau, A. Kurylov, M. Kamionkowski and P. Vogel,
`New contribution to WIMP-nucleus scattering', 
{\it Phys. Rev. Lett.} {\bf 91} (2003) 231301
[arXiV:astro-ph/0309115];

\noindent C. Savage, P. Gondolo and K. Freese,
`Can WIMP spin dependent couplings explain DAMA data, in the light
of null results from other experiments?',
arXiv:astro-ph/0408346.



\bibitem{IGEX3} A. Morales, `Searching for WIMP dark matter: the case for
Germanium ionization detectors', talk given at the 29th International Meeting
on Fundamental Physics, Sitges, Barcelona (2001), arXiv:hep-ex/0111089.

\bibitem{HDMS2} GENIUS Collaboration, H.V. Klapdor-Kleingrothaus et al.,
`GENIUS - a Supersensitive Germanium Detector System for Rare Events',
arXiv:hep-ph/9910205.

\bibitem{xenon} See e.g., E. Aprile et al.,
`The XENON dark matter search experiment',
arXiv:astro-ph/0407575.

\bibitem{old} H. Goldberg, `Constraint on the photino mass from cosmology',
{\it Phys. Rev. Lett.} {\bf 50} (1983) 1419;\\
%
\noindent J. Ellis, J.S. Hagelin, D.V. Nanopoulos and M. Srednicki, `Search for
supersymmetry at the {$\bar p$}$p$ collider', {\it Phys. Lett.} {\bf B127}
(1983) 233;\\
%
\noindent L.M. Krauss `New constraints on "INO" masses from cosmology (I).
Supersymmetric "inos"', {\it Nucl. Phys.} {\bf B227} (1983) 556;\\
%
\noindent J. Ellis, J.S. Hagelin, D.V. Nanopoulos, K.A. Olive and M. Srednicki,
`Supersymmetric relics from the Big Bang', {\it Nucl. Phys.} {\bf B238} (1984)
453.

\bibitem{mupb} J.E. Kim and H.P. Nilles, `The $\mu$ problem and the strong CP
problem' {\it Phys. Lett.} {\bf B138} (1984) 150.

\bibitem{musol} G.F. Giudice and A. Masiero, `A natural solution to the $\mu$
problem in supergravity theories', {\it Phys. Lett.} {\bf B206} (1988) 480;\\
%
\noindent J.E. Kim and H.P. Nilles, Gaugino condensation and the cosmological
implications of the hidden sector', {\it Phys. Lett.} {\bf B263} (1991) 79;\\
%
\noindent E.J. Chun, J.E. Kim and H.P. Nilles, `A natural solution of the $\mu$
problem with a composite axion in the hidden sector', {\it Nucl. Phys.} {\bf
B370} (1992) 105;\\
%
\noindent J.A. Casas and C. Mu\~noz, `A natural solution to the $\mu$ problem',
{\it Phys. Lett.} {\bf B306} (1993) 288 [arXiv:hep-ph/9302227];\\
%
\noindent G. Lopes-Cardoso, D. L\"ust and T. Mohaupt, `Moduli spaces and target
space duality symmetries in (0,2) $Z_n$ orbifold theories with continuous
Wilson lines', {\it Nucl. Phys.} {\bf B432} (1994) 68 [arXiv:hep-th/9405002];\\
%
\noindent I. Antoniadis, E. Gava, K.S. Narain and T.R. Taylor, `Effective $\mu$
term in superstring theory', {\it Nucl. Phys.} {\bf B432} (1994) 187
[arXiv:hep-th/9405024];\\
%
\noindent A. Brignole, L.E. Ib\'a\~nez and C. Mu\~noz, `Orbifold-induced $\mu$
term and electroweak symmetry breaking', {\it Phys. Lett.} {\bf B387} (1996)
769 [arXiv:hep-ph/9607405];\\
%
\noindent K. Choi, J.S. Lee and C. Mu\~noz, `Supergravity radiative effects on
soft terms and the $\mu$ term', {\it Phys. Rev. Lett.} {\bf 80} (1998) 3686
[arXiv:hep-ph/9709250].

\bibitem{NMSSM1} H.~P.~Nilles, M.~Srednicki and D.~Wyler, `Weak interaction
breakdown induced by supergravity', {\it Phys.\ Lett.\ } {\bf B120} (1983)
346;\\
%
J.~M.~Frere, D.~R.~T.~Jones and S.~Raby, `Fermion masses and induction of the
weak scale by supergravity', {\it Nucl.\ Phys.\ } {\bf B222} (1983) 11;\\
%
J.~P.~Derendinger and C.~A.~Savoy, `Quantum effects and SU(2) X U(1) breaking
in supergravity gauge theories', {\it Nucl.\ Phys.\ } {\bf B237} (1984) 307.

\bibitem{NMSSM2} J.~R.~Ellis, J.~F.~Gunion, H.~E.~Haber, L.~Roszkowski and F.~Zwirner,
`Higgs bosons in a nonminimal supersymmetric model',
{\it Phys.\ Rev.\ } {\bf D39} (1989) 844;\\
%
\noindent M.~Drees, `Supersymmetric models with extended Higgs sector', {\it
Int.\ J.\ Mod.\ Phys.\ } {\bf A4} (1989) 3635;\\
%
\noindent U.~Ellwanger, M.~Rausch de Traubenberg and C.~A.~Savoy, `Particle
spectrum in supersymmetric models with a gauge singlet', {\it Phys.\ Lett.\ }
{\bf B315} (1993) 331 [arXiv:hep-ph/9307322]; `Phenomenology of supersymmetric
models with a singlet', {\it Nucl.\ Phys.\ } {\bf B492} (1997) 21
[arXiv:hep-ph/9611251];\\
%
\noindent S.~F.~King and P.~L.~White, `Resolving the constrained minimal and
next-to-minimal supersymmetric standard models', {\it Phys.\ Rev.\ } {\bf D52}
(1995) 4183 [arXiv:hep-ph/9505326].

\bibitem{Bast} M.~Bastero-Gil, C.~Hugonie, S.~F.~King, D.~P.~Roy and
S.~Vempati, `Does LEP prefer the NMSSM?', {\it Phys.\ Lett.\ } {\bf B489}
(2000) 359 [arXiv:hep-ph/0006198].

\bibitem{Abel1} S.~A.~Abel, S.~Sarkar and P.~L.~White, `On the cosmological
domain wall problem for the minimally extended supersymmetric standard model',
{\it Nucl.\ Phys.\ } {\bf B454} (1995) 663 [arXiv:hep-ph/9506359].

\bibitem{Abel2} S.~A.~Abel, `Destabilising divergences in the NMSSM', {\it
Nucl.\ Phys.\ } {\bf B480} (1996) 55 [arXiv:hep-ph/9609323];\\
%
\noindent C.~Panagiotakopoulos and K.~Tamvakis, `Stabilized NMSSM without domain
walls', {\it Phys.\ Lett.\ } {\bf B446} (1999) 224 [arXiv:hep-ph/9809475].

\bibitem{NLEP} U.~Ellwanger and C.~Hugonie, `Topologies of the (M+1)SSM with a
singlino LSP at LEP2', {\it Eur.\ Phys.\ J.\ } {\bf C13} (2000) 681
[arXiv:hep-ph/9812427].

\bibitem{NHIGGS} U.~Ellwanger and C.~Hugonie, `Masses and couplings of the
lightest Higgs bosons in the (M+1)SSM', {\it Eur.\ Phys.\ J.\ } {\bf C25}
(2002) 297 [arXiv:hep-ph/9909260].

\bibitem{NMHDECAY} U.~Ellwanger, J.~F.~Gunion and C.~Hugonie, `NMHDECAY: A
Fortran code for the Higgs masses, couplings and decay widths in the NMSSM',
arXiv:hep-ph/0406215.

\bibitem{NLHC} U.~Ellwanger, J.~F.~Gunion, C.~Hugonie and S.~Moretti, `NMSSM
Higgs discovery at the LHC', arXiv:hep-ph/0401228.

\bibitem{Ndirdet} R. Flores, K.A. Olive and D. Thomas, `Light-neutralino
interactions in matter in an extended supersymmetric standard model', {\it
Phys. Lett.} {\bf B263} (1991) 425.
%
\bibitem{bk} V.A. Bednyakov and H.V. Klapdor-Kleingrothaus, `About direct dark
matter detection in next-to-minimal supersymmetric standard model', {\it Phys.
Rev.} {\bf D59} (1999) 023514 [arXiv:hep-ph/9802344].

\bibitem{Nrelden} B.R. Greene and P.J. Miron, `Supersymmetric cosmology with a
gauge singlet', {\it Phys. Lett.} {\bf B168} (1986) 226;\\
%
\noindent R. Flores, K.A. Olive and D. Thomas, `A new dark matter candidate in
the minimal extension of the supersymmetric standard model', {\it Phys. Lett.}
{\bf B245} (1990) 509;\\
%
\noindent K.A. Olive and D. Thomas, `A light dark matter candidate in an
extended supersymmetric model', {\it Nucl. Phys.} {\bf B355} (1991) 192;\\
%
\noindent S.A. Abel, S. Sarkar and I.B. Whittingham, `Neutralino dark matter in
a class of unified theories', {\it Nucl. Phys.} {\bf B392} (1993) 83
[arXiv:hep-ph/9209292];\\
%
\noindent A. Stephan, `Dark matter constraints on the parameter space and
particle spectra in the nonminimal SUSY standard model', {\it Phys. Lett.} {\bf
B411} (1997) 97 [arXiv:hep-ph/9704232]; `Dark matter constraints in the minimal
and nonminimal SUSY standard model', {\it Phys. Rev.} {\bf D58} (1998) 035011
[arXiv:hep-ph/9709262];\\
%
\noindent A.~Menon, D.~E.~Morrissey and C.~E.~M.~Wagner, `Electroweak
baryogenesis and dark matter in the nMSSM', arXiv:hep-ph/0404184.

\bibitem{tadp} H.~P.~Nilles, M.~Srednicki and D.~Wyler, `Constraints on the
stability of mass hierarchies in supergravity', {\it Phys.\ Lett.\ } {\bf B124}
(1983) 337;\\
%
\noindent U.~Ellwanger, `Nonrenormalizable interactions from supergravity,
quantum corrections and effective low-energy theories', {\it Phys.\ Lett.\ }
{\bf B133} (1983) 187;\\
%
\noindent J.~Bagger and E.~Poppitz, `Destabilizing divergences in supergravity
coupled supersymmetric theories', {\it Phys.\ Rev.\ Lett.\ } {\bf 71} (1993)
2380 [arXiv:hep-ph/9307317];\\
%
\noindent J.~Bagger, E.~Poppitz and L.~Randall, `Destabilizing divergences in
supergravity theories at two loops', {\it Nucl.\ Phys.\ } {\bf B455} (1995) 59
[arXiv:hep-ph/9505244].

\bibitem{Romao:jy} J.~C.~Romao, 'Spontaneous CP violation in susy models: a no
go theorem', {\it Phys.\ Lett.\ } {\bf B173} (1986) 309.

\bibitem{LEPneu} K.~Hagiwara et al., {\it Phys. Rev.} {\bf D66} (2002)
010001.

\bibitem{LEPneu2} DELPHI Collaboration,
J.~Abdallah et al., `Searches
for supersymmetric particles in e+ e- collisions up to 208-GeV and
interpretation of the results within the MSSM', {\it Eur. Phys. J.} {\bf C31}
(2004) 421 [arXiv:hep-ex/0311019];

OPAL Collaboration,
G.~Abbiendi {\it et al.}, `Search for chargino and
neutralino production at s**(1/2) = 192-GeV - 209-GeV at LEP',
{\it Eur. Phys. J.} {\bf C35} (2004) 1 [arXiv:hep-ex/0401026].

\bibitem{LEPchar} LEP \ SUSY\ Working Group, LEPSUSYWG Note/02-04.1.

\bibitem{LEPHC} LEP Higgs Working Group for Higgs boson searches Collaboration,
`Search for charged Higgs bosons: Preliminary combined results using LEP data
collected at energies up to 209-GeV', arXiv:hep-ex/0107031.

\bibitem{LEPH} LEP Higgs Working Group for Higgs boson
searches Collaboration, R.~Barate et al.,
`Search for the standard model Higgs boson at LEP',
{\it Phys. Lett.} {\bf B565} (2003) 61 [arXiv:hep-ex/0306033];\\
%
\noindent OPAL Collaboration,
G.~Abbiendi et al., `Flavour
independent search for Higgs bosons decaying into hadronic final states in e+
e- collisions at LEP', arXiv:hep-ex/0312042;
`Decay-mode
independent searches for new scalar bosons with the OPAL  detector at LEP',
{\it Eur. Phys. J.}  {\bf C27} (2003) 311 [arXiv:hep-ex/0206022];\\
%
\noindent LEP Higgs Working Group for Higgs boson searches Collaboration,
`Flavor independent search for hadronically decaying neutral Higgs  bosons at
LEP', arXiv:hep-ex/0107034;
`Searches for Higgs Bosons Decaying into Photons: Combined Results from the
LEP Experiments' LHWG Note/2002-02;
`Searches
for invisible Higgs bosons: Preliminary combined results using  LEP data
collected at energies up to 209-GeV', arXiv:hep-ex/0107032;

%
\noindent ALEPH Collaboration,
D.~Buskulic et al., `Search for a
nonminimal Higgs boson produced in the reaction e+ e- $\to$ h Z*',
{\it Phys. Lett.} {\bf B313} (1993) 312;

\noindent DELPHI Collaboration, `Search for Neutral Higgs Bosons in Extended
Models' CERN EP 2003-061 (Submitted to EPJ).


\bibitem{Gudrun} G. Hiller,
`$b$-physics signals of the lightest CP-odd Higgs in the
NMSSM at large $\tan\beta$.',
{\it Phys. Rev.} {\bf D70} (2004) 034018 
[arXiv:hep-ph/0404220]. 
%


\bibitem{ko} S. Baek, Y.~G. Kim and P. Ko,
`Neutralino dark matter scattering and  $B_s \rightarrow  \mu^+ \mu^-$
in SUSY models',
arXiv:hep-ph/0406033. 
%



\bibitem{forthc}
D.G. Cerde\~no, E. Gabrielli, D.~E.~L\'opez-Fogliani,
C.~Mu\~noz and A.~M.~Teixeira, in preparation.


\bibitem{Jungman:1995df} For a review, see G.~Jungman, M.~Kamionkowski and
K.~Griest, `Supersymmetric dark matter', {\it Phys.\ Rept.\ } {\bf 267} (1996)
195 [arXiv:hep-ph/9506380].

\bibitem{Falk:1999mq} For a summary, see also: U.~Chattopadhyay, T.~Ibrahim and
P.~Nath, `Effects of CP violation on event rates in the direct detection of
dark matter', {\it Phys.\ Rev.\ } {\bf D60} (1999) 063505
[arXiv:hep-ph/9811362];\\
%
\noindent T.~Falk, A.~Ferstl and K.~A.~Olive, `Variations of the neutralino
elastic cross-section with CP violating phases', {\it Astropart.\ Phys.\ } {\bf
13} (2000) 301 [arXiv:hep-ph/9908311].

\bibitem{Ellis:2000ds} J.~R.~Ellis, A.~Ferstl and K.~A.~Olive, `Re-evaluation
of the elastic scattering of supersymmetric dark matter', {\it Phys. Lett.}
{\bf B481} (2000) 304 [arXiv:hep-ph/0001005].



\bibitem{Rosz} See e.g., Y.G. Kim, T. Nihei, L. Roszkowski and R. Ruiz
  de Austri, `Upper and lower limits on neutralino 
WIMP mass and spin-independent scattering cross section, 
and impact of new $(g-2)_{\mu}$ measurement',
{\it J. High Energy Phys.} {\bf 10} (2004) 015 [arXiv:hep-ph/0208069].


\bibitem{cermu} See e.g., D.~G.~Cerde\~no and C. Mu\~noz, `Neutralino dark
matter in supergravity theories with non-universal scalar and gaugino
masses',
{\it J. High Energy Phys.} {\bf 07} (2002) 024 
[arXiv:hep-ph/0405057], and references therein.

\bibitem{omega} D.~G.~Cerde\~no, M.~G\'omez,
C.~Hugonie, D.~E.~L\'opez-Fogliani,
C.~Mu\~noz and A.~M.~Teixeira, in preparation.

\bibitem{wmap} WMAP Collaboration,
C.L. Bennett et al., `First year Wilkinson microwave anisotropy
probe (WMAP) observations: preliminary maps and basic results', {\it Astrophys.
J. Suppl.} {\bf 148} (2003) 1 [arXiv:astro-ph/0302207];\\
%
\noindent D.N. Spergel et al., `First year Wilkinson microwave anisotropy probe
(WMAP) observations: determination of cosmological parameters', {\it Astrophys.
J. Suppl.} {\bf 148} (2003) 175 [arXiv:astro-ph/0302209].

\bibitem{omega2} G. Belanger and C. Hugonie, in preparation.

\end{thebibliography}
\end{document}